%% file: main.tex
\documentclass[journal]{IEEEtran}

\input{config/packages}

\begin{document}

\bstctlcite{IEEEexample:BSTcontrol}

\input{config/title-page}

\input{common/abstract}



\input{secs/00-introduction}
\input{secs/02-foundations}
\input{secs/03-ai-for-iscc}
\input{secs/04-issc-for-ai}
\input{secs/05-virtuous-cycle}
\input{secs/06-synthesis}
\input{secs/07-applications}
\input{secs/08-future}
\input{secs/09-conclusion}

\InputIfFileExists{common/acknowledgment.tex}{}{}

\balance

\bibliographystyle{IEEEtran}



\bibliography{
    references/BSTcontrol,
    references/introduction_cleaned,
    references/related_surveys_cleaned,
    references/taxonomy_cleaned,
    references/ai-for-iscc_cleaned,
    references/iscc-for-ai_cleaned,
    references/applications_cleaned,
    references/future_cleaned,
    references/references_cleaned,
    references/gscholar-alert_cleaned,
    references/new,
    references/ref,
    references/f03_coverage
}




\end{document}

%% file: config/packages.tex
\usepackage{amsmath,amsfonts}
\usepackage{array}
\usepackage{multirow} 
\usepackage{textcomp}
\usepackage{stfloats}
\usepackage{url}
\usepackage{hyperref}
\usepackage{verbatim}
\usepackage{graphicx}
\usepackage{booktabs}
\usepackage{cite}
\usepackage{amsthm}

\usepackage{soul}
\usepackage[linesnumbered,ruled,vlined]{algorithm2e}
\usepackage{balance}

\usepackage{booktabs} 
\usepackage{xcolor} 
\usepackage{amssymb}
\usepackage{booktabs}
\usepackage{pifont}

\usepackage{tabularx}  
\usepackage{booktabs}  
\usepackage{enumitem}  

\usepackage{pdflscape}

\usepackage{pifont} 
\newcommand{\cmark}{\ding{51}}  
\newcommand{\xmark}{\ding{55}}  
\newcommand{\pmark}{$\sim$}     

\newcolumntype{Y}[1]{%
    >{\hsize=#1\hsize\linewidth=\hsize\raggedright\arraybackslash}X%
}

%% file: config/title-page.tex
\title{AI-Aided and AI-Oriented Integrated Sensing, Communication, and Computation: A Survey}


\author{Yaya Etiabi,~\IEEEmembership{Member,~IEEE},
Wessam Ajib,~\IEEEmembership{Senior Member,~IEEE},\\
El Mehdi Amhoud,~\IEEEmembership{Senior Member,~IEEE},
Halima Elbiaze,~\IEEEmembership{Senior Member,~IEEE}



\thanks{ Yaya Etiabi, Wessam Ajib, and Halima Elbiaze are with Department of Computer Science, Université du Québec à Montréal (UQAM), Montréal, Canada (e-mails: \{etiabi.yaya, ajib.wessam, elbiaze.halima\}@uqam.ca).}
\thanks{ El Mehdi Amhoud is with College of Computing, Mohammed VI Polytechnic University, Benguerir, Morocco (e-mail: elmehdi.amhoud@um6p.ma).}

\thanks{This work has been submitted to the IEEE for possible publication. Copyright may be transferred without notice, after which this version may no longer be accessible.}

}

\maketitle

%% file: common/abstract.tex
\begin{abstract}

The evolution from 5G to 6G networks marks a transition toward the ``intelligence of everything.'' While 5G enables enhanced connectivity, 6G aspires to embed intelligence directly into communication infrastructures, creating systems that are adaptive, context-aware, and capable of autonomous decision-making. Realizing such systems depends on harmonizing three fundamental functions that compete for shared network resources and often pursue conflicting goals: (i) sensing for environmental and contextual information acquisition, (ii) communication for reliable and timely information exchange, and (iii) computation for information processing and decision-making. The interdependencies among these functions motivate their joint design as a unified system, giving rise to the paradigm of Integrated Sensing, Communication, and Computation (ISCC).
Within this paradigm, artificial intelligence (AI) has emerged both as an enabling tool for ISCC optimization and as a workload whose requirements influence ISCC system design. This survey examines ISCC from this dual perspective by distinguishing between \emph{AI-aided ISCC}, in which learning-based methods support the joint design and optimization of sensing, communication, and computation, and \emph{AI-oriented ISCC}, in which these functions are jointly configured to support learning, inference, and distributed intelligence. Building on these two perspectives, the survey explores the co-design of AI and ISCC, highlighting their reciprocal dependencies and the potential for a virtuous cycle of mutual performance enhancement. The survey further provides a comparative synthesis of system architectures, performance metrics, and evaluation practices, examines representative application domains, identifies open research problems, and discusses emerging research directions.

\end{abstract}


%% file: secs/00-introduction.tex
\section{Introduction}
\label{sec:introduction}

\IEEEPARstart{T}{he} evolution of wireless networks beyond the fifth generation (5G) is driven in part by the convergence of sensing, communication, and computation functions into unified architectural frameworks. The sixth generation (6G) of wireless systems aims to bridge the physical and digital worlds by integrating environmental awareness, information exchange, and intelligent processing within a common infrastructure~\cite{Wang2023vision,saad2020vision}. This convergence opens up opportunities to exploit synergies among functions through the joint design and allocation of spectrum, hardware, and computational resources~\cite{dang2020six}.

The International Mobile Telecommunications framework for 2030 (IMT-2030)~\cite{itu2030framework} reflects this evolution. Alongside the evolution of 5G's enhanced mobile broadband (eMBB), massive machine-type communication (mMTC), and ultra-reliable low-latency communication (URLLC) scenarios, the framework identifies three new usage scenarios: integrated sensing and communication (ISAC), artificial intelligence (AI) and communication, and ubiquitous connectivity. The envisioned capabilities include peak data rates of up to 200~Gbit/s, radio-interface latency as low as 0.1~ms, reliability of up to $1-10^{-7}$, and centimeter-level positioning accuracy, depending on the usage scenario and deployment conditions.

At the intersection of these developments lies the paradigm of Integrated Sensing, Communication, and Computation (ISCC)~\cite{ISCC}. While ISAC brings together sensing and communication, ISCC extends this integration to include computation as a first-class function~\cite{Sun2024,Deng2024}. Fig.~\ref{fig:evolution_ai_iscc} traces this progression
from function-specific wireless systems to AI--ISCC co-design. The joint design of the three functions is particularly relevant to applications such as autonomous cyber-physical systems, large-scale digital twins, and immersive extended reality, whose performance depends on the acquisition, exchange, and processing of information within the wireless infrastructure~\cite{Letaief2022}.

\subsection{The Core Challenge: The ISCC Trilemma}
\label{subsec:challenge}

\begin{figure*}[!t]
    \centering
    \includegraphics[width=\textwidth]{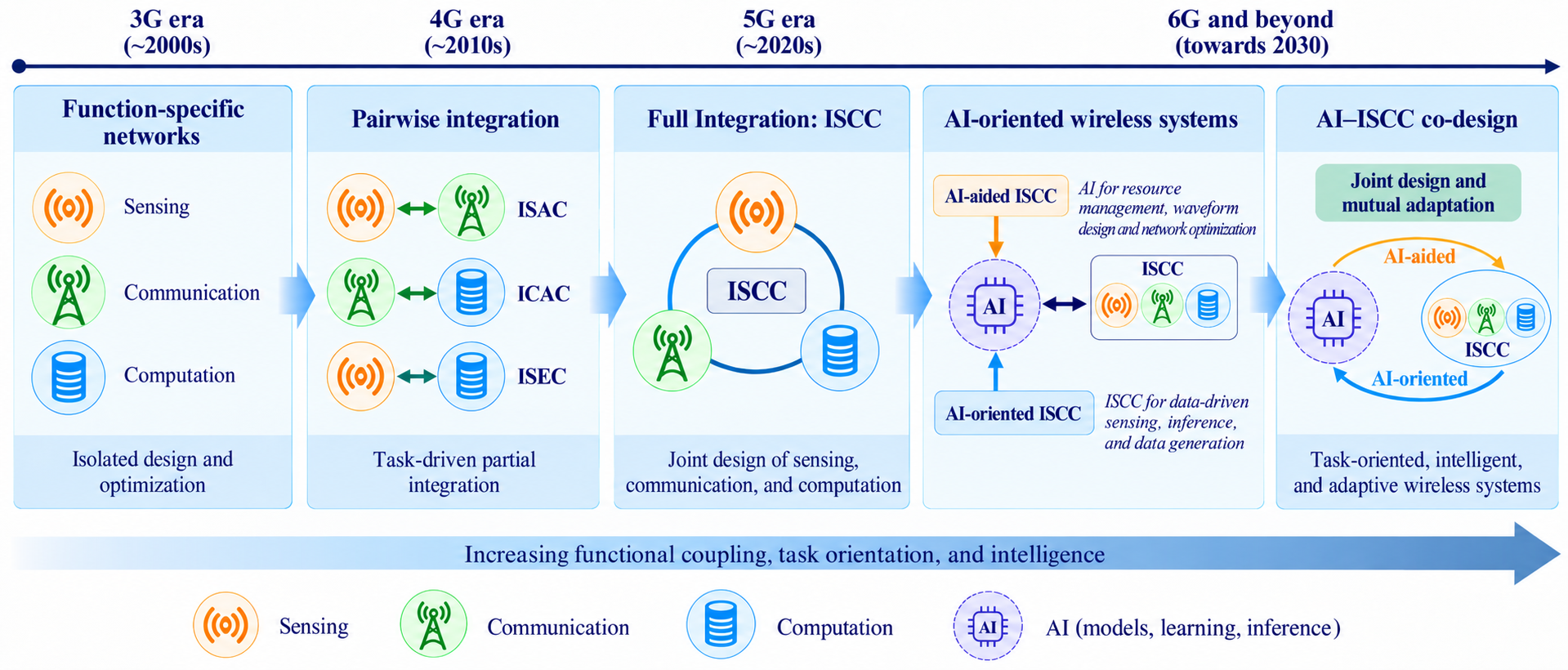}
    \caption{Conceptual progression from function-specific wireless systems toward AI--ISCC co-design. Pairwise integration includes integrated sensing and computation (ISEC), with roots in participatory sensing~\cite{mcs_01_2006,mcs_02_2011}; integrated communication and computation (ICAC), exemplified by over-the-air computation~\cite{aircomp_00_2007} and mobile edge computing~\cite{mec_00_2014}; and integrated sensing and communication (ISAC), exemplified by joint waveform design~\cite{isac_00_2011}. The integration of all three functions is articulated in~\cite{ISCC}. The periods indicate broad technological trends rather than the dates of introduction or standardization of these paradigms.}
    \label{fig:evolution_ai_iscc}
\end{figure*}

A fundamental challenge in ISCC design lies in managing the tradeoffs between its three constituent functions. This can be conceptualized as the \textit{ISCC Trilemma}, in which improving the performance of one function may require compromises in one or both of the others. These tradeoffs arise partly from competition for finite resources, primarily spectrum, time, energy, and hardware~\cite{S01}. At the physical layer, for example, the requirements of reliable and spectrally efficient data transmission must be reconciled with those of radar sensing, including delay--Doppler resolution and sidelobe levels for target detection and parameter estimation~\cite{liu2022integrated}. At higher layers, the decision to extract information locally from sensed data or offload its processing introduces a tradeoff between local computation expenditure and the resources required for wireless transmission.

The design of ISCC systems consequently involves coupled performance objectives and resource constraints, which often give rise to complex multi-objective optimization (MOO) problems~\cite{Ren2025}. The need to solve these problems as wireless channels, sensing environments, and task requirements evolve can make repeated model-based optimization computationally demanding~\cite{cheng2026d3pg_light_iscc}. This motivates the use of AI to support ISCC design and control. 
At the same time, emerging AI workloads impose new requirements on the acquisition, exchange, and processing of information, shaping the evolution of ISCC systems (Fig.~\ref{fig:evolution_ai_iscc}). These developments give rise to a dual relationship between AI and ISCC.

\subsection{The Dual Role of AI in ISCC Systems}
\label{subsec:dual_role}

AI plays two complementary and non-exclusive roles in ISCC systems.

\subsubsection{AI-Aided ISCC (Intelligence as Enabler)}

AI-aided ISCC addresses the complexity of joint sensing, communication, and computation optimization, particularly under time-varying channels, heterogeneous devices, and dynamic task requirements~\cite{2024Dong,Ren2025}. Learning-based techniques can provide computationally efficient approximations to conventional optimization or complement model-based methods when making resource allocation and control decisions~\cite{Yang202418608_PSS-Scopus-20250825-151225_0131,Cai202413584_PSS-Scopus-20250825-151225_0190}. Advances in pairwise integration also offer enabling mechanisms for AI-aided ISCC, including model-driven learning for joint ISAC waveform and beamforming design, reinforcement learning for channel sampling, and predictive beamforming based on sensory observations~\cite{Jiang2025,Mason2025,Wang2024}. Transfer and meta-learning provide complementary mechanisms for adapting sensing and inference models across environments and tasks~\cite{ahmed2024unified,yet-10628047}. The central concern of AI-aided ISCC is how learning can improve the design and operation of the integrated system.

\subsubsection{AI-Oriented ISCC (Intelligence as Workload)}

AI-oriented ISCC considers AI applications as workloads whose execution relies on the coordination of sensing, communication, and computation resources~\cite{Letaief2022}. Whereas conventional federated learning (FL) couples local model training with wireless model exchange~\cite{yang2019federated,chen2024edge}, sensing-driven federated learning also brings the acquisition of training data into the system design problem~\cite{wen2026airfeel}. Related dependencies arise in distributed and split inference, where the execution of a model is coupled with the transmission of intermediate representations~\cite{Yao2025_PSS-Scopus-20250825-151225_0072,wu2025neuromorphic_split}. In collaborative perception, multi-task inference, and intelligent industrial control, these dependencies extend to the selection and coordinated processing of observations from multiple devices~\cite{Dong20258481_PSS-Scopus-20250825-151225_0013,wang2026decentralized_multitask_iscc,tang2025seco4}. Across these workloads, the design of ISCC systems must account for how resource allocation affects learning, inference, or control performance.

\subsubsection{The Co-Design of AI and ISCC: The Virtuous Cycle}
\label{subsec:ai_iscc_codesign}

The interplay between these two roles motivates the co-design of AI and ISCC. Task requirements and ISCC constraints jointly shape decisions concerning both the integrated system and the AI workload, including its model configuration and execution strategy~\cite{Yao2025_PSS-Scopus-20250825-151225_0072,Jiao202519767_PSS-Scopus-20250825-151225_0064}. In dynamic environments, the resulting task and system performance can also inform subsequent configurations. This feedback provides a basis for the AI--ISCC virtuous cycle, in which improvements in AI-task performance and ISCC operation can reinforce one another. The mechanisms and conditions underlying this interaction are examined in Section~\ref{sec:codesign}.

\subsection{Motivation and Focus of this Survey}

The growing body of research on AI and ISCC spans different levels of integration, from learning-based resource management to the joint configuration of AI workloads and wireless systems. Although these problems share underlying functional dependencies, differences in system models, design objectives, and performance criteria make their results difficult to interpret within a common analytical framework.

This survey addresses these dependencies through the complementary perspectives of AI-aided ISCC, AI-oriented ISCC, and AI--ISCC co-design. The literature was initially identified through Scopus and subsequently expanded through targeted searches in IEEE Xplore and the Association for Computing Machinery (ACM) Digital Library, with coverage extending to August 2026,
supplemented by more recent contributions. The review primarily focuses on AI-enabled ISCC optimization, ISCC design for AI workloads, and the reciprocal dependencies between AI and ISCC. It also draws on advances in pairwise integration that underpin the foundations and enabling mechanisms of ISCC.

\input{secs/01-related-surveys}

\subsection{Contributions}
\label{subsec:contributions}

The main contributions of this survey are summarized as follows.
\begin{itemize}
    \item \textit{A unified analysis of AI-aided and AI-oriented ISCC.} We organize the literature according to the role of AI in integrated wireless systems and examine the design problems, optimization strategies, and workload-specific requirements underlying each perspective.
    \item \textit{An analysis of the co-design of AI and ISCC.} Building on these two perspectives, we examine how reciprocal dependencies between AI requirements and ISCC constraints expand the design space, and distinguish joint optimization, state-dependent adaptation, and feedback mechanisms that can give rise to the virtuous cycle.
    \item \textit{A comparative synthesis of architectures, performance evaluation, and experimental evidence.} We compare system architectures and the metrics, datasets, benchmarks, and experimental practices used to evaluate them, identifying the factors that influence the interpretation and comparability of reported results.
    \item \textit{An analysis of application domains and future research directions.} We examine how application-specific requirements shape AI--ISCC design and identify unresolved research problems and emerging directions for integrated wireless systems.
\end{itemize}

\subsection{Survey Organization}

\begin{figure*}[!t]
    \centering
    \includegraphics[width=\textwidth]{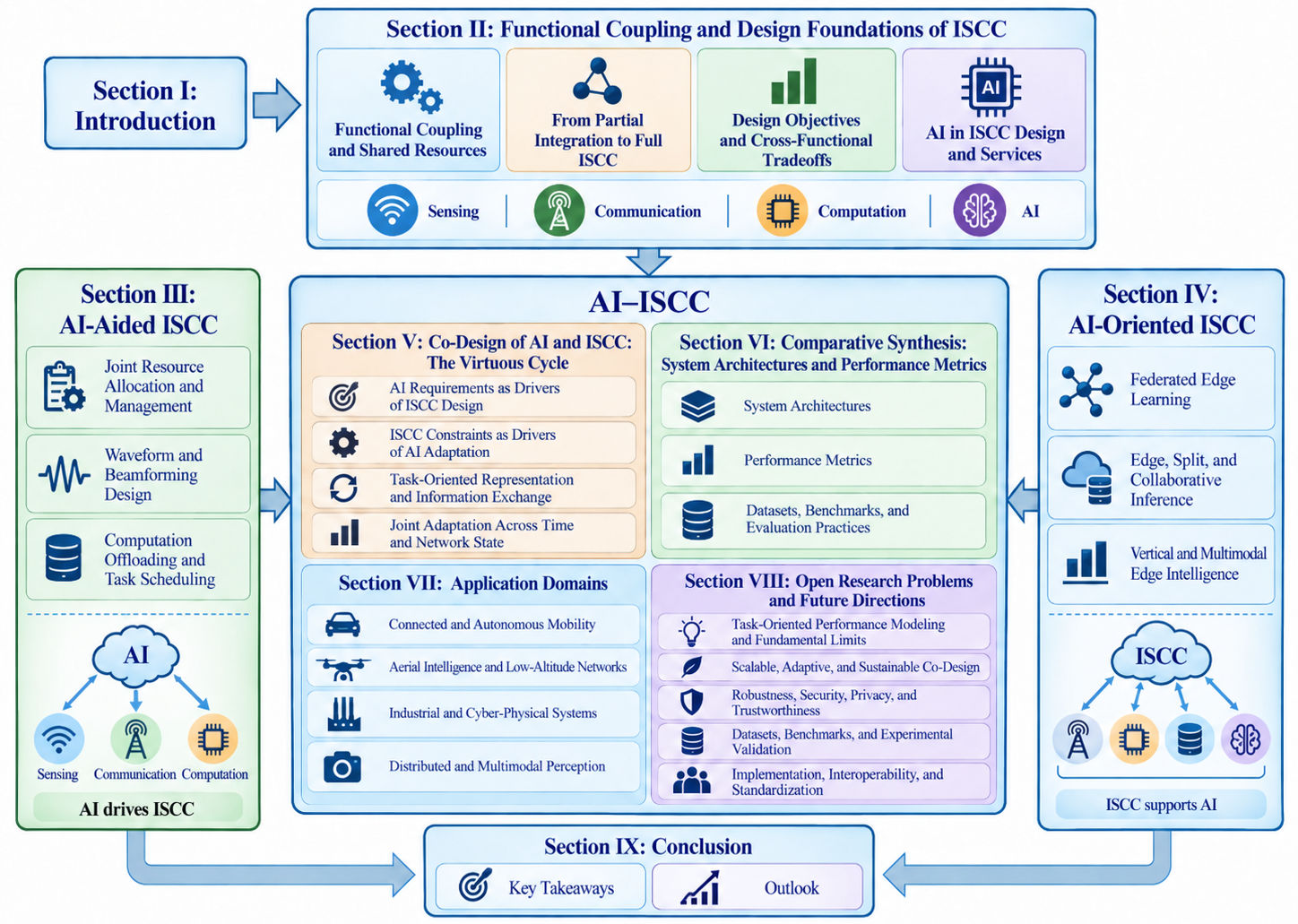}
    \caption{Organization and structure of the survey.}
    \label{fig:survey_organization}
\end{figure*}

The remainder of this survey is organized as illustrated in Fig.~\ref{fig:survey_organization}.

Section~\ref{sec:iscc_foundations} establishes the functional coupling and design foundations of ISCC. Sections~\ref{sec:ai-for-iscc} and~\ref{sec:iscc-for-ai} examine AI-aided and AI-oriented ISCC, respectively, before Section~\ref{sec:codesign} develops their interplay through co-design. Section~\ref{sec:comparative_synthesis} compares system architectures and evaluation practices, and Section~\ref{sec:applications} relates these developments to application-specific requirements. Section~\ref{sec:future} discusses open research problems and future directions, and Section~\ref{sec:conclusion} concludes the survey.

%% file: secs/01-related-surveys.tex
\subsection{Related Surveys}
\label{sec:related}

\begin{table*}[!t]
\caption{Comparison of Related Surveys on ISCC and AI Integration}
\label{tab:survey_comparison}
\centering
\renewcommand{\arraystretch}{1.15}
\scriptsize
\begin{tabular}{|p{1.0cm}|c|p{2.5cm}|p{6.8cm}|c|c|c|c|c|c|c|}
\hline
\textbf{Ref.} & \textbf{Year} & \textbf{Primary Focus} & \textbf{Coverage} &
\rotatebox{90}{\textbf{AI-Aided}} &
\rotatebox{90}{\textbf{AI-Oriented}} &
\rotatebox{90}{\textbf{Dual Perspective}} &
\rotatebox{90}{\textbf{AI--ISCC Co-design}} &
\rotatebox{90}{\textbf{Performance Metrics}} &
\rotatebox{90}{\textbf{System Architectures}} &
\rotatebox{90}{\textbf{Application Domains}} \\
\hline
\hline

D. Wen et al.~\cite{S01} & 2025 & Comprehensive ISCC fundamentals and enabling technologies &
- Analysis of partial integration (ICC, ISC, ISAC) and their limitations
\newline - Signal design and waveform optimization for joint integration
\newline - Network resource management and task-oriented ISCC design
\newline - ISCC for federated learning, edge inference, and other applications &
\pmark & \pmark & \xmark & \pmark & \pmark & \pmark & \cmark \\
\hline

G. Zhu et al.~\cite{S02} & 2023 & ISCC paradigm for enabling edge intelligence and AIoT in 6G &
- Task-oriented ISCC design principles for edge AI
\newline - Centralized and federated edge learning
\newline - Edge inference architectures and resource management
\newline - Design challenges and enabling techniques &
\xmark & \cmark & \xmark & \pmark & \pmark & \cmark & \pmark \\
\hline

G. Wu et al.~\cite{ai_iscc} & 2025 & AI-enabled ISCC &
- Review of sensing, communication, and computation technologies and partial integrations
\newline - Analysis of the interactions between AI and ISCC
\newline - Intelligent resource management approaches
\newline - ISCC support for AI applications &
\pmark & \pmark & \cmark & \pmark & \pmark & \pmark & \pmark \\
\hline

Y. Zhang et al.~\cite{zhang2026ai_empowered_tscc} & 2026 & AI-empowered task-oriented sensing, communication, and computation &
- AI-based and AI-assisted methods for task-oriented SCC
\newline - Task classification by modality, scenario, and task attributes
\newline - Model-driven and task-execution performance metrics
\newline - Applications and technical challenges of AI-empowered TSCC &
\cmark & \pmark & \pmark & \pmark & \cmark & \pmark & \cmark \\
\hline

Z. Liu et al.~\cite{liu2026integrated_sensing_edge_ai} & 2026 & Integrated sensing and edge AI for intelligent perception &
- Integration of sensing, communication, and AI computation
\newline - Design principles and architectures for intelligent perception
\newline - Performance metrics and cross-functional tradeoffs
\newline - Applications, enabling technologies, and standardization &
\pmark & \cmark & \pmark & \pmark & \cmark & \cmark & \cmark \\
\hline

C. Li et al.~\cite{survey_ioV_iscc} & 2026 & ISCC for Internet of Vehicles and connected autonomous driving &
- Three-phase ISCC evolution (Collaborative, Fusion, Integrated)
\newline - Physical-layer signal integration techniques
\newline - Network-level task-oriented resource management
\newline - Safety-critical and real-time requirements for vehicular systems &
\pmark & \pmark & \xmark & \pmark & \pmark & \cmark & \cmark \\
\hline

X. Wang et al.~\cite{S05} & 2024 & ISCC technologies enabling immersive metaverse experiences &
- Digital twin integration for metaverse applications
\newline - Edge computing architectures for metaverse services
\newline - Metaverse system architecture and requirements
\newline - Latency reduction and QoS considerations &
\xmark & \xmark & \xmark & \xmark & \pmark & \cmark & \cmark \\
\hline

L. Zhang et al.~\cite{S11} & 2023 & Integration of communications, sensing, computing, and storage for the metaverse &
- I-CSCS framework
\newline - 6G role and requirements for metaverse enablement
\newline - ISAC technologies in the metaverse context
\newline - Spectrum resources, coverage, security, and system architecture &
\xmark & \xmark & \xmark & \xmark & \pmark & \cmark & \cmark \\
\hline

M. Dai et al.~\cite{S03} & 2022 & ISCC for smart oceans via space-air-ground-sea integrated networks &
- Multi-domain network integration (space, air, ground, sea)
\newline - Marine sensing, communication, and computing paradigms
\newline - Resource exploration and environmental monitoring
\newline - Challenges in maritime and underwater environments &
\xmark & \xmark & \xmark & \xmark & \pmark & \cmark & \cmark \\
\hline

M. Zhao et al.~\cite{S06} & 2023 & Federated learning design and optimization within the ISCC framework for 6G &
- FL architectural design strategies in ISCC systems
\newline - Device scheduling and bandwidth allocation for FL
\newline - Resource-constrained FL optimization techniques
\newline - Integration with 6G network infrastructure &
\xmark & \cmark & \xmark & \pmark & \pmark & \cmark & \pmark \\
\hline

M. Nemati et al.~\cite{S04} & 2022 & Joint sensing, communication, computation, and localization for IoT ecosystems &
- Radar sensing and localization integration
\newline - Diverse IoT application scenarios
\newline - Sensing-communication coupling mechanisms
\newline - Implementation considerations and practical aspects &
\xmark & \xmark & \xmark & \xmark & \pmark & \pmark & \cmark \\
\hline
\hline

\textbf{This survey} & -- & \textbf{AI-aided and AI-oriented ISCC with AI--ISCC co-design} &
- \textbf{Unified analysis of AI-aided and AI-oriented ISCC}
\newline - \textbf{AI--ISCC co-design across sensing, communication, computation, and AI workload variables}
\newline - \textbf{Comparative synthesis of system architectures, performance metrics, and evaluation practices}
\newline - \textbf{Emerging applications and their ISCC design requirements} &
\cmark & \cmark & \cmark & \cmark & \cmark & \cmark & \cmark \\
\hline

\end{tabular}

\vspace{0.15cm}
\begin{flushleft}
\footnotesize
\textbf{Legend:} \cmark~Extensive and structured coverage; \pmark~Partial or limited coverage; \xmark~No substantive coverage. \\
\textbf{AI-Aided:} AI techniques for ISCC system design, optimization, or resource management. \\
\textbf{AI-Oriented:} ISCC infrastructure and resource design for supporting AI learning, inference, perception, or related workloads. \\
\textbf{Dual Perspective:} Analysis of AI-aided and AI-oriented ISCC and their interplay within a common perspective. \\
\textbf{AI--ISCC Co-design:} Joint consideration of ISCC decisions and AI model or workload variables within the system design. \\
\textbf{Performance Metrics:} Analysis of performance criteria and metrics used to evaluate the surveyed approaches. \\
\textbf{System Architectures:} Analysis of architectures used to realize integrated sensing, communication, computation, and AI functions. \\
\textbf{Application Domains:} Analysis of application domains and their system design requirements.
\par\smallskip
{\scriptsize\textit{Abbreviations:} AIoT, artificial intelligence of things; FL, federated learning; ICC, integrated communication and computation; I-CSCS, integrated communications, sensing, computing, and storage; IoT, Internet of Things; ISAC, integrated sensing and communication; ISC, integrated sensing and computation; QoS, quality of service; SCC, sensing, communication, and computation; TSCC, task-oriented sensing, communication, and computation.\par}
\end{flushleft}
\end{table*}

Existing surveys approach ISCC from different perspectives, as summarized in Table~\ref{tab:survey_comparison}. Wen et al.~\cite{S01} establish its functional and technological foundations, progressing from pairwise integration to joint signal design and resource management for the three functions. Their treatment of federated learning and edge inference complements the task-oriented perspective of Zhu et al.~\cite{S02}, whose survey centers on the design of integrated wireless systems to support edge intelligence. Together, these works provide a basis for examining ISCC both as a general system design paradigm and as an infrastructure for AI workloads.

The relationship between AI and ISCC is addressed more explicitly by Wu et al.~\cite{ai_iscc}, who examine both intelligent resource management and ISCC support for AI applications, recognizing their mutual reinforcement. Subsequent surveys extend this line of research through different analytical frameworks. Zhang et al.~\cite{zhang2026ai_empowered_tscc} distinguish AI-based and AI-assisted methods within task-oriented sensing, communication, and computation, while Liu et al.~\cite{liu2026integrated_sensing_edge_ai} investigate the integration of sensing and edge AI for intelligent perception. The former emphasizes the characteristics of supported tasks and their performance metrics, and the latter examines architectures and tradeoffs associated with the perception pipeline.

A complementary body of surveys examines ISCC in particular application domains or for specific workloads. Research on the metaverse~\cite{S05,S11}, smart-ocean networks~\cite{S03}, and connected autonomous driving~\cite{survey_ioV_iscc} establishes how application requirements influence integrated system design. More specialized perspectives are provided by surveys of federated learning within ISCC~\cite{S06} and joint sensing, communication, computation, and localization for Internet of Things (IoT) systems~\cite{S04}.

Building on these advances, the present survey places the reciprocal dependencies between AI and ISCC at the center of its analysis. The dual role of AI provides an organizing principle for examining the interplay between sensing, communication, computation, and AI workload configurations. Particular attention is devoted to their co-design, encompassing joint optimization, state-dependent adaptation, and feedback mechanisms underlying the AI--ISCC virtuous cycle. This analytical perspective provides a common basis for comparing system architectures, performance criteria, and experimental evidence across different workloads and application domains.

%% file: secs/02-foundations.tex
\section{Functional Coupling and Design Foundations of ISCC}
\label{sec:iscc_foundations}

The design of ISCC systems depends on the resources shared by sensing, communication, and computation, as well as the functional dependencies established through the information they acquire, exchange, and process. This section examines the resulting coupling mechanisms and how they shape the design space of integrated wireless systems.

\subsection{Functional Coupling and Shared Resources}

The sensing function acquires task-relevant information from the physical environment, for example through radio-frequency (RF)-based localization, tracking, and environmental mapping~\cite{liu2022integrated,S01}. Communication transfers raw data, intermediate representations, model information, and control signals among devices and infrastructure, while computation performs operations such as signal processing, feature extraction, inference, and model training~\cite{he2023integrated,S01}. Depending on the architecture, these operations may be distributed across devices, edge servers, and cloud infrastructure.

\begin{figure}[!t]
    \centering
    \includegraphics[width=\linewidth]{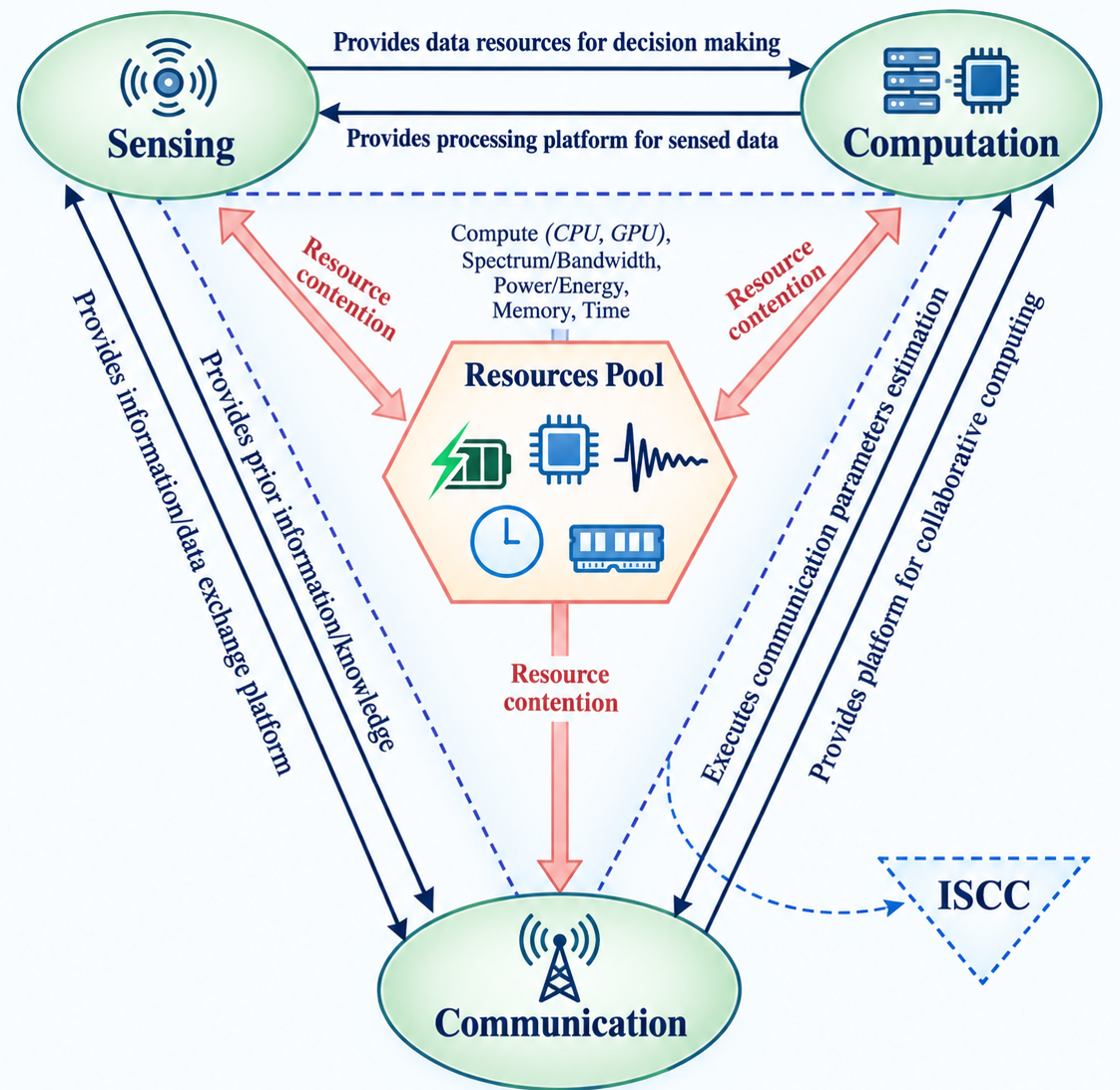}
    \caption{Functional coupling and resource competition among sensing, communication, and computation. CPU: central processing unit; GPU: graphics processing unit.}
    \label{fig:iscc_framework}
\end{figure}

As illustrated in Fig.~\ref{fig:iscc_framework}, the interaction between the three functions arises through both resource sharing and information dependencies. Spectrum, transmission power, time, spatial degrees of freedom, energy, and computational capacity govern the performance attainable by each function. Sensing quality also determines the information available for subsequent transmission and processing, while communication constraints influence which information can be delivered and at what fidelity. Computational capabilities, in turn, determine how and where the information can be processed, and computation results can guide subsequent sensing and communication decisions. Thus, functional coupling persists even when resources are allocated separately to individual functions.

\subsection{From Partial Integration to Full ISCC}

Pairwise integration exploits specific aspects of these dependencies. ISAC exploits shared or jointly configured spectrum, waveforms, hardware, and signal-processing resources~\cite{liu2022integrated,S01}. Integrated communication and computation (ICAC, also referred to as ICC) couples information transmission with computational tasks through paradigms such as mobile edge computing and over-the-air computation (AirComp)~\cite{mao2017survey,lyu2018multiuser,S01}. Integrated sensing and computation (ISEC, also referred to as ISC) links data acquisition with processing, particularly when sensing decisions determine the quality, volume, or cost of the data used by downstream computational tasks~\cite{S01}.

Although each form of pairwise integration addresses particular functional dependencies, the interaction with the remaining function can also influence the resulting design. For example, sharing a waveform between sensing and communication affects both the quality of acquired observations and the resources available to deliver them for computation. The integration can extend to the signal level when the superposition of transmitted waveforms over the wireless channel enables the computation of functions of distributed data. In this setting, the joint design of transmit and aggregation beamformers involves minimizing computation distortion under sensing-accuracy and transmit-power constraints~\cite{li2023iscco_mimo}. Functional dependencies also arise at the architectural level, where the placement of computation changes the communication requirements and can influence how observations are selected or acquired. Full ISCC therefore extends pairwise integration by coordinating the three functions within a common design framework~\cite{S01,ISCC}. The potential gains from integration arise from both the shared use of system resources and the joint optimization of interdependent functional decisions.

\subsection{Design Objectives and Cross-Functional Tradeoffs}

The functional dependencies described above influence the relationship between the design variables and the performance objectives of ISCC systems. These objectives may involve sensing accuracy or detection probability, communication rate and reliability, computation latency, and energy expenditure. Optimizing any one of them can affect the feasible operating region of the others~\cite{S01}. For tasks involving learning, inference, or perception, the relevant tradeoffs also depend on how impairments and resource constraints propagate through the information-processing pipeline. Sensing noise affects the quality of the acquired data, communication distortion and quantization alter the information delivered for processing, and computation constraints govern the complexity and latency of the processing operations~\cite{Zhuang20243205_PSS-Scopus-20250825-151225_0067,wen2026airfeel,Yao2025_PSS-Scopus-20250825-151225_0072}.

The choice of optimization objective therefore depends on the service supported by the integrated system. In some formulations, individual functional requirements define the constraints under which resource efficiency is optimized. In others, an analytical characterization of downstream task performance provides the objective for joint resource allocation. These alternatives are particularly relevant to AI workloads, for which learning convergence or inference accuracy can depend on the combined effects of sensing, communication, and computation decisions. Section~\ref{sec:comparative_synthesis} examines the implications of these different formulations for system evaluation and comparison.

\subsection{AI in ISCC Design and Services}

AI extends the scope of ISCC design in two ways. On the system side, the choice of a learning-based optimization or control mechanism introduces considerations such as the availability of state information, the cost of executing the policy, and the timescale at which decisions can be updated~\cite{Yang202418608_PSS-Scopus-20250825-151225_0131,Cai202413584_PSS-Scopus-20250825-151225_0190}. On the workload side, the design variables may extend beyond physical-resource allocation to include the representation of sensed information, model partitioning, local training batch size, and aggregation strategy~\cite{wen2026airfeel,wen2024task_oriented_edge_ai}. These variables determine both the resources consumed during task execution and the attainable task performance, making their selection closely dependent on the ISCC configuration.

\begin{figure*}[!t]
    \centering
    \includegraphics[width=\linewidth]{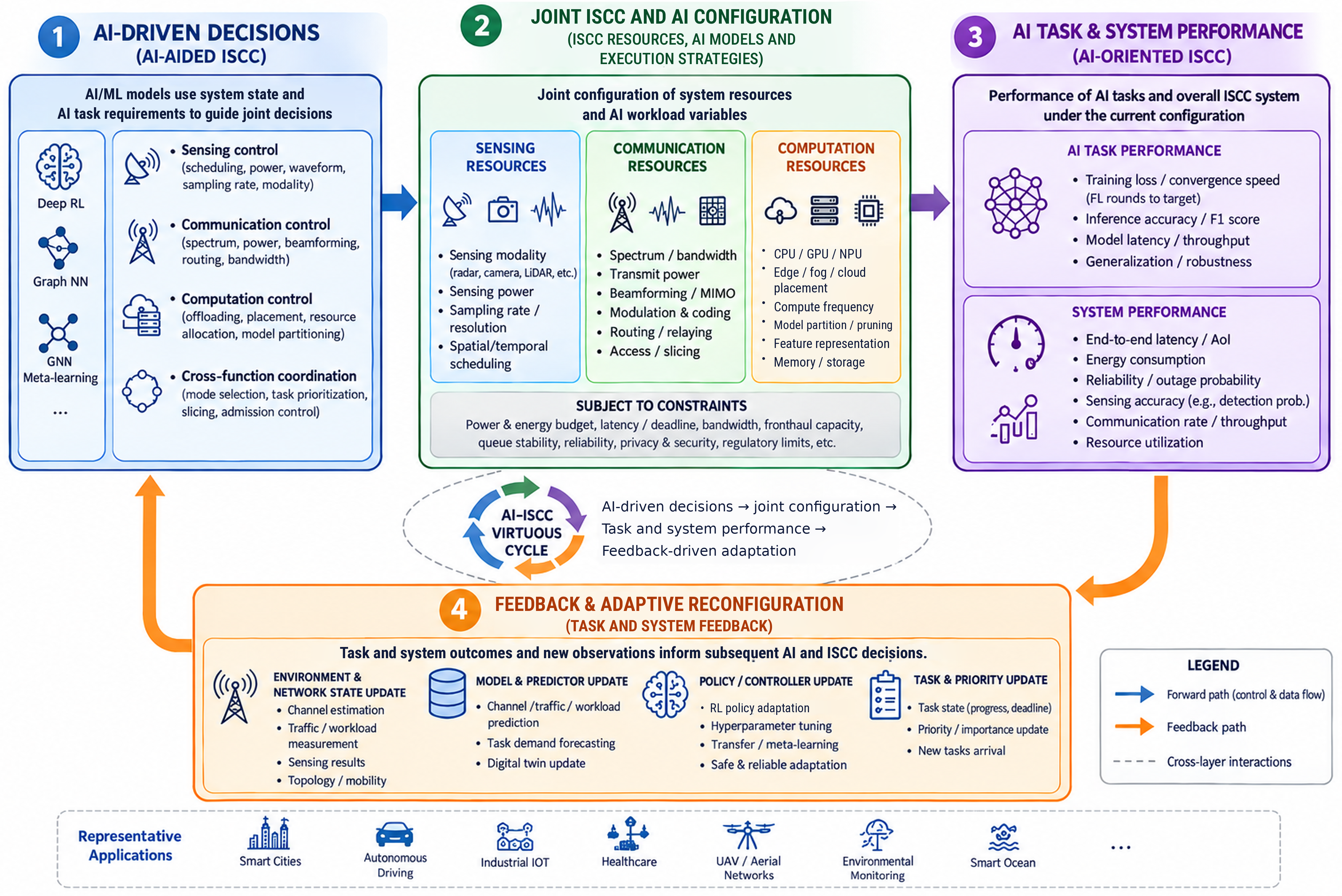}
    \caption{Interplay between AI-aided ISCC, AI-oriented ISCC, and AI--ISCC co-design. Learning-based mechanisms can support ISCC decisions, ISCC configurations determine the execution conditions of AI workloads, and the resulting task and system performance can inform subsequent decisions in dynamic settings. AoI: age of information; CPU/GPU/NPU: central/graphics/neural processing unit; FL: federated learning; GNN: graph neural network; IoT: Internet of Things; MIMO: multiple-input multiple-output; ML: machine learning; NN: neural network; RL: reinforcement learning; UAV: unmanned aerial vehicle.}
    \label{fig:virtouscycle}
\end{figure*}

Fig.~\ref{fig:virtouscycle} illustrates the resulting relationships between AI-aided ISCC, AI-oriented ISCC, and their co-design. The following sections examine the design approaches associated with each role and the mechanisms through which they can be coupled, including adaptation based on measured task and system performance.

%% file: secs/03-ai-for-iscc.tex
\section{AI-Aided ISCC}
\label{sec:ai-for-iscc}

\begin{table*}[!t]
\centering
\caption{Learning-based optimization approaches for ISCC design and control.}
\label{tab:ai_aided}
\footnotesize
\renewcommand{\arraystretch}{1.08}
\setlength{\tabcolsep}{2.1pt}

\begin{tabularx}{\textwidth}{
>{\centering\arraybackslash}p{0.85cm}
>{\raggedright\arraybackslash}p{1.75cm}
>{\raggedright\arraybackslash}p{1.65cm}
>{\raggedright\arraybackslash}X
>{\raggedright\arraybackslash}X
>{\raggedright\arraybackslash}X}
\toprule
\textbf{Ref.} &
\textbf{Scenario} &
\textbf{Method} &
\textbf{Decision Variables} &
\textbf{Objective / Main Constraints} &
\textbf{Results} \\
\midrule

\multicolumn{6}{l}{\textit{Joint resource allocation and management}}\\
\midrule

\cite{Chen202327_PSS-Scopus-20250825-151225_0158}
&
Heterogeneous vehicular intelligent tasks
&
DDQN
&
Communication-resource, computation-resource, and cache allocation for real-time and delay-tolerant tasks
&
Optimize task performance under heterogeneous resource and latency requirements
&
Approximately 5\% or greater improvement in exploration efficiency with DDQN and an increase of approximately 0.15 in mAP through resource allocation under resource constraints
\\

\cite{zhao2024iov_iscc}
&
Collaborative sensing in IoV
&
DRL
&
Uplink-bandwidth allocation, downlink-bandwidth allocation, and edge-computing allocation
&
Maximize collaborative sensing-task completion rate subject to service-delay and resource constraints
&
Higher sensing-task completion rate than the evaluated conventional allocation schemes
\\

\cite{learning-based-iscc}
&
Delay-aware ISCC
&
Unsupervised primal--dual learning
&
Joint sensing, communication, and computation resource allocation
&
Minimize end-to-end task delay under resource, delay, and sensing mutual information constraints
&
Lower end-to-end delay and improved convergence relative to the evaluated baselines
\\

\cite{cheng2026d3pg_light_iscc}
&
Single-UAV ISCC
&
D3PG-Light
&
Communication-bandwidth fraction $b_t$, sensing-time fraction $s_t$, and CPU-frequency fraction $c_t$
&
Jointly account for latency, sensing performance, and energy consumption through a multi-objective reward
&
95th-percentile latency of 23.6~ms, reduced by 2.7~ms, 33.7~ms, and over 80\% relative to DDPG, TD3, and D3PG, respectively. 73.05\% fewer parameters than D3PG, with slightly higher energy consumption than D3PG
\\

\cite{ji2026uav_emergency_iscc}
&
UAV-assisted emergency IoV
&
ACBC-TD3
&
Vehicle scheduling order, UAV motion, and task-offloading ratio
&
Maximize collaborative sensing-task completion under latency and emergency-link constraints
&
Faster convergence and higher task-completion performance than standard TD3 in simulated emergency scenarios
\\

\cite{ran_slicing}
&
ISCC service slicing
&
DDPG
&
Service-level slice resource allocation
&
Jointly optimize sensing accuracy, transmission latency, and processing delay for heterogeneous services
&
Lower latency and higher sensing accuracy than the evaluated DRL baselines
\\

\midrule
\multicolumn{6}{l}{\textit{Waveform and beamforming design}}\\
\midrule

\cite{Yang202418608_PSS-Scopus-20250825-151225_0131}
&
Vehicular ISCC
&
SDR, Gaussian randomization, and DDPG
&
Aggregation beamformer, transmit beamforming vectors, and power allocation
&
Maximize achievable data rate while satisfying sensing and computation requirements
&
Higher achievable sum rate and improved convergence relative to the evaluated benchmarks
\\

\cite{Ren202528201_PSS-Scopus-20250825-151225_0034}
&
RIS-assisted vehicular ISCC
&
Dual-critic DRL
&
RSU beamformer and RIS reflection matrix
&
Maximize sum rate subject to sensing, transmit-power, RIS-gain, and computation-related vehicle-service constraints
&
Higher converged reward than classical DDPG, and higher sum rate and sensing SINR across the evaluated RIS-size and user-number settings
\\

\cite{qi2026uav_iscc_efficiency}
&
UAV-enabled ISCC
&
LSTM-TD3
&
UAV beamforming vector, CPU frequency, and UAV trajectory
&
Maximize successfully processed bits per unit of UAV energy consumption subject to a sensing-performance threshold
&
Computing efficiency at convergence of $7.87\times10^{4}$~bits/J, representing gains of $14.73\%$ and $33.82\%$ over TD3 and DDPG, respectively
\\

\midrule
\multicolumn{6}{l}{\textit{Computation offloading and task scheduling}}\\
\midrule

\cite{Cai202413584_PSS-Scopus-20250825-151225_0190}
&
Dependency-aware vehicular tasks
&
Meta-DRL
&
Execution scheduling of dependency-constrained ISCC tasks under time-varying resource availability
&
Reduce processing delay while accounting for inter-task dependencies and vehicular mobility
&
Reduced dependent-task processing delay and rapid adaptation to new vehicular network environments
\\

\cite{Dong20258481_PSS-Scopus-20250825-151225_0013}
&
Vehicular cooperative perception
&
PPO with problem decomposition
&
Sensing-subregion allocation, offloading decisions, bandwidth allocation, and computation-resource allocation
&
Maximize information value while reducing execution delay and energy under coverage and resource constraints
&
Higher information value together with lower task-execution delay and energy consumption than the selected baselines
\\

\cite{huang2025collaborative_sensing_offloading}
&
Collaborative vehicle--RSU sensing
&
Iterative DQN and convex optimization
&
Offloading destination $\eta_{un}$ and per-vehicle bandwidth ratio $\beta_{un}$
&
Minimize total sensing-task processing delay subject to latency, bandwidth, and RSU-computation constraints
&
In a network of three RSUs and 25 vehicles, delay reductions of $13.1\%$, $25.3\%$, and $37.2\%$ over average bandwidth allocation, processing without collaborative computing, and traditional raw-data offloading, respectively, at $18$~MB/task. Corresponding reductions of $9.4\%$, $18.1\%$, and $34.6\%$ at $180$~Gcycles/s
\\

\bottomrule
\end{tabularx}
\par\smallskip
\parbox{\textwidth}{\scriptsize\raggedright\textit{Abbreviations:} ACBC-TD3, adaptive TD3 with cosine annealing and behavior cloning; CPU, central processing unit; DDPG, deep deterministic policy gradient; DDQN, distributed deep Q-network; DQN, deep Q-network; DRL, deep reinforcement learning; IoV, Internet of Vehicles; LSTM, long short-term memory; mAP, mean average precision; PPO, proximal policy optimization; RIS, reconfigurable intelligent surface; RSU, roadside unit; SDR, semidefinite relaxation; SINR, signal-to-interference-plus-noise ratio; TD3, twin delayed DDPG; UAV, unmanned aerial vehicle.}
\end{table*}

The functional dependencies established in Section~\ref{sec:iscc_foundations} give rise to resource allocation and scheduling problems in which decisions concerning one function affect the operating conditions of the others. Learning-based approaches provide a means of solving these problems under varying channel, sensing, and computation conditions. We examine how the coupling among system functions shapes the design of learning-based optimizers and control policies, from resource management and radio-domain optimization to the scheduling and placement of computation tasks. Table~\ref{tab:ai_aided} summarizes representative formulations, optimization methods, and reported results.

\subsection{Joint Resource Allocation and Management}

The requirements of the supported services determine which resources must be coordinated and how their allocation affects task performance. Real-time and delay-tolerant vehicular tasks, for instance, require different allocations of communication, computation, and caching resources. Chen et al.~\cite{Chen202327_PSS-Scopus-20250825-151225_0158} formulate separate allocation problems for these two classes of tasks and use a distributed deep Q-network (DDQN) to learn the allocation policy. When sensing itself constitutes a service delivered collaboratively, resource allocation must also account for the completion of sensing tasks. This motivates the formulation in~\cite{zhao2024iov_iscc}, where uplink and downlink bandwidth and edge-computing resources are allocated to maximize the completion rate of stochastic collaborative sensing tasks under service-delay constraints. Meeting such delay requirements involves accounting for the dependencies among sensing, signal processing, data delivery, and subsequent computation. Yang et al.~\cite{learning-based-iscc} model these interactions using tandem and parallel queues and derive a delay bound through stochastic network calculus. The resulting joint resource allocation problem minimizes end-to-end delay under resource, latency, and sensing mutual information constraints using unsupervised primal--dual learning with a multilevel decomposition and residual network architecture. This formulation illustrates how analytical models of functional dependencies can guide the design of learning-based optimizers alongside approaches that learn resource allocation policies through interaction with the environment.

The coordination of resources becomes particularly important in unmanned aerial vehicle (UAV)-assisted systems, where the platform's own computation and energy budgets can constrain sensing and communication, while its mobility can affect service availability. In the single-UAV design of Cheng et al.~\cite{cheng2026d3pg_light_iscc}, communication bandwidth, sensing duration, and onboard central processing unit (CPU) frequency are jointly controlled through the D3PG-Light policy. The resulting tradeoffs are apparent in simulations driven by measured air--ground channel data. The policy achieves a 95th-percentile latency of 23.6~ms and the highest normalized sensing utility among the evaluated schemes, at slightly higher energy expenditure than the original D3PG, but lower expenditure than deep deterministic policy gradient (DDPG) and twin delayed DDPG (TD3). Its parameter count is also reduced by 73.05\% relative to D3PG, which is of particular relevance to the deployment of learned controllers on resource-constrained platforms. When the UAV assists collaborative sensing in post-disaster vehicular networks, the allocation problem extends to vehicle scheduling and platform motion, which jointly determine the opportunities for task offloading. Ji et al.~\cite{ji2026uav_emergency_iscc} address this problem using ACBC-TD3, combining behavior cloning, adaptive parameter adjustment, and cosine-annealed learning rates to improve convergence and sensing-task completion relative to standard TD3 in simulated emergency scenarios.

The distribution of control decisions becomes another design consideration when sensing and computation involve several cooperating aerial and ground nodes. In this context, Liu et al.~\cite{Liu2025_PSS-Scopus-20250825-151225_0010} formulate joint task scheduling and resource allocation in terms of sensing and computation-offloading penalties, subject to UAV mobility and energy constraints. By considering both centralized deep reinforcement learning (DRL) control and independent decisions based on local information, the study highlights the importance of information availability in determining how the coordination of the three functions can be distributed among network entities.

When heterogeneous services share the same infrastructure, the design of a learned controller also depends on how their performance requirements are represented in a common objective. Huang et al.~\cite{huang2026dt_llm_iscc} address concurrent communication and target classification in vehicular ISCC through a value-of-service metric relating data rate and classification accuracy. Their approach uses a digital twin to provide predictive information for proximal policy optimization (PPO), and a large language model (LLM) to refine the reward function. AI thus contributes to both policy learning and reward design. Beyond individual resource assignments, heterogeneous services can also be accommodated at a higher level of resource management through radio access network (RAN) slicing~\cite{ran_slicing}. In this setting, the learned policy partitions resources among slices according to their sensing-accuracy, transmission-latency, and processing-delay requirements. The shift from task-level control to service-level partitioning highlights the role of control granularity in ISCC, particularly when several services with distinct functional requirements coexist within the same infrastructure.

\subsection{Waveform and Beamforming Design}

The integration of sensing, communication, and computation introduces additional considerations into the design of wireless signals and their spatial configuration. In mobile crowd ISCC systems, for instance, the waveform precoding problem involves minimizing energy consumption and computation cost under sensing-accuracy constraints~\cite{li2024mean}. The complexity of such design problems motivates the exploration of learning-based optimization methods, particularly when waveform and beamforming variables are closely coupled. Advances in ISAC provide a basis for these developments, as illustrated by Jiang et al.~\cite{Jiang2025}, who unfold iterative optimization algorithms to jointly design symbol-level waveforms and RIS passive beamforming for target detection and direction-of-arrival estimation. Incorporating computation requirements into such learning-based formulations extends the design problem to account for the processing of the information acquired and exchanged through the resulting waveforms.

In existing learning-based beamforming approaches for ISCC, the role of computation is reflected in both the optimization objectives and the constraints imposed on radio-resource allocation. One class of formulations retains communication throughput as the objective and incorporates sensing and computation requirements into the constraints of the radio-domain problem. Yang et al.~\cite{Yang202418608_PSS-Scopus-20250825-151225_0131}, for example, maximize the achievable data rate in vehicular ISCC through joint beamforming and power allocation. The structure of the problem allows the aggregation beamformer to be obtained through semidefinite relaxation and Gaussian randomization, with DDPG determining the transmit beamforming and resource-allocation variables. The introduction of a reconfigurable intelligent surface (RIS) provides additional radio-domain degrees of freedom, as illustrated by~\cite{Ren202528201_PSS-Scopus-20250825-151225_0034}, where the roadside unit (RSU) beamformer and RIS reflection matrix are jointly optimized subject to sensing, power, RIS-gain, and vehicle-service constraints. A dual-critic DRL architecture mitigates Q-value overestimation in this continuous control problem. While the latter design expands the radio-domain decision space, computation-related requirements still enter through the service constraints.

When computation performance becomes the primary objective, the radio-domain decisions can instead be coupled explicitly with those governing processing and mobility. Qi and Wu~\cite{qi2026uav_iscc_efficiency} jointly optimize the beamforming vector, onboard CPU frequency, and trajectory of a UAV to maximize the number of successfully processed computation bits per unit of total UAV energy consumption, subject to a sensing-performance threshold. In this setting, the trajectory affects the conditions for information exchange, while beamforming and CPU-frequency allocation govern the transfer and processing of the workload. Their long short-term memory (LSTM)-based TD3 algorithm exploits historical state--action information to coordinate these decisions. The reported computing efficiency at convergence reaches $7.87\times10^{4}$~bits/J, corresponding to gains of $14.73\%$ and $33.82\%$ over TD3 and DDPG, respectively. The study also reports average gains of $109.40\%$ and $235.65\%$ over random- and fixed-trajectory baselines, respectively, across different task durations. These comparisons complement those against alternative learning algorithms by examining the performance of joint control relative to predefined mobility strategies.

\subsection{Computation Offloading and Task Scheduling}

\begin{figure}[!t]
    \centering
    \includegraphics[width=\linewidth]{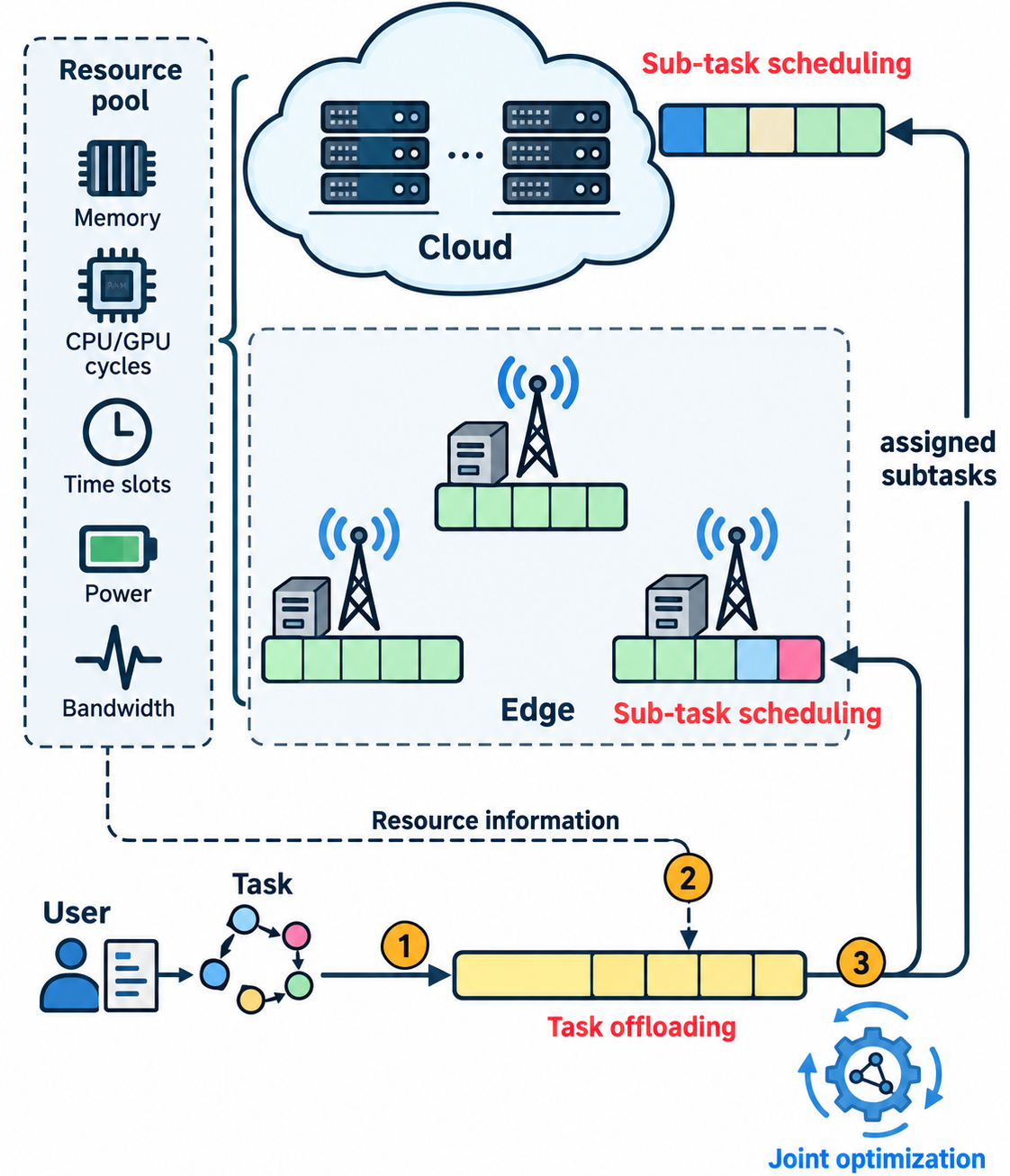}
    \caption{Joint task offloading and subtask scheduling across edge and cloud computing resources. CPU: central processing unit; GPU: graphics processing unit.}
    \label{fig:task_offloading}
\end{figure}

In ISCC, the placement and scheduling of computation tasks are closely related to the structure of the sensing workload. Dependencies between operations determine when processing can begin, while the usefulness and redundancy of observations determine which data warrant transmission and computation. Fig.~\ref{fig:task_offloading} illustrates how resource-state information can be jointly exploited to determine task offloading and subtask scheduling across edge and cloud computing resources.

Task dependencies are particularly important when sensing and computation are scheduled in a time-varying network. A computation task may only start once the corresponding sensing task has been completed, such that its execution schedule depends on both sensing progress and the availability of processing resources. Cai et al.~\cite{Cai202413584_PSS-Scopus-20250825-151225_0190} incorporate these precedence relations into vehicular scheduling and employ meta-DRL to adapt the execution schedule to changing network environments. The reported reductions in dependent-task processing delay illustrate the relevance of accounting for the order of operations in addition to their individual resource requirements.

The scheduling problem can be further extended by considering the informational content of the sensing workload. In cooperative perception, different sensing subregions can contribute differently to the perception task, and observations collected by multiple nodes can contain overlapping information. These considerations motivate two complementary approaches to computation control~\cite{Dong20258481_PSS-Scopus-20250825-151225_0013,huang2025collaborative_sensing_offloading}. Dong et al. use an information-value function to guide sensing-subregion assignment, followed by PPO-based offloading and resource allocation under coverage, latency, and resource constraints. Huang et al. exploit redundancy between vehicle and RSU observations before offloading to reduce the amount of data that must be transferred and processed. The resulting optimization lends itself to a decomposition in which a deep Q-network (DQN) selects the processing destination and convex optimization determines the bandwidth allocation. In simulations with three RSUs and 25 vehicles, the joint scheme reduces total task-processing delay by $37.2\%$ compared with traditional raw-data offloading at a task size of $18$~MB. Together, these studies illustrate how accounting for the value and redundancy of sensing observations can alter the computation workload itself, alongside the allocation of resources to execute it.

\subsection*{\textbf{Lessons Learned}}

\begin{enumerate}[leftmargin=*]

    \item \textit{The learned control space should reflect functional dependencies.} The strength of the dependencies among decision variables determines the extent to which they need to be controlled jointly. Hierarchical control provides opportunities to coordinate tightly coupled decisions while decomposing the broader resource management problem into different levels of control. Such a decomposition can limit the dimensionality of the learned action space while preserving essential cross-functional interactions.

    \item \textit{The structure of the optimization problem can guide the role of learning.}
    Analytical models of functional dependencies can guide the formulation of learning objectives, as illustrated by the delay-aware approach in~\cite{learning-based-iscc}. The hybrid methods in~\cite{Yang202418608_PSS-Scopus-20250825-151225_0131,huang2025collaborative_sensing_offloading} exploit the problem structure by assigning structured subproblems to analytical optimization and using learned policies for the remaining decisions. Such a decomposition can reduce the dimensionality of the learned action space, but its suitability for online operation depends on the combined latency and computational cost of policy inference and model-based optimization.

    \item \textit{Service objectives should be complemented by functional requirements.}
    As illustrated by the latency--energy tradeoff in~\cite{cheng2026d3pg_light_iscc}, improvements in task or service performance may come at the expense of individual functions or resources. Explicit functional constraints and complementary performance metrics are therefore important for ensuring that a learned policy operates within the requirements of the supported service.

    \item \textit{Adaptation should be commensurate with the dynamics of the system.} The practical value of a learned controller depends on its ability to maintain service performance as channels, workloads, and topology change. The generalization capabilities of the learned policy determine the extent to which these variations can be accommodated without further learning. When online adaptation is required, its latency and computational cost must be compatible with the dynamics of the controlled processes.

\end{enumerate}

%% file: secs/04-issc-for-ai.tex
\section{AI-Oriented ISCC}
\label{sec:iscc-for-ai}

AI-oriented ISCC addresses the joint design and allocation of sensing, communication, and computation resources to support AI workloads. The resulting optimization problems depend on how sensed information is acquired, exchanged, and processed throughout the learning or inference pipeline. This section examines these dependencies across federated edge learning, distributed inference, and vertical and multimodal edge intelligence. Table~\ref{tab:iscc_for_ai} summarizes representative studies in terms of their functional designs, optimization objectives, and reported results.

\subsection{Federated Edge Learning}

\begin{figure*}[!t]
    \centering
    \includegraphics[width=\textwidth]{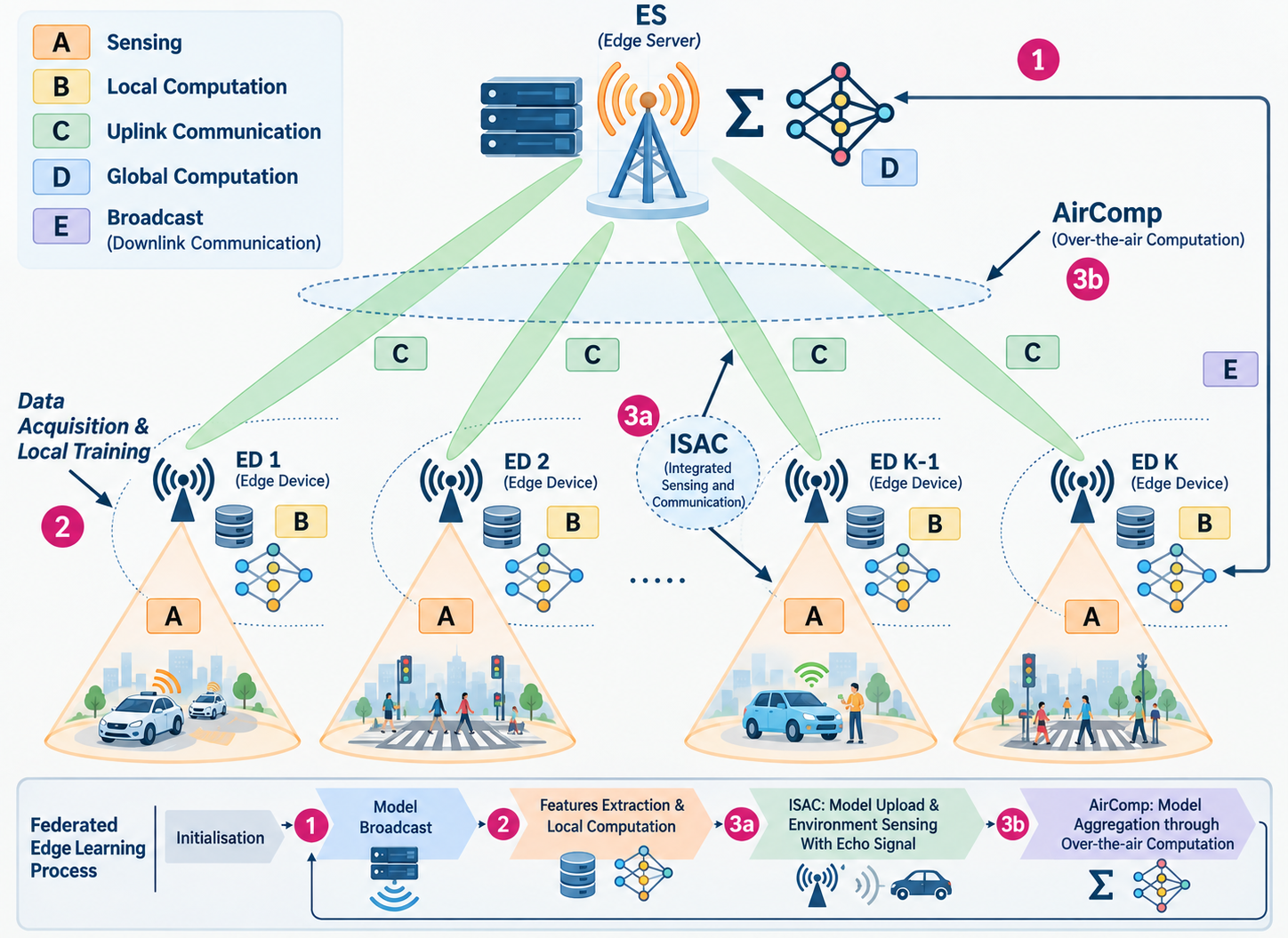}
    \caption{Illustration of ISCC-enabled federated edge learning with over-the-air aggregation.}
    \label{fig:feel_aircomp}
\end{figure*}

In ISCC-enabled federated edge learning (FEEL), the integration of sensing into the learning framework introduces the quantity and quality of acquired training data as additional dimensions of resource optimization. Liu et al.~\cite{Liu2023158_PSS-Scopus-20250825-151225_0128} characterize these dependencies for human-motion recognition by identifying a sensing-power threshold beyond which further increases in power yield little improvement in sample quality. Under this sensing model, joint resource allocation determines the total number of samples that can be acquired within the training time and energy budgets, while the convergence analysis guides their allocation across successive learning rounds. The resulting batch-size rule favors larger batches as the training loss decreases. With AirComp, Liang et al.~\cite{Liang2023592_PSS-Scopus-20250825-151225_0173} further relate the average training error to the number of sensed samples and aggregation errors. The architecture illustrated in Fig.~\ref{fig:feel_aircomp} embodies these dependencies, which are jointly characterized in the convergence analysis of Wen et al.~\cite{wen2026airfeel}.

\begin{table*}[!t]
\centering
\caption{Functional configurations supporting learning and inference workloads.}
\label{tab:iscc_for_ai}
\scriptsize
\renewcommand{\arraystretch}{1.15}
\setlength{\tabcolsep}{1.5pt}

\begin{tabularx}{\textwidth}{@{}
>{\centering\arraybackslash}p{0.75cm}
>{\raggedright\arraybackslash}p{1.50cm}
>{\hsize=0.70\hsize\linewidth=\hsize\raggedright\arraybackslash}X
>{\hsize=0.85\hsize\linewidth=\hsize\raggedright\arraybackslash}X
>{\hsize=0.95\hsize\linewidth=\hsize\raggedright\arraybackslash}X
>{\hsize=0.85\hsize\linewidth=\hsize\raggedright\arraybackslash}X
>{\hsize=1.65\hsize\linewidth=\hsize\raggedright\arraybackslash}X
@{}}
\toprule
\textbf{Ref.} &
\textbf{AI Workload} &
\textbf{Sensing Design} &
\textbf{Communication Design} &
\textbf{Computation / Model Design} &
\textbf{Objective / Metric} &
\textbf{Results} \\
\midrule

\multicolumn{7}{l}{\textit{Federated edge learning}}\\
\midrule

\cite{wen2026airfeel}
&
Over-the-air FEEL
&
Sensing power and batch size
&
Transmit power and AirComp receive scaling
&
Device computation frequency and local batch size
&
Per-round loss reduction and convergence
&
Larger batches lead to lower optimal sensing power. For a given batch size, the energy-efficient computation frequency is the minimum satisfying the latency constraint. Joint optimization improves learning performance over fixed-parameter designs.
\\

\cite{tang2025uav_feel}
&
UAV-assisted FEEL
&
UAV deployment and sensing batch size
&
Per-UAV bandwidth allocation
&
Batch-dependent local training workload
&
Minimum training time under a target learning optimality gap
&
Joint bandwidth, batch-size, and deployment optimization improves convergence and testing accuracy within a given training time.
\\

\cite{alert-ai-10959106}
&
Multi-UAV over-the-air FL
&
UAV deployment and sensing batch size
&
Downlink power control and uplink AirComp transceiver design
&
Local training and global model aggregation
&
Learning convergence
&
Convergence characterization jointly accounts for sensing sample size, downlink broadcast distortion, and uplink aggregation errors.
\\

\cite{hou2025semi_supervised_iscc}
&
UAV-assisted split federated learning
&
Position-dependent target sensing
&
Intermediate activation exchange and periodic client-model aggregation
&
Split layer, aggregation frequency, and semi-supervised training
&
Minimum UAV energy under a convergence requirement
&
Joint deployment and split-training design reduces modeled UAV energy expenditure for a target accuracy relative to the evaluated baselines across three modeled urban environments.
\\

\cite{zheng2026signal_level_iscc}
&
Signal-level AirComp-FL
&
Joint target detection through shared uplink waveforms
&
Transmit powers and receive scaling
&
Over-the-air model aggregation
&
Minimum aggregation distortion subject to target-detection requirements
&
Analytically characterized optimal power control for model aggregation and target detection using a common waveform.
\\

\cite{liang2026fedavg_fedsgd_iscc}
&
FedAVG-ISCC / FedSGD-ISCC
&
Sample acquisition and sensing errors
&
Wireless model exchange and communication errors
&
Local model updates and learning-protocol comparison
&
Convergence and testing accuracy under IID/non-IID data
&
FedAVG-ISCC benefits from multiple local updates under IID data, whereas FedSGD-ISCC exhibits greater robustness to data heterogeneity and communication errors in the evaluated regimes.
\\

\cite{zhang2026online_fed_learning_limited_cache}
&
Online FEEL
&
Newly sensed samples and sensing resources
&
OFDM resources for model aggregation
&
Cache management, historical-data selection, and local training
&
Per-round loss reduction and convergence
&
Loss-aware caching improves learning performance over alternative caching and resource-allocation schemes while accounting for concept drift.
\\

\midrule
\multicolumn{7}{l}{\textit{Edge, split, and collaborative inference}}\\
\midrule

\cite{wen2024task_oriented_edge_ai}
&
Multi-view edge inference
&
Sensing power and sensing time
&
Transmit power, communication time, and feature quantization
&
Fixed local feature extractors and edge-side classifiers
&
Maximum average pairwise discriminant gain
&
Optimal allocation improves SVM and MLP classification accuracy over power-aware, time-aware, and quantization-aware alternatives.
\\

\cite{Zhuang20243205_PSS-Scopus-20250825-151225_0067}
&
Over-the-air edge inference
&
Per-device sensing power
&
Feature-wise transmit precoders and receive beamformers
&
Local feature extraction and AirComp aggregation
&
Maximize the minimum pairwise discriminant gain
&
Task-oriented aggregation improves classification accuracy by accounting for the discriminability of the least separable class pair.
\\

\cite{Wang202440814_PSS-Scopus-20250825-151225_0022}
&
Multi-view multi-task inference
&
Device scheduling and sensory-view selection
&
Shared communication-resource allocation
&
Shared feature computation and multi-task inference
&
Inference performance under shared resource constraints
&
Sensory-view reuse reduces redundant acquisition, feature computation, and communication across concurrent inference tasks.
\\

\cite{wang2026decentralized_multitask_iscc}
&
Decentralized multi-task inference
&
Per-device sensing power
&
Subcarrier assignment, transmit precoding, and receive beamforming
&
Distributed feature aggregation and task-specific inference
&
Multi-objective discriminant-gain optimization
&
Joint resource coordination supports concurrent inference tasks while accounting for full-duplex self-interference.
\\

\cite{Yao2025_PSS-Scopus-20250825-151225_0072}
&
Adaptive split inference
&
Sensing power
&
Transmit power and feature quantization
&
DNN split point, pruning ratio, and device computational capacity
&
Minimum device energy under accuracy and latency constraints
&
Energy reductions of up to 40\% over existing methods. Some baselines become infeasible at a latency deadline of 0.6~s.
\\

\cite{wen2026uav_assisted_edge_inference}
&
UAV-assisted edge inference
&
Sensing power
&
UAV trajectory, device access sequence, and transmission parameters
&
Local feature extraction and computation frequency
&
Minimum inference delay under energy and accuracy constraints
&
Joint mobility and resource optimization reduces end-to-end inference delay relative to the evaluated baselines.
\\

\midrule
\multicolumn{7}{l}{\textit{Vertical and multimodal edge intelligence}}\\
\midrule

\cite{cao2026vfeel}
&
Vertical FEEL
&
Batch size and per-device sensing power
&
Transmit powers and AirComp denoising factors
&
Local embedding computation and global model training
&
Learning convergence under per-round latency and overall energy constraints
&
Increasing device participation can amplify the impact of aggregation errors on convergence. An optimized batch size of approximately 150 outperforms fixed sizes of 100, 200, and 400.
\\

\cite{chen2026multimodal_edge_intelligence}
&
Multimodal edge inference
&
Wi-Fi, RFID, and mmWave with fixed acquisition settings
&
Transmission time and feature-level quantization
&
Pretrained MCR$^2$ feature extractors and classifier
&
MCR$^2$-based objective under delay and energy constraints
&
Improved inference accuracy over the evaluated baselines. Under one channel configuration, the weakest link receives 51 of 90~ms with mmWave sensing versus 21~ms in the all-Wi-Fi setting.
\\

\bottomrule
\end{tabularx}
\par\smallskip
\parbox{\textwidth}{\scriptsize\raggedright\textit{Abbreviations:} AirComp, over-the-air computation; DNN, deep neural network; FEEL, federated edge learning; FedAVG, federated averaging (named method); FedSGD, federated stochastic gradient descent; FL, federated learning; IID, independent and identically distributed; MCR$^2$, maximal coding rate reduction; MLP, multilayer perceptron; mmWave, millimeter-wave; OFDM, orthogonal frequency-division multiplexing; RFID, radio-frequency identification; SVM, support vector machine; UAV, unmanned aerial vehicle.}
\end{table*}

By accounting jointly for sensing noise, AirComp distortion, and local batch sizes, Wen et al. derive a bound on the per-round loss reduction that guides the allocation of resources across the three functions. A particularly important outcome is the tradeoff between the quantity and quality of training samples. Under latency and energy constraints, increasing the batch size leads to a lower optimal sensing power. For a given batch size, the energy-efficient computation frequency is the minimum satisfying the latency constraint under the adopted energy model. Human-motion recognition simulations with six edge devices corroborate the convergence analysis. Thus, the learning benefit of acquiring additional samples depends on the resources available to acquire, process, and aggregate them. Controllable model dropout~\cite{Jiao202519767_PSS-Scopus-20250825-151225_0064} extends this design space by allowing devices to train submodels with different numbers of active parameters. Since the dropout rate influences the computation load as well as convergence, model complexity and batch size can be adjusted jointly with sensing and communication resources.

The relationship between training progress and resource expenditure also depends on where data are acquired and how models are exchanged. In UAV-assisted FEEL, deployment determines both sensing quality and communication conditions. Tang et al.~\cite{tang2025uav_feel} therefore optimize UAV positions, bandwidth allocation, and batch size to minimize training time for a target learning optimality gap. For multi-UAV over-the-air learning, Fu et al.~\cite{alert-ai-10959106} further account for downlink model-broadcast distortion in addition to uplink aggregation errors and sample availability. When local training itself is distributed between UAVs and an edge server, the model partition and the frequency of client-side aggregation become additional design variables. In the split federated learning framework of Hou et al.~\cite{hou2025semi_supervised_iscc}, these variables are optimized alongside UAV deployment to minimize UAV energy consumption subject to a convergence requirement. The framework also addresses limited label availability by incorporating confidence-based pseudo-labeling of unlabeled samples into the training procedure. Extending federated learning to a space-air-ground hierarchy introduces further dependencies between data acquisition and the routes through which model updates are exchanged. Yang et al.~\cite{yang2025hierarchical_feel_sagin} address sensing variance and transmission energy through a two-stage design that includes route selection between UAVs and a low-Earth orbit satellite.

Resource coordination also involves selecting which devices participate in each round~\cite{Zheng2024804_PSS-Scopus-20250825-151225_0197}. This choice depends in part on the quality of the links available for model exchange. With a RIS, the propagation environment can be configured jointly with device participation and sensing batch sizes. Wu et al.~\cite{wu2026ris_feel_iscc} use a convergence bound accounting for model transmission errors to guide this joint design together with transmit-power and bandwidth allocation.

Sensing can also serve objectives other than training-data acquisition within the same federated learning system. Zheng et al.~\cite{zheng2026signal_level_iscc} exploit the same uplink waveforms for model aggregation and joint target detection. In this case, the resources available to minimize aggregation distortion are directly constrained by the detection requirement of the shared waveform.

Resource conditions also influence the relative efficiency of alternative learning protocols. The comparison of federated averaging (FedAvg) and federated stochastic gradient descent (FedSGD) for ISCC (FedAVG-ISCC and FedSGD-ISCC, respectively) in~\cite{liang2026fedavg_fedsgd_iscc} accounts for sensing, local computation, communication errors, and data heterogeneity under latency and energy constraints. The convergence analysis and simulations on MNIST and Fashion-MNIST identify distinct operating regimes. FedAVG-ISCC benefits from multiple local updates under independent and identically distributed (IID) data, whereas FedSGD-ISCC exhibits greater robustness to data heterogeneity and communication errors under the evaluated conditions. The tradeoffs in FEEL thus extend beyond physical-resource allocation to the distribution of computation and communication over the learning process.

The temporal coupling of sensing and learning becomes more pronounced when devices retain observations across training rounds. In the accumulated-data framework of Zhou et al.~\cite{zhou2026accumulated_feel}, newly acquired samples augment the local datasets used for subsequent training. Resource allocation therefore needs to account for the contribution of current sensing decisions to future learning rounds. The authors employ Lyapunov optimization to coordinate new data acquisition, computation frequency, bandwidth, and transmission power with a long-term learning objective. In online learning under finite storage and changing data distributions, retaining observations also requires deciding which historical samples remain useful. The limited-cache formulation in~\cite{Hu2025_PSS-Scopus-20250825-151225_0062}, extended to an orthogonal frequency-division multiplexing (OFDM) system by Zhang et al.~\cite{zhang2026online_fed_learning_limited_cache}, coordinates the acquisition of new samples and the reuse of cached historical data. Reusing observations can reduce sensing costs, but concept drift may diminish their contribution to learning. By accounting for this effect in a first-order convergence analysis, the latter work integrates data selection and cache management into the optimization of per-round loss reduction under latency, energy, and cache constraints. The reported human-motion recognition simulations demonstrate the benefits of loss-aware caching over the evaluated caching and resource-allocation baselines.

\subsection{Edge, Split, and Collaborative Inference}

\begin{figure}[!t]
    \centering
    \includegraphics[width=\linewidth]{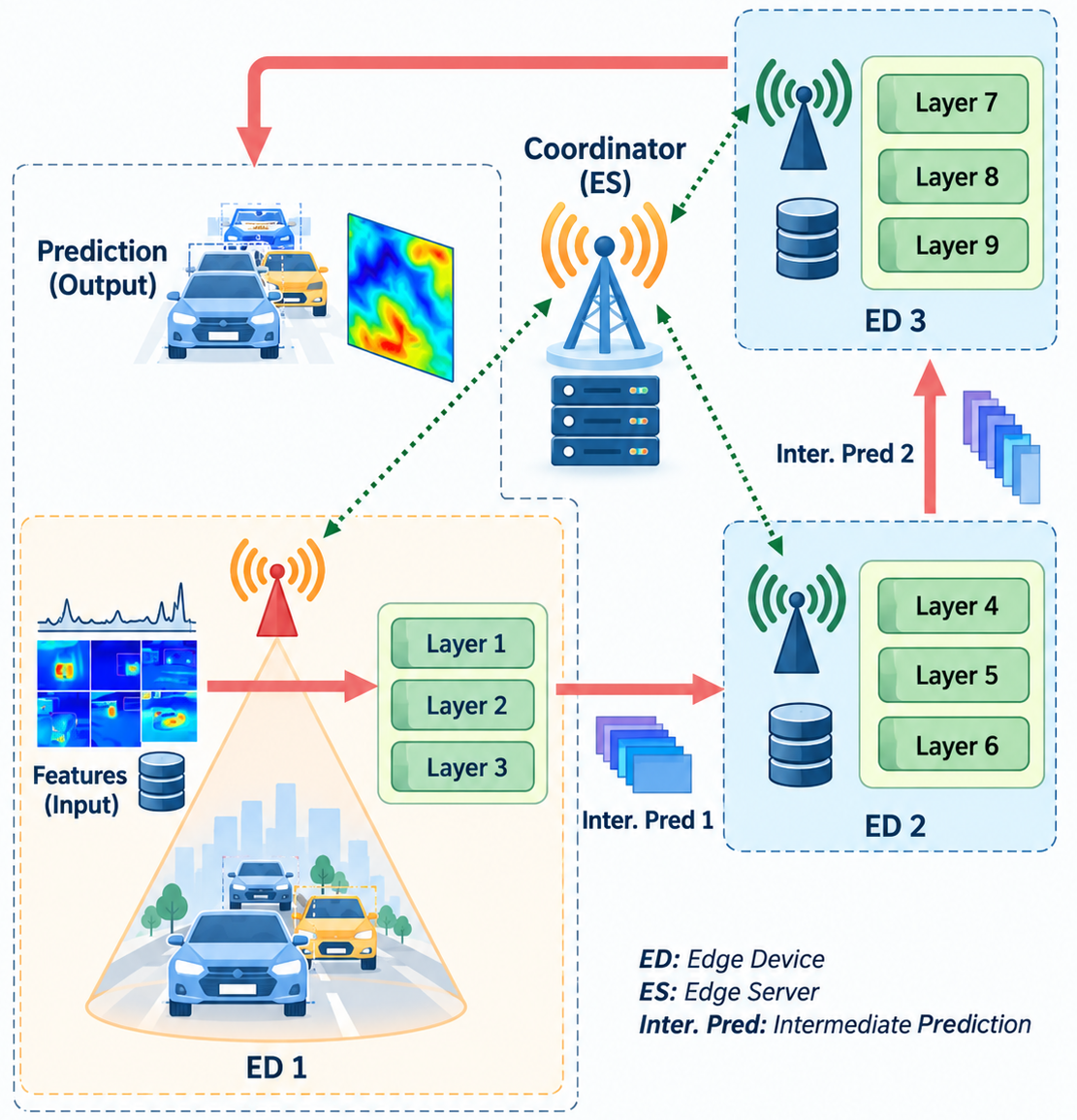}
    \caption{Collaborative split inference across edge devices coordinated by an edge server.}
    \label{fig:split_inference}
\end{figure}

Edge inference changes the optimization objective from learning progress over multiple rounds to the quality, latency, and energy cost of individual or concurrent inference tasks. Distributed execution introduces several forms of functional coupling. Multiple devices can provide complementary observations for a common inference task, while successive portions of a model can be executed at different nodes, as illustrated in Fig.~\ref{fig:split_inference}. In both cases, the quality and size of exchanged representations are closely related to the resources required to meet inference-performance and latency requirements.

For classification tasks, discriminant gain provides an analytical link between the distortions incurred in acquiring and exchanging features and the separability of the underlying classes. Two formulations illustrate how the transmission mechanism and the choice of task-oriented objective influence resource allocation~\cite{wen2024task_oriented_edge_ai,Zhuang20243205_PSS-Scopus-20250825-151225_0067}. In the multi-view inference framework of Wen et al., separately quantized radar features are transmitted to an edge server, and sensing and communication powers, communication time, and quantization resolution are jointly optimized to maximize the average pairwise discriminant gain. The sum-of-ratios formulation enables an optimal solution to the non-convex problem. Zhuang et al. instead aggregate distributed features directly over the wireless channel through AirComp, coupling sensing-power allocation with feature-wise transmit precoding and receive beamforming. Their objective maximizes the minimum pairwise discriminant gain, giving priority to the least separable class pair rather than average class separability. Human-motion recognition experiments in both studies demonstrate the benefits of task-oriented allocation, with Wen et al. also establishing a positive relationship between discriminant gain and the accuracy of support vector machine (SVM) and multilayer perceptron (MLP) classifiers.

The quality and dimensionality of these features can also be shaped by the feature extraction model itself. Wu et al.~\cite{wu2026dae_vae_iscc} combine a denoising autoencoder with a controlled latent-space variational autoencoder to obtain compact sensing representations for UAV-assisted edge inference. The discriminant gain of the resulting features links the learned representation to the joint optimization of sensing, communication, and computation resources.

When several inference tasks share a sensing infrastructure, resource efficiency also depends on whether the same observations can serve multiple tasks. In the centralized multi-view multi-task framework of Wang et al.~\cite{Wang202440814_PSS-Scopus-20250825-151225_0022}, joint device scheduling and resource allocation exploit this opportunity to reduce repeated sensing, feature computation, and transmission. In decentralized inference~\cite{wang2026decentralized_multitask_iscc}, the absence of a central aggregator changes the way observations are exchanged and combined. Resource allocation must then accommodate task-specific feature aggregation and, in full-duplex device-to-device (D2D) networks, the self-interference arising from concurrent transmission and reception. The latter study evaluates concurrent human-motion, gender, and age-group classification, reporting improved inference performance over a successive convex approximation (SCA) benchmark.

In addition to selecting and exchanging observations, distributed inference offers the possibility of adjusting the execution of the model itself. Yao et al.~\cite{Yao2025_PSS-Scopus-20250825-151225_0072} illustrate how the energy cost of split inference depends jointly on model partitioning, pruning, feature quantization, and the allocation of sensing, communication, and computation resources. Their formulation approximates inference accuracy as a function of sensing, quantization, and pruning distortions, enabling device-energy minimization under accuracy and end-to-end latency constraints. Human-motion recognition simulations report energy reductions of up to 40\% over existing methods. At a deadline of $0.6$~s, some baselines become infeasible, demonstrating how the choice of model configuration can affect the ability to satisfy task deadlines.

The spatial distribution of sensing and processing nodes also affects the time required to collect intermediate representations. In UAV-assisted inference~\cite{wen2026uav_assisted_edge_inference}, a UAV sequentially collects locally extracted features and relays them to an edge server, making inference latency dependent on the access sequence and trajectory as well as the resources assigned to each device. Joint mobility and resource optimization reduces end-to-end inference delay relative to the evaluated baselines while satisfying energy and accuracy constraints.

\subsection{Vertical and Multimodal Edge Intelligence}

The distribution of sensing information across devices influences how their observations contribute to learning and inference. Vertical FEEL extends sample-partitioned learning to settings in which devices acquire different features of common samples. In the framework of Cao et al.~\cite{cao2026vfeel}, distributed devices sense the same targets from different perspectives, compute local feature embeddings, and transmit them through AirComp for global training. The resulting convergence analysis reveals that increasing device participation can slow convergence by amplifying the effects of aggregation errors on local model updates. This result complements the sample-quantity and sensing-quality tradeoff discussed earlier for horizontal FEEL. With feature-partitioned data, the learning benefit of additional observations also depends on the fidelity of the jointly aggregated representation. Under per-round latency and overall energy constraints, Cao et al. optimize batch size, sensing and transmission powers, and AirComp denoising factors. In their human-motion recognition simulations, an optimized batch size of approximately 150 yields better learning performance than fixed sizes of 100, 200, and 400. The study also reports higher testing accuracy than the evaluated conventional ISCC-enabled horizontal FEEL baseline.

The contributions of distributed observations can become even more heterogeneous when different sensing modalities are involved. Chen et al.~\cite{chen2026multimodal_edge_intelligence} consider joint edge inference from Wi-Fi, radio-frequency identification (RFID), and millimeter-wave (mmWave) observations, using maximal coding rate reduction (MCR$^2$) to learn compact discriminative features and characterize task-oriented sensing performance. The resulting objective guides the allocation of transmission time and feature-level quantization under delay, energy, and communication constraints. An illustrative result on the XRF55 human-activity dataset concerns the relationship between channel quality and modality-specific information utility. Under one channel configuration, the device with the weakest communication link receives 51~ms of the available 90~ms transmission budget when providing mmWave features, compared with 21~ms when all devices use Wi-Fi. This allocation illustrates why channel quality alone is insufficient to determine transmission priorities when the information conveyed by different sensing modalities contributes differently to inference.

\subsection*{\textbf{Lessons Learned}}

\begin{enumerate}[leftmargin=*]

    \item \textit{Resource efficiency should be assessed over the execution horizon of the AI workload.}
    In learning, resource allocation affects both convergence and the cost of each training round. An improvement in convergence measured in rounds does not necessarily translate into a reduction in training time or energy. For inference, resource efficiency similarly depends on the accuracy achieved within the task deadline. These considerations make the choice of performance target essential to the assessment of ISCC designs.

    \item \textit{The value of sensing information is conditional on the processing and aggregation mechanisms.}
    Increasing the number of observations or participating devices can improve information availability while intensifying competition for resources or amplifying aggregation errors. With shared observations, cached data, and heterogeneous sensing modalities, the contribution of an additional observation further depends on the information already available to the AI task. Resource allocation should therefore account for the marginal task utility of sensing information throughout the processing pipeline.

    \item \textit{Task-oriented surrogates must reflect the performance requirements of the supported AI workload.} Average and minimum pairwise discriminant gain encode different priorities in resource allocation~\cite{wen2024task_oriented_edge_ai,Zhuang20243205_PSS-Scopus-20250825-151225_0067}. While the former captures overall class separability, the latter emphasizes the least separable class pair. The choice of a surrogate should therefore account for the task-performance requirements and the effects of impairments introduced throughout the sensing, communication, and computation functions.

\end{enumerate}

%% file: secs/05-virtuous-cycle.tex
\section{Co-Design of AI and ISCC: The Virtuous Cycle}
\label{sec:codesign}

The preceding sections have examined how AI supports ISCC design and how integrated systems accommodate the resource requirements of AI workloads. Bringing these perspectives together reveals a broader design space in which the requirements of an AI task influence the configuration of ISCC resources, while the resulting sensing, communication, and computation conditions determine how the task can be executed. The reciprocal dependencies illustrated in Fig.~\ref{fig:virtouscycle} establish the basis of AI--ISCC co-design. They can be exploited through joint optimization for a given operating state or through adaptation as that state evolves. When the outcomes of these decisions inform subsequent configurations, the interaction can develop into a virtuous cycle in which improvements in AI performance and ISCC operation reinforce one another. We examine the mechanisms underlying these interactions, summarized in Table~\ref{tab:codesign_virtuous_cycle}.

\begin{table*}[!t]
\centering
\caption{Reciprocal dependencies and mechanisms of AI--ISCC co-design.}
\label{tab:codesign_virtuous_cycle}
\scriptsize
\renewcommand{\arraystretch}{1.15}
\setlength{\tabcolsep}{1.5pt}

\begin{tabularx}{\textwidth}{@{}
>{\centering\arraybackslash}p{0.75cm}
>{\hsize=0.95\hsize\linewidth=\hsize\raggedright\arraybackslash}X
>{\hsize=0.95\hsize\linewidth=\hsize\raggedright\arraybackslash}X
>{\hsize=1.20\hsize\linewidth=\hsize\raggedright\arraybackslash}X
>{\hsize=0.90\hsize\linewidth=\hsize\raggedright\arraybackslash}X
@{}}
\toprule
\textbf{Ref.} &
\textbf{AI Requirement / Adaptation} &
\textbf{ISCC Decisions} &
\textbf{Coupling Mechanism} &
\textbf{Results} \\
\midrule

\multicolumn{5}{l}{\textit{AI requirements driving ISCC design}}\\
\midrule

\cite{du2026flisc3}
&
FEEL convergence under IID and non-IID data
&
Transmit beamforming and sensing/communication time allocation
&
Learning convergence determines the allocation of resources between training-data acquisition and model transmission.
&
Improved training performance under limited resources for both IID and non-IID data.
\\

\cite{Chen2024_PSS-Scopus-20250825-151225_0041}
&
Learning-performance gain of multimodal federated perception
&
Time, frequency, and computation-resource scheduling
&
The marginal learning benefit of resource expansion is balanced against data acquisition, transmission, and training costs.
&
Higher service workload and normalized learning gain than single-process optimization.
\\

\midrule
\multicolumn{5}{l}{\textit{ISCC constraints driving AI adaptation}}\\
\midrule

\cite{Deng2024}
&
DNN split point for UAV--edge target classification
&
User association, target assignment, beamforming, computation allocation, and UAV locations
&
Model splitting redistributes computation and communication loads, while resource conditions and task requirements determine feasible partitions.
&
Higher average sum rate than Edge, MaxODG, and Random partitioning. Fully local execution violates task-latency requirements.
\\

\cite{energy_iscc_edge}
&
On-device, on-server, or cooperative inference
&
Mode-dependent sensing, communication, and computation allocation
&
Inference modes exploit different sensing observations and incur different resource costs. Available resources determine the energy-efficient execution mode.
&
Practical-platform experiments validate the mode-selection analysis.
\\

\cite{alert-ai-Liu2025TowardIE}
&
DNN partition across device, MEC, and cloud tiers
&
ISAC beamforming and device/MEC computation allocation
&
Partitioning distributes feature-transfer and computation loads across heterogeneous tiers according to sensing and resource constraints.
&
Lower inference latency than the evaluated two-tier schemes across different MEC computation capacities.
\\

\cite{Jiao202519767_PSS-Scopus-20250825-151225_0064}
&
Controllable model dropout and batch size
&
Sensing power, communication power, and local computation resources
&
Model complexity determines local training demand, while device-resource availability governs the feasible dropout rate and batch size.
&
Improved learning performance over the evaluated baseline schemes.
\\

\cite{wu2026early_exit_iscc}
&
Early-exit selection and inference accuracy
&
Sensing power, communication power, and UAV hovering location
&
The exit point determines computational depth and inference accuracy, while available resources determine feasible execution depth.
&
Energy-efficient inference through joint early-exit and resource optimization.
\\

\midrule
\multicolumn{5}{l}{\textit{Task-oriented representation and information exchange}}\\
\midrule

\cite{alert-ai-perf10994307}
&
Action-detection threshold and detection time step
&
Sensing sampling rate, selective offloading, and edge computation allocation
&
Detector configuration balances the loss of relevant observations against unnecessary transmission and computation.
&
Higher sensing accuracy than two baselines in commodity-device experiments.
\\

\cite{he2026sensem}
&
Semantic video-block selection and edge inference
&
CSI-based spatial sensing and transmit-power allocation
&
Sensing-derived spatial information guides semantic representation selection, transmission, and downstream inference.
&
Improved video analytics accuracy under stringent latency constraints in WARP experiments.
\\

\cite{alert-ai-beam-xu2026edge}
&
Discriminant gain of the edge classifier
&
Transmit and receive precoding
&
Class separability is expressed as a function of the precoding coefficients and directly determines transceiver design.
&
At $-10$~dB, accuracy gains of approximately 15 and 10 percentage points on synthetic and ModelNet data, respectively, over phase-aligned transmission with matched combining.
\\

\cite{yang2025movable_antennas_iscc}
&
Minimum pairwise discriminant gain
&
Sensing power, transmit precoding, receive beamforming, and antenna positions
&
Task-oriented feature aggregation determines both transceiver parameters and the spatial configuration of the receiving interface.
&
Higher inference accuracy than the evaluated conventional antenna configurations.
\\

\midrule
\multicolumn{5}{l}{\textit{Joint adaptation across time and network state}}\\
\midrule

\cite{alert-ai-uav-11023099}
&
Distributed NN-layer splitting after UAV failures
&
UAV trajectories, resource allocation, target scheduling, and beamforming
&
Topology changes trigger redistribution of inference workloads and simultaneous adaptation of physical resources.
&
Improved group-target sensing performance over some benchmarks under minimum communication and sensing requirements.
\\

\cite{tian2026sense2com}
&
Adaptive early, intermediate, or late fusion
&
Communication scheduling and computation-resource coordination
&
Network and resource conditions determine the fusion strategy, which in turn determines the information exchanged and computation demand.
&
Improved cooperative perception service performance under varying network conditions.
\\

\cite{xu2026task_priority_multimodal_iscc}
&
Task priorities, semantic fusion, and task execution
&
UAV-assisted resource allocation and computation offloading
&
Traffic demand and information freshness alter task utility, guiding adaptation of semantic processing and physical resources.
&
Higher priority-weighted semantic utility, particularly for high-priority tasks.
\\

\cite{zhang2026online_fed_learning_limited_cache}
&
Historical-data selection and online model updates
&
Sensing-resource allocation, OFDM aggregation, cache management, and local computation
&
Concept drift changes the learning value of cached observations and the balance between historical-data reuse and fresh acquisition.
&
Loss-aware caching improves learning performance over the evaluated alternatives.
\\

\bottomrule
\end{tabularx}
\par\smallskip
\parbox{\textwidth}{\scriptsize\raggedright\textit{Abbreviations:} CSI, channel state information; DNN, deep neural network; FEEL, federated edge learning; IID, independent and identically distributed; ISAC, integrated sensing and communication; MaxODG, maximum output data greedy; MEC, mobile edge computing; NN, neural network; OFDM, orthogonal frequency-division multiplexing; UAV, unmanned aerial vehicle; WARP, Wireless Open Access Research Platform.}
\end{table*}

\subsection{AI Requirements as Drivers of ISCC Design}

The influence of AI requirements on ISCC design depends on how task performance is related to resource expenditure. In federated learning, convergence analysis provides a means of establishing this relationship across successive stages of the training process. For instance, the FLISC$^3$ formulation of Du et al.~\cite{du2026flisc3} optimizes transmit beamforming and time allocation to accelerate convergence while satisfying sensing and communication requirements. The sensing-quality constraints in this formulation specify acquisition requirements, whereas the convergence analysis of Air-FEEL~\cite{wen2026airfeel}, discussed in Section~\ref{sec:iscc-for-ai}, incorporates the effects of sensing noise directly into the learning objective. Jiang et al.~\cite{jiang2026isac_for_ai} further characterize the FL optimality gap as a function of sensing signal-to-noise ratio (SNR), local dataset size, and model-upload success probability. Although their resource optimization primarily concerns sensing and communication, this formulation illustrates how the learning objective can guide the balance between data-acquisition quality and model-transfer reliability. The extent to which these effects are captured in convergence analysis thus determines which physical-layer tradeoffs can be evaluated in terms of their consequences for learning.

For inference, the corresponding design problem is to translate the informational requirements of the task into resource-allocation priorities. The discriminant-gain and multimodal feature formulations examined in Section~\ref{sec:iscc-for-ai} provide examples in which the discriminability and complementarity of sensory representations determine the relative value of allocating resources to different observations~\cite{wen2024task_oriented_edge_ai,Zhuang20243205_PSS-Scopus-20250825-151225_0067,chen2026multimodal_edge_intelligence}. The role of task performance also depends on the service objective. Maximizing inference performance under a resource budget gives rise to a different formulation from minimizing energy expenditure under prescribed accuracy and latency requirements~\cite{energy_iscc_edge,Yao2025_PSS-Scopus-20250825-151225_0072}. In the latter case, determining whether a more economical resource configuration is feasible may require adapting the execution of the model itself.

The allocation problem becomes more involved when the same infrastructure supports several AI services. Besides accounting for the individual performance of each task, the system must consider the benefits of sharing observations and the relative utility of devoting resources to different workloads. Multi-view multi-task inference exploits the former by coordinating the acquisition and processing of common observations~\cite{Wang202440814_PSS-Scopus-20250825-151225_0022}. In multimodal federated perception, Chen et al.~\cite{Chen2024_PSS-Scopus-20250825-151225_0041} address the latter through a social-welfare objective that balances learning-performance gains against the cost of acquiring data, transmitting models, and performing training. Their numerical evaluations report higher service workload and normalized learning-performance gain than single-process resource optimization. When the urgency of a task also varies with traffic demand or information freshness, service utility becomes a time-dependent criterion for coordinating semantic fusion, task execution, and resource allocation~\cite{xu2026task_priority_multimodal_iscc}. As the allocation of resources responds to these different notions of task value, the AI execution strategy itself becomes an important degree of freedom for accommodating the resulting resource constraints.

\subsection{ISCC Constraints as Drivers of AI Adaptation}

The reciprocal direction of co-design concerns the adaptation of AI execution to the conditions under which sensory information is acquired and processed. In distributed inference, the choice of execution architecture affects not only the computation load at individual nodes but also the observations available to the model and the information exchanged between nodes. Liu et al.~\cite{energy_iscc_edge} illustrate this coupling by considering on-device, on-server, and cooperative inference with complementary sensing views acquired at the device and edge server. The energy-efficient execution mode depends on resource availability and the required inference accuracy, a relationship examined experimentally on a practical ISCC platform. Finer-grained partitioning offers additional flexibility within distributed architectures. In the multi-UAV target-classification system of Deng et al.~\cite{Deng2024}, jointly optimizing the deep neural network (DNN) split point with association, target assignment, beamforming, computation allocation, and UAV placement yields a higher average communication sum rate than the evaluated fixed-partitioning schemes. Fully local execution achieves a higher sum rate but violates the sensing-task latency requirements under the considered UAV computation capacity. This result illustrates how the apparent communication benefit of a model configuration can be offset by its computation demands.

The dependence of model partitioning on the availability of computation resources becomes more pronounced in heterogeneous infrastructures. A three-tier architecture comprising devices, mobile edge computing (MEC), and the cloud~\cite{alert-ai-Liu2025TowardIE} allows segments of a pre-trained DNN to be distributed across tiers with different computing capacities. Joint partitioning, beamforming, and computation-resource allocation reduce sensing-inference latency relative to the evaluated two-tier schemes over different MEC capacities. The partitioning decision also influences the resources available to other services, as illustrated by the study of concurrent conventional offloading and sensing-inference workloads in~\cite{liu2026multitier_iscc}. Here, edge--cloud partitioning is jointly optimized with computation allocation and hybrid beamforming to manage the tradeoff between computation latency and sensing performance. The distribution of model execution can therefore serve as a mechanism for managing resource contention across workloads as well as for accommodating the demands of an individual inference task.

Whereas partitioning primarily redistributes model execution, altering the model complexity can change the amount of computation required. The joint split-point, pruning, and quantization design in~\cite{Yao2025_PSS-Scopus-20250825-151225_0072} illustrates how these degrees of freedom complement one another by controlling the computation load, the volume of intermediate representations, and their fidelity. Early-exit mechanisms add the possibility of varying the depth of inference itself. Wu et al.~\cite{wu2026early_exit_iscc} relate discriminant gain and inference accuracy to the selected exit point, optimizing it jointly with sensing power, communication power, and UAV hovering location to reduce energy expenditure while supporting task completion. Related choices arise during training, where controllable model dropout adjusts the computation demand of local updates~\cite{Jiao202519767_PSS-Scopus-20250825-151225_0064}, while the choice between FedAvg and FedSGD influences how computation and communication are distributed over the learning process~\cite{liang2026fedavg_fedsgd_iscc}. These alternatives extend resource optimization beyond meeting the demands of a predetermined workload to determining the workload that should be executed under the prevailing system conditions.

\subsection{Task-Oriented Representation and Information Exchange}

The adaptability of the AI workload also extends to the information on which it operates. Decisions concerning observation selection, representation, and transmission influence the computational and communication demands of the task as well as the performance that can be achieved with the retained information. Their interdependence is particularly apparent when task-relevant information can be identified before it reaches the downstream inference model.

At the sensing front end, relevance can vary both over time and across distributed observations. Chen et al.~\cite{alert-ai-perf10994307} exploit temporal relevance through device-side action detection that triggers the offloading of sensing windows for edge recognition. The detector configuration determines the balance between unnecessary transmissions caused by false detections and the loss of relevant observations due to missed actions. Its threshold and detection time step are therefore optimized alongside the sensing sampling rate and edge computation allocation, with commodity-device experiments demonstrating higher sensing accuracy than the evaluated baselines. In a distributed inference setting, Liu and Huang~\cite{liu2026semantic_sensor_selection} exploit spatially distributed information by jointly considering sensor relevance and channel conditions when selecting participating sensors and determining the number of features to transmit. Their evaluations on synthetic data and ModelNet report higher inference accuracy than channel-based selection, relevance-based selection without channel awareness, and all-inclusive feature fusion. Together, these mechanisms illustrate why the decisions governing information acquisition and delivery should account for the contribution of retained observations to inference as well as the resources consumed in processing them.

The relationship between sensing and representation design becomes more direct when information acquired through one sensing mechanism guides the processing of another. The SenSem framework of He et al.~\cite{he2026sensem}, for example, uses spatial information embedded in channel state information (CSI) to identify regions of interest in video frames. Informative video blocks and transmit power are jointly selected, and the sensing-derived spatial information is also incorporated into edge inference. Experiments on the Wireless Open Access Research Platform (WARP) demonstrate improved video analytics accuracy under stringent latency constraints. In this setting, wireless sensing contributes both to the inference task and to the selection of the visual information that is conveyed to the edge server. The same principle of retaining task-relevant information extends to the feature-level quantization strategies discussed in Section~\ref{sec:iscc-for-ai}, where the fidelity of individual feature elements is adjusted according to their discriminative contribution~\cite{wen2024task_oriented_edge_ai,chen2026multimodal_edge_intelligence}.

Whether features are constructed through learned encoders or obtained through conventional signal processing, their statistical properties can in turn guide the design of the physical interface. Building on the use of discriminant gain for over-the-air inference~\cite{Zhuang20243205_PSS-Scopus-20250825-151225_0067}, Xu et al.~\cite{alert-ai-beam-xu2026edge} derive an explicit relationship between class separability and the precoding coefficients, enabling joint transmit and receive precoder optimization. At an equivalent SNR of $-10$~dB with ten classes and sixteen receive antennas, the resulting design improves classification accuracy from approximately 65\% to 80\% on synthetic Gaussian mixture data and from 50\% to 60\% on ModelNet, relative to phase-aligned transmission and matched combining. Yang et al.~\cite{yang2025movable_antennas_iscc} extend task-oriented transceiver design to the spatial configuration of the receiver through movable antennas. By jointly optimizing antenna positions, sensing powers, and transceiver parameters, their formulation maximizes the minimum pairwise discriminant gain and achieves higher inference accuracy than the evaluated conventional schemes. The spatial configuration of the physical interface thus becomes another means of controlling the quality of the representation available to the classifier. Across these stages, the selection of relevant observations determines the potential contribution of the exchanged information, while representation and transceiver design govern how that contribution is preserved under resource constraints.

\subsection{Joint Adaptation Across Time and Network State}

The preceding mechanisms establish reciprocal dependencies that can be exploited in designing an AI--ISCC system for a given operating state. During service execution, however, changes in the physical network, task demand, and data distribution can alter the relative merits of the available configurations. The scope of adaptation then depends on which part of the coupled system is affected by these changes.

Changes in connectivity and available computation capacity may require reconfiguring the distributed processing pipeline. In the multi-UAV group-target awareness framework of Yin et al.~\cite{alert-ai-uav-11023099}, the inference workload of unavailable UAVs is partitioned into neural-network blocks and redistributed among the remaining platforms. A collaborative multi-agent reinforcement-learning algorithm coordinates layer splitting with UAV trajectories, resource allocation, sensing-target scheduling, and beamforming, improving group-target sensing performance over some of the evaluated benchmarks under minimum communication and sensing requirements. In vehicular cooperative perception, connectivity fluctuations can instead alter the relative benefits of exchanging raw observations, intermediate features, or inference outputs. The Sense2Com framework~\cite{tian2026sense2com} therefore coordinates communication scheduling with the selection of early, intermediate, or late fusion through cooperative multi-agent reinforcement learning. In both cases, changes in the network determine how the AI workload is distributed, and the resulting execution strategy alters the traffic and computation loads that the network must accommodate.

Even without a change in topology, evolving service demand can require adjusting the resources and information processing devoted to individual tasks. The UAV-swarm controller of Hou et al.~\cite{hou2025air_iscc_uav} exemplifies AI-aided adaptation of system resources to the network state, using PPO to balance service success and energy expenditure. In task-prioritized multimodal perception~\cite{xu2026task_priority_multimodal_iscc}, a soft actor-critic algorithm additionally adapts semantic fusion and task execution alongside computation offloading and resource allocation. The reported gains in priority-weighted semantic utility, particularly for high-priority tasks, illustrate the value of adapting information processing to changing task priorities under delay constraints. The distinction is important because a learning-based controller can adapt ISCC resources without reconfiguring the AI workload it supports.

In online learning, the consequences of resource decisions also extend beyond the current execution period. Even in the absence of concept drift, accumulating observations across rounds couples current acquisition decisions with subsequent training workloads~\cite{zhou2026accumulated_feel}. As discussed in Section~\ref{sec:iscc-for-ai}, the cache-assisted FEEL formulations in~\cite{Hu2025_PSS-Scopus-20250825-151225_0062,zhang2026online_fed_learning_limited_cache} coordinate historical-data reuse, fresh acquisition, and local training and model exchange under resource constraints. Concept drift can diminish the relevance of cached observations to the current model update. In turn, the choice of observations used for an update influences the model from which subsequent learning proceeds. The value of adaptation must therefore be considered over the learning horizon, rather than solely in relation to the resources saved in the current round.

These forms of adaptation clarify how the reciprocal dependencies in Fig.~\ref{fig:virtouscycle} can be exploited during operation. Joint optimization establishes a coordinated configuration, and state-dependent adaptation revises it as the operating conditions change. A further feedback mechanism arises when measured AI-task outcomes inform subsequent ISCC decisions, whose effects on the task can then be observed and used for further adaptation. Such a mechanism can support a virtuous cycle when the resulting improvements justify the cost of the updates. The benefits demonstrated through joint optimization or adaptation to observed network states provide evidence of functional coupling, while sustained reciprocal improvements depend on the availability, timeliness, and use of task-level feedback.

\subsection*{\textbf{Lessons Learned}}

\begin{enumerate}[leftmargin=*]

    \item \textit{Reciprocal dependencies determine the scope of co-design.}
    The choice of AI configuration can change the feasible set of resource allocations, just as resource constraints can change the feasible set of AI configurations. Identifying these dependencies is essential for determining which model, representation, and system variables should be optimized jointly. Variables with weak coupling may be coordinated separately to avoid unnecessary optimization complexity.

    \item \textit{Information-processing decisions have consequences throughout the integrated system.}
    The loss of task-relevant information at one stage may limit the performance achievable through subsequent resource allocation. Coordinating observation selection, representation, transmission, and execution requires accounting for these downstream effects as well as the marginal resource costs of preserving information.

    \item \textit{The timescales of AI and ISCC adaptation should be coordinated.} Changes in network conditions, task requirements, and data distributions can affect the coupled AI--ISCC system over different timescales. Coordinating their adaptation requires accounting for the time needed to observe the effects of a configuration on task performance and to exploit this feedback in subsequent decisions. The persistence of the resulting benefits must also justify the communication and computation costs of reconfiguration.

\end{enumerate}

%% file: secs/06-synthesis.tex
\section{Comparative Synthesis: System Architectures and Performance Metrics}
\label{sec:comparative_synthesis}

The preceding sections classify ISCC designs according to the role of AI and the interactions between the AI workload and the physical system. Across these categories, architectural choices determine where observations are acquired, what information is exchanged, and how computation is distributed. Their implications for system performance depend on the quantities being optimized and the conditions under which they are evaluated. This section compares these architectural and evaluation dimensions across the surveyed literature.

\subsection{System Architectures}
\label{subsec:system_architectures}

A fundamental architectural distinction concerns the distribution of computation and the organization of information exchange. As summarized in Table~\ref{tab:system_architectures}, centralized device--edge systems share a common physical organization but differ substantially in the information exchanged and the processing assigned to each node. Model updates in FEEL, embeddings in vertical FEEL, and features in edge inference impose different requirements on aggregation and synchronization, even when the same wireless infrastructure is employed~\cite{wen2026airfeel,cao2026vfeel,wen2024task_oriented_edge_ai,Zhuang20243205_PSS-Scopus-20250825-151225_0067}. Hierarchical device--edge--cloud architectures extend this design space through additional execution locations~\cite{alert-ai-Liu2025TowardIE}, whose benefits depend on inter-tier transmission costs and the contention among workloads sharing computation resources~\cite{liu2026multitier_iscc}. Decentralized device-to-device (D2D) architectures, in turn, distribute information fusion and inference among cooperating devices, making peer connectivity and interference integral to the coordination problem~\cite{wang2026decentralized_multitask_iscc,tian2026sense2com}. Decentralized execution does not necessarily imply decentralized resource optimization, particularly when decisions require network-wide channel or task-state information. Conversely, the same air--ground architecture can support centralized or decentralized resource controllers, depending on the information available to each decision-making entity~\cite{Liu2025_PSS-Scopus-20250825-151225_0010}.

Mobility and geographical distribution introduce further architectural dimensions. A UAV may serve as a sensing and learning device, as a mobile relay within a centralized inference architecture, or as a platform colocating sensing, communication, and edge computation~\cite{tang2025uav_feel,wen2026uav_assisted_edge_inference,alert-uav-11298206}. In these configurations, mobility affects different stages of the processing chain, from the acquisition of observations to feature delivery and task execution. Space--air--ground integrated networks (SAGINs) extend the available computation paths by incorporating satellite resources~\cite{zhang2025sagin_iscc}. For hierarchical FEEL, the aerial and satellite links also determine the paths along which model information is exchanged, coupling the learning architecture with inter-tier connectivity and transmission costs~\cite{yang2025hierarchical_feel_sagin}. Geographically distributed terrestrial infrastructures offer analogous opportunities for resource sharing through cooperation among RSUs or distributed access points. While cooperating RSUs can redistribute sensing information and computation tasks~\cite{huang2025collaborative_sensing_offloading}, cell-free access additionally introduces spatial coordination at the radio interface~\cite{peng2026cellfree_iccs}. Increasing the number of cooperating nodes can expand the available sensing, communication, or computation resources, but the resulting benefits depend on the cost of exchanging information and coordinating their use.

These architectural dimensions are therefore complementary rather than mutually exclusive. Distributed wireless access, for example, may coexist with centralized computation, just as mobile sensing may coexist with hierarchical execution. A comparison based solely on the physical network topology would consequently overlook differences in the functional dependencies that must be coordinated. The organization of information exchange and computation provides the basis for identifying where bottlenecks arise and which coordination mechanisms are required.

\begin{table*}[!t]
\centering
\caption{Architectural organization and coordination requirements in ISCC systems.}
\label{tab:system_architectures}
\footnotesize
\renewcommand{\arraystretch}{1.15}
\setlength{\tabcolsep}{1.5pt}

\begin{tabularx}{\textwidth}{@{}
>{\raggedright\arraybackslash}p{1.90cm}
>{\hsize=0.70\hsize\linewidth=\hsize\raggedright\arraybackslash}X
>{\hsize=1.10\hsize\linewidth=\hsize\raggedright\arraybackslash}X
>{\hsize=1.05\hsize\linewidth=\hsize\raggedright\arraybackslash}X
>{\hsize=1.15\hsize\linewidth=\hsize\raggedright\arraybackslash}X
>{\raggedright\arraybackslash}p{1.35cm}
@{}}
\toprule
\textbf{Architecture} &
\textbf{Sensing Location} &
\textbf{Information Flow} &
\textbf{Computation Organization} &
\textbf{Coordination Requirements} &
\textbf{References} \\
\midrule

Centralized device--edge
&
Distributed edge devices
&
Features, embeddings, or model updates sent to a common edge server, optionally via AirComp
&
Local feature extraction or training; edge aggregation, inference, or global-model update
&
Device participation, aggregation quality, and resource allocation across devices
&
\cite{wen2024task_oriented_edge_ai,
Zhuang20243205_PSS-Scopus-20250825-151225_0067,
wen2026airfeel,
cao2026vfeel}
\\

Hierarchical device--edge--cloud
&
Devices and/or edge infrastructure
&
Intermediate representations and computation tasks across tiers
&
Partitioned model execution and task processing across devices, MEC, and cloud
&
Model partitioning, inter-tier transmission, computation allocation, and contention among workloads
&
\cite{alert-ai-Liu2025TowardIE,
liu2026multitier_iscc}
\\

Decentralized D2D
&
Cooperating sensing devices
&
Peer exchange and aggregation of sensed data or intermediate features
&
Local feature extraction, distributed fusion, and device-side inference
&
Peer connectivity, interference management, fusion strategy, and distributed information exchange
&
\cite{wang2026decentralized_multitask_iscc,
tian2026sense2com}
\\

Aerial and mobile
&
Ground devices and aerial platforms
&
Ground--air, air--ground, or air--edge exchange of sensing data, features, or model updates
&
Processing distributed among ground devices, UAVs, and edge servers
&
Mobility, sensing geometry, access scheduling, aerial energy budgets, and computation placement
&
\cite{tang2025uav_feel,
wen2026uav_assisted_edge_inference,
alert-uav-11298206}
\\

Space--air--ground
&
Aerial platforms and terrestrial sensing nodes
&
Task offloading and hierarchical model exchange across terrestrial, aerial, and satellite links
&
Tier-distributed computation and hierarchical federated learning
&
Inter-tier connectivity, offloading, interference, model routing, and heterogeneous computation resources
&
\cite{zhang2025sagin_iscc,yang2025hierarchical_feel_sagin}
\\

Distributed multi-edge
&
Vehicles and infrastructure sensing nodes
&
Sensing information and computation tasks exchanged between devices and cooperating edge servers
&
Local execution or processing at interconnected edge servers
&
Collaborative sensing, redundancy reduction, task redistribution, and inter-edge coordination
&
\cite{huang2025collaborative_sensing_offloading}
\\

Cell-free distributed
&
Multi-antenna terminals
&
Task offloading through distributed access points
&
Local, distributed edge, or centralized cloud execution
&
Access-point cooperation, beamforming, offloading decisions, and execution scheduling
&
\cite{peng2026cellfree_iccs}
\\

\bottomrule
\end{tabularx}
\par\smallskip
\parbox{\textwidth}{\scriptsize\raggedright\textit{Abbreviations:} AirComp, over-the-air computation; D2D, device-to-device; MEC, mobile edge computing; UAV, unmanned aerial vehicle.}
\end{table*}

\subsection{Performance Metrics}
\label{subsec:performance_metrics}

The architectural distinctions discussed above also influence the scope of performance evaluation. Functional metrics, such as sensing error, achievable rate, and computation latency, characterize individual stages of the ISCC processing chain. System-level metrics instead capture their combined effects on service performance and resource efficiency. Even metrics expressed in the same units may characterize different system properties. For instance, the computing efficiency defined by Qi and Wu~\cite{qi2026uav_iscc_efficiency} relates successfully processed computation bits to total UAV energy consumption, whereas conventional communication energy efficiency relates transmitted information to communication energy expenditure. Their comparison requires specifying both the useful output and the resources included in the measurement.

When ISCC supports AI workloads, the quality of intermediate representations provides an additional link between functional performance and the downstream task. Discriminant gain and maximal coding rate reduction, for example, characterize properties of representations that can be exploited in resource optimization~\cite{wen2024task_oriented_edge_ai,Zhuang20243205_PSS-Scopus-20250825-151225_0067,chen2026multimodal_edge_intelligence}. Their role becomes clearer when considered alongside the corresponding learning or inference metrics. As Table~\ref{tab:metric_comparison} illustrates, quantities optimized in one formulation can serve as constraints in another. Energy minimization under inference-accuracy and latency requirements~\cite{Yao2025_PSS-Scopus-20250825-151225_0072,energy_iscc_edge} and latency minimization under accuracy and energy constraints~\cite{wen2026uav_assisted_edge_inference} explore different operating tradeoffs. Reporting only the optimized quantity would therefore provide an incomplete basis for comparison. The accompanying constraints and achieved task performance are needed to establish whether the solutions operate under comparable requirements.

Comparable evaluations must also account for how measurements are aggregated across processing stages, users, and time. End-to-end task delay and individual stage delays characterize different portions of the execution path, while mean, maximum, and percentile latency reflect different aspects of service performance~\cite{peng2026cellfree_iccs,cheng2026d3pg_light_iscc}. The evaluation horizon discussed for learning workloads in Section~\ref{sec:iscc-for-ai} similarly determines whether efficiency refers to individual training rounds or the resources consumed throughout training. In systems employing learning-based controllers, controller convergence must further be distinguished from convergence of the supported learning task. Together, these distinctions explain why common metric names and units are insufficient to establish quantitative comparability across studies.

\begin{table*}[!t]
\centering
\caption{Performance criteria across functional, representation, task, and system levels in ISCC.}
\label{tab:metric_comparison}
\footnotesize
\renewcommand{\arraystretch}{1.15}
\setlength{\tabcolsep}{1.5pt}

\begin{tabularx}{\textwidth}{@{}
>{\centering\arraybackslash}p{0.75cm}
>{\hsize=1.05\hsize\linewidth=\hsize\raggedright\arraybackslash}X
>{\hsize=0.85\hsize\linewidth=\hsize\raggedright\arraybackslash}X
>{\hsize=0.90\hsize\linewidth=\hsize\raggedright\arraybackslash}X
>{\hsize=1.20\hsize\linewidth=\hsize\raggedright\arraybackslash}X
@{}}
\toprule
\textbf{Ref.} &
\textbf{Functional Metrics} &
\textbf{Representation Metrics} &
\textbf{AI / Service Metrics} &
\textbf{End-to-End / System Metrics} \\
\midrule

\cite{cheng2026d3pg_light_iscc}
&
Normalized sensing utility and energy
&
--
&
Controller reward and training stability
&
Multi-objective reward (O); mean and percentile system latency; energy consumption
\\

\cite{qi2026uav_iscc_efficiency}
&
Sensing SINR (C); processed computation bits
&
--
&
Successful computation-task processing
&
Computing efficiency (bits/J) (O); total UAV energy
\\

\cite{Dong20258481_PSS-Scopus-20250825-151225_0013}
&
Communication and computation resource consumption
&
Information value of selected data
&
Region-of-interest coverage (C)
&
Joint information value, execution delay, and energy consumption (O)
\\

\cite{peng2026cellfree_iccs}
&
Achievable rate, sensing requirements (C), and execution latency
&
--
&
--
&
Maximum overall task latency (O)
\\

\midrule

\cite{tang2025uav_feel}
&
Sensing success probability and communication rate
&
--
&
Learning optimality gap (C); training loss and testing accuracy
&
Total federated-training time (O)
\\

\cite{wen2026airfeel}
&
Sensing noise variance
&
AirComp aggregation distortion
&
Per-round loss-reduction bound (O); convergence and testing accuracy
&
Per-round latency and device energy budgets (C)
\\

\cite{cao2026vfeel}
&
Sensing noise and latency/energy components
&
Embedding-aggregation distortion
&
Loss-reduction bound (O); convergence and testing accuracy
&
Per-round latency and overall training energy (C)
\\

\cite{Zhuang20243205_PSS-Scopus-20250825-151225_0067}
&
Sensing distortion and energy budgets (C)
&
Aggregation distortion; minimum pairwise discriminant gain (O)
&
Classification accuracy
&
--
\\

\cite{chen2026multimodal_edge_intelligence}
&
Transmission latency and communication energy
&
Maximal coding rate reduction (O)
&
Recognition accuracy
&
End-to-end latency and device energy budgets (C)
\\

\cite{Yao2025_PSS-Scopus-20250825-151225_0072}
&
Sensing, communication, and computation energy; latency components
&
Feature-quantization distortion and pruning-induced degradation
&
Inference accuracy (C)
&
Total device energy (O); end-to-end inference latency (C)
\\

\cite{wen2026uav_assisted_edge_inference}
&
Sensing, transmission, and execution delays
&
--
&
Inference accuracy (C)
&
End-to-end task delay (O); energy consumption (C)
\\

\midrule

\cite{energy_iscc_edge}
&
Sensing, communication, and computation energy
&
--
&
Inference accuracy (C)
&
Total energy across inference modes (O)
\\

\cite{Deng2024}
&
Communication sum rate (O); received sensing-echo power
&
--
&
Classification accuracy requirement (C)
&
End-to-end sensing-task latency (C)
\\

\cite{xu2026task_priority_multimodal_iscc}
&
Communication and computation latency
&
Semantic information utility
&
Task priorities and information freshness
&
Priority-weighted semantic utility (O); task delay (C)
\\

\bottomrule
\end{tabularx}

\par\smallskip
\parbox{\textwidth}{\scriptsize
(O): optimization objective or its analytical surrogate;
(C): explicitly constrained performance quantity;
--: no quantity selected for this category.
Unmarked quantities are modeled or evaluated performance measures.}

\par\smallskip
\parbox{\textwidth}{\scriptsize\raggedright\textit{Abbreviations:} AirComp, over-the-air computation; SINR, signal-to-interference-plus-noise ratio; UAV, unmanned aerial vehicle.}
\end{table*}

\subsection{Datasets, Benchmarks, and Evaluation Practices}
\label{subsec:datasets_benchmarks_evaluation}

\begin{table*}[!t]
\centering
\caption{Datasets and benchmarking resources for AI--ISCC evaluation.}
\label{tab:iscc_datasets}
\footnotesize
\renewcommand{\arraystretch}{1.13}
\setlength{\tabcolsep}{1.5pt}

\begin{tabularx}{\textwidth}{@{}
>{\raggedright\arraybackslash}p{2.15cm}
>{\raggedright\arraybackslash}p{1.65cm}
>{\hsize=1.00\hsize\linewidth=\hsize\raggedright\arraybackslash}X
>{\hsize=0.85\hsize\linewidth=\hsize\raggedright\arraybackslash}X
>{\hsize=0.90\hsize\linewidth=\hsize\raggedright\arraybackslash}X
>{\hsize=1.25\hsize\linewidth=\hsize\raggedright\arraybackslash}X
@{}}
\toprule
\textbf{Dataset / Resource} &
\textbf{Source} &
\textbf{Signals / Workloads} &
\textbf{Supported Tasks} &
\textbf{System Information} &
\textbf{Evaluation Scope and Limitations} \\
\midrule

\multicolumn{6}{l}{\textit{Measured sensing and communication datasets}}\\
\midrule

XRF55~\cite{wang2024xrf55}
&
Measured
&
Wi-Fi, RFID, mmWave radar, and Kinect
&
Human activity recognition and multimodal perception
&
Synchronized observations; 55 activities, 39 subjects, and four scenes
&
Multimodal and cross-domain sensing; separate models for feature transfer and computation.
\\

MM-Fi~\cite{yang2023mmfi}
&
Measured
&
Wi-Fi CSI, mmWave radar, LiDAR, and RGB-D
&
Activity recognition, pose estimation, and multimodal learning
&
Synchronized observations from 40 subjects with pose annotations
&
Cross-subject and multimodal evaluation; no integrated resource-allocation or edge-execution characterization.
\\

DeepSense 6G~\cite{alkhateeb2023deepsense6g}
&
Measured
&
Wireless communication, radar, LiDAR, cameras, and positioning data
&
Beam prediction, positioning, and environment-aware communication
&
Scenario-dependent synchronized sensing, channel, and mobility information
&
Measured sensing--communication coupling; AI execution and computation costs require separate characterization.
\\

DISC~\cite{pegoraro2025disc}
&
Measured
&
60-GHz IEEE 802.11ay channel impulse responses
&
Activity recognition, gait analysis, tracking, and sparse sensing
&
Communication-native CIRs under different packet-transmission patterns
&
Communication-dependent sensing; limited to recorded configurations and environments.
\\

ISAC-UAV~\cite{beuster2023uav_testbed}
&
Measured
&
Time-varying OFDM channel transfer functions
&
UAV localization and tracking
&
Distributed receiver observations and synchronized UAV ground truth
&
Multistatic aerial sensing; no computation or offloading characterization; access by request.
\\

mmHSense~\cite{bhat2026mmhsense}
&
Measured
&
Wi-Fi and 5G sensing signals
&
Gesture recognition, identification, pose estimation, and localization
&
Six labeled datasets across multiple testbeds and sensing configurations
&
Communication-native human sensing; no joint benchmark of AI execution and resource costs.
\\

\midrule
\multicolumn{6}{l}{\textit{Synthetic sensing and communication resources}}\\
\midrule

DeepMIMO~\cite{alkhateeb2019deepmimo}
&
Ray tracing
&
Parameterized MIMO channels and multipath components
&
Beam prediction, channel learning, and localization
&
Tx/Rx positions, path gains, delays, AoAs, and AoDs
&
Reproducible site-specific channels; fidelity depends on propagation model, scenario, and configuration.
\\

M$^3$SC~\cite{cheng2023m3sc}
&
Synthetic
&
RGB, depth, LiDAR, mmWave radar, and communication CIRs
&
Multimodal perception and communication
&
Aligned physical-scene and electromagnetic-domain data
&
Scene-consistent multimodal evaluation; synthetic-to-real validation and separate computation models required.
\\

SynthSoM~\cite{cheng2025synthsom}
&
Synthetic
&
RF channels, path loss, radar, RGB, depth, and LiDAR
&
Multimodal perception and air--ground sensing--communication tasks
&
Aligned data across scenarios, weather conditions, and frequency bands
&
Diverse simulated conditions; separate models for hardware-dependent computation and closed-loop resource responses.
\\

\midrule
\multicolumn{6}{l}{\textit{Complementary benchmarking frameworks}}\\
\midrule

SDP~\cite{huang2026sensing_dataset_protocol}
&
Benchmark protocol
&
Heterogeneous wireless-sensing signals
&
Detection, recognition, and vital-sign estimation
&
Common data representation, preprocessing, and evaluation procedures
&
Reproducible sensing benchmarks; no integrated communication and computation evaluation.
\\

FedScale~\cite{lai2022fedscale}
&
Data and runtime
&
Federated-learning datasets and device-system profiles
&
Distributed learning and client coordination
&
Client data partitions, computation speed, connectivity, and availability
&
Statistical and system heterogeneity; no explicit sensing acquisition or physical-layer aggregation model.
\\

MLPerf Tiny~\cite{banbury2021mlperf_tiny}
&
Hardware benchmark
&
Standardized embedded inference workloads
&
Classification, keyword spotting, and anomaly detection
&
Measured inference accuracy, latency, and optional energy
&
Hardware-dependent AI execution; no integrated sensing and wireless resource-allocation characterization.
\\

OpenRAN Gym~\cite{bonati2023openran_gym}
&
Experimental platform
&
RAN measurements and programmable control policies
&
Learning-based network control
&
Network telemetry, radio configurations, and experimental control interfaces
&
Closed-loop network control; extensions needed for sensing and computation-aware AI workloads.
\\

\bottomrule
\end{tabularx}
\par\smallskip
\parbox{\textwidth}{\scriptsize\raggedright\textit{Abbreviations:} AoA, angle of arrival; AoD, angle of departure; CIR, channel impulse response; CSI, channel state information; ISAC, integrated sensing and communication; LiDAR, light detection and ranging; MIMO, multiple-input multiple-output; mmWave, millimeter-wave; OFDM, orthogonal frequency-division multiplexing; RAN, radio access network; RF, radio frequency; RFID, radio-frequency identification; RGB, red, green, blue; RGB-D, RGB and depth; SDP, sensing dataset protocol; Tx/Rx, transmitter/receiver; UAV, unmanned aerial vehicle.}
\end{table*}

The empirical basis of current ISCC studies remains heterogeneous, ranging from paper-specific sensing simulations and internally generated observations~\cite{wen2026airfeel,cao2026vfeel,tang2025uav_feel,Yao2025_PSS-Scopus-20250825-151225_0072} to publicly available multimodal data. The latter provide a common basis for task-level evaluation, as illustrated by Chen et al.~\cite{chen2026multimodal_edge_intelligence}, who use eight activity classes from XRF55~\cite{wang2024xrf55} with simulated wireless links and prescribed computation parameters. A shared dataset thus improves reproducibility at the sensing and task levels, while the evaluation of integrated operation still depends on how the remaining stages are modeled.

Table~\ref{tab:iscc_datasets} compares resources that could support broader ISCC evaluation. Measured sensing--communication datasets and synthetic multimodal resources are particularly valuable when their observations and channel information correspond to the same physical environment. This correspondence helps preserve dependencies between sensing quality and communication conditions that would otherwise be difficult to reproduce with independently generated data. Complementary frameworks for distributed learning, embedded inference, and network control address other aspects of the processing chain, including client heterogeneity, hardware-dependent execution, and the response of programmable networks to control decisions. Their combination with sensing--communication resources offers a basis for extending evaluation beyond individual functions.

Despite their complementary coverage, none of the resources in Table~\ref{tab:iscc_datasets} provides a unified benchmark for resource-dependent sensing, communication, computation, and AI execution. Recorded observations and channel traces generally correspond to predetermined configurations, so evaluating changes in sensing powers, sampling rates, waveforms, or antenna configurations requires additional modeling. This limitation is particularly consequential for adaptive ISCC, where control decisions affect subsequent system states and task performance. Synchronized computation traces are also generally absent, leaving hardware-dependent execution latency and energy consumption to separate models. Unless these models preserve dependencies with the sensing and communication stages, the resulting evaluations may overlook the very interactions that motivate joint design.

The scope of generalization and empirical validation must likewise be interpreted in relation to the experimental setting. Random sample-level partitioning may place observations from the same subject, session, or trajectory in both training and testing sets, potentially yielding optimistic estimates of generalization to unseen subjects or environments. For distributed learning, client-level heterogeneity further influences the interpretation of convergence and testing accuracy~\cite{liang2026fedavg_fedsgd_iscc}, while learning-based controllers must contend with changes in network and resource conditions. Practical experiments address a complementary question concerning the fidelity of the system model. He et al.~\cite{he2023integrated} validate sensing-performance characterization using a Universal Software Radio Peripheral (USRP) B210 testbed, while Chen et al.~\cite{alert-ai-perf10994307} evaluate action detection and recognition using commodity wireless devices. The evidence provided by such experiments depends on which stages of the integrated processing chain are implemented and measured.

Within a common experimental setting, baseline and ablation design determine the extent to which performance gains can be attributed to integration. Comparisons in split inference, Air-FEEL, vertical FEEL, and multimodal edge intelligence examine the contributions of model, representation, and resource configurations~\cite{Yao2025_PSS-Scopus-20250825-151225_0072,wen2026airfeel,cao2026vfeel,chen2026multimodal_edge_intelligence}. Fixed-configuration baselines help assess the benefits of adapting individual variables, whereas independently optimized or sequential designs provide a basis for assessing joint-optimization gains. Both require comparable resource budgets, task requirements, and available system information, as well as repeated trials when performance varies across runs. Establishing such equivalence across independently conducted studies is more difficult because dataset partitions, model configurations, computation assumptions, and measurement boundaries vary. The surveyed literature consequently supports structural and methodological comparisons more reliably than direct numerical rankings of ISCC solutions.

\subsection*{\textbf{Lessons Learned}}

\begin{enumerate}[leftmargin=*]

    \item \textit{Architectural comparisons should distinguish resource placement from coordination.}
    To isolate the benefits of a coordination mechanism, architectural comparisons should distinguish changes in the available resources from changes in how those resources are used. Comparisons that modify both the infrastructure and the coordination strategy may otherwise conflate their respective contributions.

    \item \textit{Performance tradeoffs should be compared under equivalent service requirements.}
    Comparisons at common task-performance targets or resource budgets can reveal tradeoffs that isolated objective values obscure. Where multiple objectives compete, evaluating performance across a range of feasible operating conditions is more informative than comparing a single operating point.

    \item \textit{Integrated benchmarking requires preserving cross-functional dependencies.}
    A benchmark should preserve the effects of resource decisions on the information available to the AI workload and the cost of its execution. Otherwise, even accurate models of individual functions may yield misleading conclusions about joint optimization.

\end{enumerate}

%% file: secs/07-applications.tex
\section{Application Domains}
\label{sec:applications}

\begin{figure*}[!t]
    \centering
    \includegraphics[width=\textwidth]{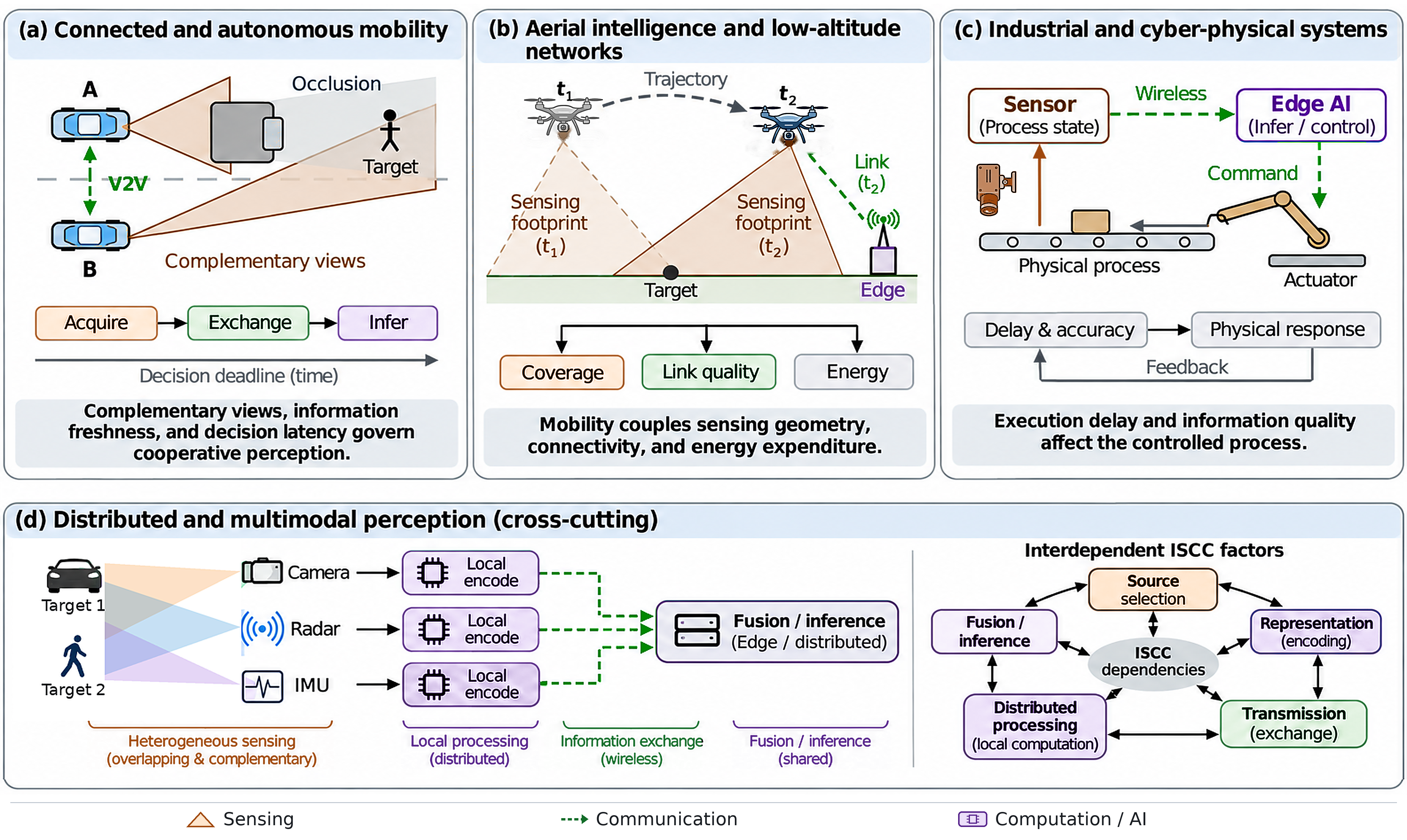}
    \caption{Application domains and their requirements for the joint design of sensing, communication, computation, and AI. IMU: inertial measurement unit; V2V: vehicle-to-vehicle.}
    \label{fig:application_domains}
\end{figure*}

The relative importance of the tradeoffs identified in the preceding sections depends on the physical environment in which ISCC operates and the requirements of the supported tasks. Connected vehicles, aerial networks, and industrial systems exemplify different forms of coupling among sensing, communication, computation, and AI. Distributed and multimodal perception constitutes a complementary perspective that cuts across these domains, highlighting the role of informational dependencies among sensing sources. Fig.~\ref{fig:application_domains} summarizes the requirements and ISCC couplings considered in these settings.

\subsection{Connected and Autonomous Mobility}

Connected and autonomous vehicles require timely perception of dynamic environments. Limitations in sensing range and visibility motivate cooperation among vehicles and infrastructure nodes, whose observations can provide complementary views of the surroundings. However, the utility of this cooperation depends on whether the additional information can be exchanged and processed before changes in the environment render it obsolete. The relevance and freshness of observations thus become central considerations in deciding which sensing tasks to assign, which information to exchange, and where to perform perception~\cite{Dong20258481_PSS-Scopus-20250825-151225_0013,huang2025collaborative_sensing_offloading,tian2026sense2com}.

These considerations extend to the temporal coordination of the perception pipeline. In vehicular task scheduling, the completion of a sensing or processing operation may condition the execution of subsequent tasks~\cite{Cai202413584_PSS-Scopus-20250825-151225_0190}. For traffic monitoring and vehicular digital twins, the perceived state must also be updated in accordance with the requirements of the supported services~\cite{xu2026task_priority_multimodal_iscc,lu2025cooperative_digital_twin}. Meanwhile, vehicle mobility continually alters the connectivity and resource availability on which these operations depend, motivating adaptive resource allocation and beamforming~\cite{Chen202327_PSS-Scopus-20250825-151225_0158,zhao2024iov_iscc,Ren202528201_PSS-Scopus-20250825-151225_0034}. Consequently, evaluating ISCC for autonomous mobility ultimately calls for examining the accuracy and timeliness of the resulting perception in relation to driving decisions and control.

\subsection{Aerial Intelligence and Low-Altitude Networks}

In aerial networks, the spatial configuration of the system is itself a design variable. Unlike fixed infrastructure, UAVs can change their positions to improve sensing geometry, connectivity, or access to distributed computation, although the associated mobility consumes energy. The relative importance of these factors depends on the role of the aerial platform. In UAV-assisted federated learning, aerial deployment influences the availability and quality of training data and the exchange of model updates~\cite{tang2025uav_feel,alert-ai-10959106}. In edge inference, mobility determines the opportunities for collecting features from ground devices within the task's latency budget~\cite{wen2026uav_assisted_edge_inference}. Platforms that also provide target sensing, wireless access, and onboard computation must accommodate the concurrent demands of these functions~\cite{alert-uav-11298206}.

Cooperation among UAVs can extend sensing coverage and distribute computation across platforms~\cite{alert-ai-edge-uav-11095979}, which is particularly valuable in emergency scenarios where terrestrial infrastructure may be unavailable~\cite{ji2026uav_emergency_iscc}. In such settings, the design problem extends beyond efficient operation to the continuity of the supported tasks. The loss of an aerial platform, for instance, may necessitate redistributing both sensing assignments and the associated inference workload~\cite{alert-ai-uav-11023099}. Coordinating mobility and distributed AI execution under energy constraints is therefore essential to sustaining mission performance.

\subsection{Industrial and Cyber-Physical Systems}

Industrial monitoring and recognition applications combine stringent timing and accuracy requirements with the limited energy and processing capabilities of many sensing devices. Resource-efficient sensing and offloading~\cite{n_prim_p018_2025_design_energy_efficient_data}, adaptive inference execution~\cite{energy_iscc_edge,Yao2025_PSS-Scopus-20250825-151225_0072}, and event-driven selective processing~\cite{alert-ai-perf10994307} offer complementary mechanisms for accommodating these constraints. Their suitability depends on the nature of the monitoring task, particularly the extent to which observations or processing operations can be omitted without compromising the detection of relevant changes in the physical environment.

When sensing and inference support closed-loop control, their performance must also be considered in relation to the dynamics of the controlled process. The SeCo4 framework~\cite{tang2025seco4} illustrates this relationship by coordinating sensing, communication, and computing within a cloud--fog architecture to minimize control costs. Related industrial IoT designs incorporate AI and edge computation into resource coordination according to control requirements~\cite{yang2025intelligent_control_iiot}. Since delays and inaccuracies in the information available to the controller can affect the subsequent evolution of the physical process, predictable execution and reliable information delivery assume particular significance in these applications. The quality of the resulting physical-system response thus provides an additional criterion for assessing ISCC designs.

\subsection{Distributed and Multimodal Perception}

Across the preceding domains, perception may rely on observations acquired from multiple viewpoints, devices, or sensing modalities. While redundant observations can provide resilience to noise or missing information, complementary observations can resolve ambiguities that cannot be addressed by individual sources. The balance between redundancy and complementarity determines the informational value of acquiring and exchanging additional observations, and varies with the task and the available sensing configuration.

In multi-view and multi-task inference, for example, a given observation may contribute differently to concurrent perception tasks, creating dependencies between device scheduling and the allocation of resources across tasks~\cite{Wang202440814_PSS-Scopus-20250825-151225_0022,wang2026decentralized_multitask_iscc}. When observations comprise different feature partitions or sensing modalities, their contribution also depends on the representations available for learning and fusion, as discussed in Section~\ref{sec:iscc-for-ai} for vertical FEEL and multimodal inference~\cite{cao2026vfeel,chen2026multimodal_edge_intelligence}. Information acquired through one modality can even guide the selection of information from another, as illustrated by sensing-assisted video analytics~\cite{he2026sensem}. These relationships make the collective utility of the available observations, alongside their individual acquisition and processing costs, a central consideration in the design of distributed perception systems.

%% file: secs/08-future.tex
\section{Open Research Problems and Future Directions}
\label{sec:future}

The integration of AI with ISCC raises research questions at several levels, from the fundamental relationship between physical resources and task performance to the coordination of heterogeneous components in operational systems. Building on the comparative synthesis in Section~\ref{sec:comparative_synthesis}, we examine these questions and the opportunities offered by emerging AI and computing paradigms.

\subsection{Task-Oriented Performance Modeling and Fundamental Limits}
\label{subsec:future_task_limits}

The task-oriented formulations examined in Section~\ref{sec:iscc-for-ai} relate physical resources to learning and inference performance through model- and task-specific quantities such as feature discriminability, aggregation distortion, and learning error~\cite{wen2024task_oriented_edge_ai,Zhuang20243205_PSS-Scopus-20250825-151225_0067,wen2026airfeel,Yao2025_PSS-Scopus-20250825-151225_0072}. A more general characterization requires identifying which properties of the acquired information and its subsequent processing determine attainable task performance. Information-theoretic analyses of ISAC characterize communication--sensing tradeoffs through mutual-information and estimation-theoretic regions~\cite{yu2025isac_fundamental_limits}, while task-oriented semantic communication uses the information bottleneck to relate inference performance to communication latency~\cite{zhao2025task_semantic_ib}. Extending these analytical foundations to encompass computation constraints and AI-task performance remains an open problem.

Such an extension requires characterizing how sensing uncertainty, communication errors, representation compression, computation delay, and model approximation jointly affect task loss or decision utility. Information theory, statistical learning theory, and complexity-aware computation models could provide complementary foundations for deriving attainable performance regions for inference, learning, and closed-loop decision tasks under shared resource constraints. In multimodal and multi-task settings, these regions should also reflect the marginal utility of acquiring and processing additional observations, accounting for informational redundancy and complementarity. A fundamental question is whether, and under what conditions, the attainable region of an integrated system extends beyond that of systems with independently designed functions under equivalent resource budgets.

\subsection{Scalable, Adaptive, and Sustainable Co-Design}
\label{subsec:future_scalable}

Jointly optimizing model configuration, sensing, radio resources, computation, mobility, and topology introduces high-dimensional problems combining discrete and continuous decisions~\cite{Yao2025_PSS-Scopus-20250825-151225_0072,chen2026multimodal_edge_intelligence,wang2026decentralized_multitask_iscc,tang2025uav_feel,zhang2026online_fed_learning_limited_cache}. Their scalability depends partly on whether the design variables can be partitioned into tractable subproblems without losing the benefits of their coupling. Hierarchical, distributed, and multi-timescale optimization offer complementary approaches to exploiting differences in the spatial scope and temporal dynamics of these decisions. An associated research problem is to determine the information that must be exchanged across subproblems and the frequency of coordination needed to maintain performance. For adaptive designs, this requires weighing the expected benefits of reconfiguration against the costs of model repartitioning, migration, retraining, and controller updates.

The ability to generalize across operating conditions is also central to adaptive co-design. Changes in topology, sensing environments, device capabilities, and data distributions can invalidate learned resource-control policies as well as the models used to predict AI-task performance. Transfer learning, meta-learning, continual learning, and uncertainty-aware optimization offer potential mechanisms for adapting under limited data and computation budgets. When adaptation occurs at both the controller and workload levels, however, changes to one may alter the operating conditions of the other. This interaction raises questions concerning the stability of coupled updates, the efficiency of adaptation, and the preservation of previously acquired knowledge.

The costs of adaptation also enter the broader question of sustainability. As illustrated by energy-aware inference and aerial ISCC designs~\cite{energy_iscc_edge,Yao2025_PSS-Scopus-20250825-151225_0072,alert-uav-11298206}, resource savings in one function may come at the expense of another. End-to-end energy models should account for the acquisition, exchange, and processing of information, including memory access, mobility, and the training and adaptation of AI models where applicable. Coupling these models with the duration and frequency of task execution would help determine whether the benefits of more resource-efficient configurations persist over the system's operating horizon. Longer-term sustainability assessments should additionally consider the environmental costs associated with infrastructure deployment and operation.

\subsection{Robustness, Security, Privacy, and Trustworthiness}
\label{subsec:future_trust}

The dependencies underlying ISCC introduce pathways through which failures and adversarial perturbations can affect multiple functions. Existing research addresses vulnerabilities at different stages of the processing chain, including privacy-aware device scheduling~\cite{Wang2023957_PSS-Scopus-20250825-151225_0138}, secure and covert transmission in UAV-assisted ISCC~\cite{lei2025secure_uav,zhang2026covert_uav_iscc}, and trustworthy data processing under shared computation-resource constraints~\cite{huang2025trustworthy_iot}. The broader edge-intelligence literature further identifies reliability, security, transparency, and sustainability as interacting dimensions of trustworthiness~\cite{wang2025trustworthy_edge_intelligence}.

An integrated treatment requires characterizing the propagation of vulnerabilities across the processing chain and, in adaptive systems, across successive control decisions. Corrupted sensing observations, distorted intermediate representations, or compromised computation and aggregation can degrade the AI task, while unreliable outputs from an AI-aided controller can affect subsequent resource configurations. Cross-functional threat modeling could help establish where uncertainty-aware control, robust distributed learning, privacy-preserving information exchange, and safe fallback mechanisms are most effective. Their design requires characterizing security--latency, privacy--utility, and robustness--resource tradeoffs in relation to task-specific requirements. Standardization activities concerning ISAC security, privacy, and trustworthiness provide an initial foundation for extending these considerations to computation and AI execution~\cite{threedgpp2026tr33777_isac_security,etsi2026gr_isc004}.

\subsection{Datasets, Benchmarks, and Experimental Validation}
\label{subsec:future_benchmarks}

The limitations identified in Section~\ref{sec:comparative_synthesis} motivate the development of benchmarking environments that preserve the dependencies between resource decisions and AI-task performance. A key requirement is to support controlled changes in system configurations while maintaining consistency among sensing observations, wireless conditions, and computation states.

Synchronized multimodal observations and task labels could be combined with channel and topology traces, hardware-dependent computation profiles, memory availability, energy measurements, and model configurations. Resource-responsive simulation environments could then extend these empirical foundations by modeling how sensing, transmission, and computation decisions influence subsequent observations and task outcomes. Common evaluation protocols should specify which variables are adapted, which resource and task constraints are held constant, and how training and testing conditions are separated. Such protocols would facilitate comparisons across architectures and the attribution of performance gains to individual or jointly adapted design variables.

The fidelity of these environments must ultimately be assessed against practical implementations. Experimental ISAC platforms and emerging implementations combining ISAC with machine learning provide foundations for extending validation to the complete AI--ISCC processing chain~\cite{luo2026isac_testbeds,tirchas2025ml_isac_experiment}. Hardware-in-the-loop and end-to-end testbeds could help characterize synchronization costs, controller overhead, runtime variability, and hardware-dependent energy consumption, and thereby establish the conditions under which modeled benefits translate into practical gains. This would support progressive validation from reproducible simulation to implementation under comparable task and resource requirements.

\subsection{Implementation, Interoperability, and Standardization}
\label{subsec:future_standardization}

Practical co-design requires interfaces through which sensing, networking, computing, and AI components can expose the information needed for joint decisions. Depending on the workload, this information may include sensing uncertainty, task deadlines, representation dimensions, model split points, aggregation quality, accelerator availability, or learning progress. The problem-specific forms in which these quantities are represented in the surveyed literature limit the portability of joint optimization strategies across workloads and implementations.

The IMT-2030 framework~\cite{itu2030framework} identifies AI and Communication and Integrated Sensing and Communication among its usage scenarios. In the 3rd Generation Partnership Project (3GPP), ISAC standardization activities extend from Release~19 service requirements to Release~20 studies of system architecture and the New Radio (NR) air interface~\cite{threedgpp2024ts22137_isac,threedgpp2026tr2370014_isac,threedgpp2026tr38765_isac}, alongside studies of AI and machine learning (ML) for the NR air interface initiated in Release~18~\cite{threedgpp2023tr38843_ai_ml_nr}. The European Telecommunications Standards Institute (ETSI) group report GR ISC~005~\cite{etsi2026gr_isc005} further investigates the integration of computing with ISAC, identifying key challenges, potential requirements, and solution approaches. These developments provide complementary foundations for ISCC standardization. An important research direction is the development of workload-aware abstractions and interoperable interfaces that expose task intent, sensing quality, representation and model state, communication conditions, computing availability, and timing requirements across heterogeneous devices and infrastructure. Such abstractions could enable coordinated resource orchestration without dependence on a particular AI architecture or hardware platform.

\subsection*{Outlook: Emerging Application and System Frontiers}

Emerging AI and computing paradigms offer new opportunities for addressing these problems and introduce additional dependencies among sensing, communication, and computation.

\paragraph{Foundation Models and Frontier AI.}
Large language, vision, multimodal, and world models introduce substantial communication, memory, and computation requirements, motivating research on distributed execution and resource-efficient deployment at the network edge~\cite{jiang2026large_ai_models,qu2025edge_llm}. Their application to ISCC spans both system optimization and the execution of AI workloads. The use of an LLM to refine the reward of a digital-twin-assisted ISCC controller~\cite{huang2026dt_llm_iscc} illustrates the former, raising questions about the validity of learned or generated reward functions across heterogeneous service requirements and operating conditions. For foundation models deployed as workloads, execution could be coordinated with the acquisition and selection of contextual information, so that the cost of sensing and exchanging additional observations is assessed alongside the cost of processing them. Agentic architectures, which have also been explored for autonomous orchestration of network control and user-plane services~\cite{li2025agent_as_service}, extend this coupling to task-driven interaction with the physical environment. A research direction is to jointly determine what information an agent should acquire, how and where it should process that information, and when the resulting decisions should trigger further sensing or control actions.

\paragraph{Digital Twins, Extended Reality, and Embodied Intelligence.}
Digital twins provide representations of network and physical-system states that can support prediction and control, as illustrated by mobile augmented reality systems linking traffic generated by simultaneous localization and mapping (SLAM) to edge communication management~\cite{zhou2026dt_mobile_ar} and by the reconstruction of delayed network state for ISAC control~\cite{tiwari2026dt_isac}. In extended reality (XR) and embodied AI, the utility of such representations depends on their freshness and fidelity relative to the requirements of localization, inference, rendering, and physical action. The coupling becomes particularly important when the actions of AI agents influence their subsequent sensing and connectivity conditions, as in wireless-aware planning and collaborative navigation~\cite{li2025embodied_nearfield}. Their convergence with ISCC opens research opportunities in jointly managing the acquisition of observations, maintenance of virtual states, information exchange, and execution of time-sensitive perception and control tasks.

\paragraph{Neuromorphic ISCC.}
Neuromorphic sensing, communication, and computation provide a common event-driven foundation for ISCC. NeuroComm~\cite{chen2023neuromorphic_wireless_cognition} illustrates this compatibility through a remote inference architecture integrating spike-based sensing, spiking neural network (SNN) processing, and impulse-radio transmission. The potential for preserving event-driven operation across different stages of the processing chain is further illustrated by neuromorphic semantic communication with wake-up radios~\cite{chen2024neuromorphic_wakeup}, wireless split computing with multi-level spikes and software-defined-radio validation~\cite{wu2025neuromorphic_split}, and neuromorphic multi-task edge inference over shared wireless access~\cite{li2025neuromorphic_multitask}. A broader Neuromorphic ISCC paradigm would exploit the dependencies between event generation, spike representation, waveform design, radio activation, SNN partitioning, and edge computation. A central research question is how to preserve event-driven sparsity and task-relevant temporal information throughout the processing chain while exploiting shared physical resources for sensing and communication. Establishing the resulting system-level efficiency requires accounting for the costs of spike encoding, wireless transmission, and neuromorphic execution.

\paragraph{Quantum-Enhanced ISCC.}
Quantum optical ISAC introduces joint communication and sensing formulations in which communication reliability is optimized under estimation-accuracy constraints~\cite{krikidis2026quantum_isac}. A longer-term question is whether quantum-assisted computation can extend the benefits of such integration to the joint management of sensing, communication, and computation resources. Research on quantum deep reinforcement learning for wireless resource allocation~\cite{astiti2026qdrl_wireless} offers a complementary starting point for investigating quantum-assisted optimization. Establishing system-level advantages over classical approaches requires accounting for quantum hardware limitations, quantum--classical interfaces, and implementation overhead.

%% file: secs/09-conclusion.tex
\section{Conclusion}
\label{sec:conclusion}

This survey has examined the interplay between AI and ISCC from two complementary perspectives, namely AI-aided ISCC and AI-oriented ISCC. Through a review of existing approaches, we have explored how AI enables the joint optimization of sensing, communication, and computation, and how their integration supports diverse learning and inference tasks. Building on these two perspectives, we have analyzed the reciprocal dependencies between AI and ISCC, highlighting their potential to foster a virtuous cycle of mutual performance enhancement.
The comparative analysis of system architectures, performance metrics, and evaluation practices has revealed the dependence of integration gains on task requirements, functional coupling, and resource constraints. The discussion of representative application domains further illustrates how these dependencies vary across operating environments. Despite the progress reported in the literature, fundamental performance limits, scalable and adaptive co-design, and experimental validation remain open research challenges.

%% file: common/acknowledgment.tex
\section*{Acknowledgment}

The authors used ChatGPT (OpenAI) to assist in refining the manuscript, including grammar checking, reformulation, and improvements to clarity and coherence, as well as refining the illustrations in Figs.~2, 4, and 8.

%% file: references/BSTcontrol.bib
@IEEEtranBSTCTL{IEEEexample:BSTcontrol,
  CTLuse_url = "no",
  CTLuse_doi = "no",
  CTLuse_forced_etal       = "yes",
  CTLmax_names_forced_etal = "6",
  CTLnames_show_etal       = "1",
  CTLdash_enumeration = "no",
}


%% file: references/ai-for-iscc_cleaned.bib
@article{Dong20258481_PSS-Scopus-20250825-151225_0013,
  author = {Dong, Mengyuan and Fu, Yuchuan and Li, Changle and Tian, Mengqiu and Richard Yu, F. and Cheng, Nan},
  journal = {IEEE Trans. Intell. Transp. Syst.},
  number = {6},
  pages = {8481 -- 8496},
  title = {Task Offloading and Resource Allocation in Vehicular Cooperative Perception With Integrated Sensing, Communication, and Computation},
  volume = {26},
  year = {2025},
  doi = {10.1109/TITS.2025.3542365}
}

@article{Wang202440814_PSS-Scopus-20250825-151225_0022,
  author = {Wang, Diao and Wen, Dingzhu and He, Yinghui and Chen, Qimei and Zhu, Guangxu and Yu, Guanding},
  journal = {IEEE Internet Things J.},
  number = {24},
  pages = {40814 -- 40830},
  title = {Joint Device Scheduling and Resource Allocation for {ISCC}-Based Multiview-Multitask Inference},
  volume = {11},
  year = {2024}
}

@article{Chen2024_PSS-Scopus-20250825-151225_0041,
  author  = {Chen, Ning and Cheng, Zhipeng and Fan, Xuwei
             and Liu, Zhang and Huang, Bangzhen and Zhao, Yifeng
             and Huang, Lianfen and Du, Xiaojiang and Guizani, Mohsen},
  title   = {Integrated Sensing, Communication, and Computing
             for Cost-Effective Multimodal Federated Perception},
  journal = {ACM Trans. Multimedia Comput. Commun. Appl.},
  volume  = {20},
  number  = {8},
  note    = {Art. no. 237},
  year    = {2024},
  doi     = {10.1145/3661313}
}

@article{Yang202418608_PSS-Scopus-20250825-151225_0131,
  author  = {Yang, Liu and Wei, Yifei and Feng, Zhiyong
             and Zhang, Qixun and Han, Zhu},
  title   = {Deep Reinforcement Learning-Based Resource Allocation
             for Integrated Sensing, Communication,
             and Computation in Vehicular Network},
  journal = {IEEE Trans. Wireless Commun.},
  volume  = {23},
  number  = {12},
  pages   = {18608--18622},
  year    = {2024},
  doi     = {10.1109/TWC.2024.3470873}
}

@inproceedings{Wang2023957_PSS-Scopus-20250825-151225_0138,
  author    = {Wang, Diao and Wen, Dingzhu
               and He, Yinghui and Yu, Guanding},
  title     = {Device Scheduling for Privacy-Aware Integrated
               Sensing, Computation, and Communication Systems},
  booktitle = {Proc. IEEE GLOBECOM Workshops},
  pages     = {957--962},
  year      = {2023},
  doi       = {10.1109/GCWkshps58843.2023.10464877}
}

@article{Chen202327_PSS-Scopus-20250825-151225_0158,
  author = {Chen, Jiujiu and Guo, Caili and Lin, Runtao and Feng, Chunyan},
  journal = {China Commun.},
  number = {3},
  pages = {27 -- 42},
  title = {Tasks-oriented joint resource allocation scheme for the Internet of vehicles with sensing, communication and computing integration},
  volume = {20},
  year = {2023}
}

@article{Cai202413584_PSS-Scopus-20250825-151225_0190,
  author = {Cai, Xuelian and Fan, Yixin and Yue, Wenwei and Fu, Yuchuan and Li, Changle},
  journal = {IEEE Trans. Veh. Technol.},
  number = {9},
  pages = {13584 -- 13599},
  title = {Dependency-Aware Task Scheduling for Vehicular Networks Enhanced by the Integration of Sensing, Communication and Computing},
  volume = {73},
  year = {2024}
}

@article{Ren202528201_PSS-Scopus-20250825-151225_0034,
  author  = {Ren, Ruixing and Zhao, Junhui and Zhang, Qingmiao
             and Wang, Dongming and Li, Jiamin},
  title   = {{DRL} Beamforming in {RIS}-Aided {IoV} for
             Integrated-Sensing-Communication-Computation},
  journal = {IEEE Internet Things J.},
  volume  = {12},
  number  = {14},
  pages   = {28201--28213},
  year    = {2025},
  doi     = {10.1109/JIOT.2025.3566380}
}

@inproceedings{ran_slicing,
  author    = {Wang, Ranyin and Friderikos, Vasilis},
  title     = {{RAN} Slicing Design Policy for Integrated
               Sensing, Communication, and Computation},
  booktitle = {Proc. Joint Eur. Conf. Netw. Commun.
               \& 6G Summit (EuCNC/6G Summit)},
  pages     = {309--314},
  year      = {2025},
  doi       = {10.1109/EuCNC/6GSummit63408.2025.11036910}
}


%% file: references/f03_coverage.bib
@article{Liu2023158_PSS-Scopus-20250825-151225_0128,
  author    = {Liu, Peixi and Zhu, Guangxu and Wang, Shuai and Jiang, Wei and Luo, Wu and Poor, H. Vincent and Cui, Shuguang},
  title     = {Toward Ambient Intelligence: Federated Edge Learning With Task-Oriented Sensing, Computation, and Communication Integration},
  journal   = {IEEE J. Sel. Topics Signal Process.},
  volume    = {17},
  number    = {1},
  pages     = {158--172},
  year      = {2023},
  doi       = {10.1109/JSTSP.2022.3226836}
}

@article{learning-based-iscc,
  author    = {Yang, Mengxin and Gu, Yixiao and Hu, Han and Zeng, Dan},
  title     = {Learning-Based Resource Allocation for Integrated Sensing, Communication, and Computation Networks: A Delay-Aware Approach},
  journal   = {IEEE Internet Things J.},
  volume    = {13},
  number    = {4},
  pages     = {6613--6625},
  year      = {2026},
  doi       = {10.1109/JIOT.2025.3636551}
}

@article{huang2026dt_llm_iscc,
  author    = {Huang, Bangzhen and Liu, Zhang and Huang, Lianfen and Fan, Xuwei and Chen, Ning and Wang, Xianbin},
  title     = {Digital Twin-Assisted Resource Allocation in {ISCC}-Enabled {IoV}: When {DRL} Meets {LLMs}},
  journal   = {IEEE Trans. Netw. Sci. Eng.},
  volume    = {13},
  pages     = {10479--10495},
  year      = {2026},
  doi       = {10.1109/TNSE.2026.3713561}
}

@article{zhou2026accumulated_feel,
  author    = {Zhou, Yong and An, Qiaochu and Wang, Zhibin and Shan, Hangguan and Shi, Yuanming and Zhou, Haibo},
  title     = {Integrated Sensing, Computation, and Communication Enabled Federated Edge Learning},
  journal   = {IEEE Trans. Wireless Commun.},
  volume    = {25},
  pages     = {7117--7131},
  year      = {2026},
  doi       = {10.1109/TWC.2025.3628961}
}

@inproceedings{hou2025semi_supervised_iscc,
  author    = {Hou, Xiangwang and Wang, Jingjing and Wang, Jiacheng and Du, Jun and Zhang, Zekai and Jiang, Chunxiao and Ren, Yong},
  title     = {Energy-Efficient Federated Semi-Supervised Learning for {UAV}-Enabled Integrated Sensing, Computation, and Communication},
  booktitle = {Proc. IEEE Int. Conf. Commun. (ICC)},
  pages     = {6680--6686},
  year      = {2025},
  doi       = {10.1109/ICC52391.2025.11161423}
}

@article{Liu2025_PSS-Scopus-20250825-151225_0010,
  author    = {Liu, Yaxi and Mao, Wencan and Li, Xulong and Huangfu, Wei and Ji, Yusheng and Xiao, Yu and Long, Keping},
  title     = {Joint Task Scheduling and Resource Allocation for {UAV}-Assisted Air-Ground Collaborative Integrated Sensing, Computation, and Communication},
  journal   = {IEEE Trans. Commun.},
  volume    = {73},
  number    = {11},
  pages     = {10929--10945},
  year      = {2025},
  doi       = {10.1109/TCOMM.2025.3591748}
}

@inproceedings{yang2025hierarchical_feel_sagin,
  author    = {Yang, Zhanpeng and Zhang, Pengfei and Zhu, Jingyang and Wen, Dingzhu and Shi, Yuanming and Chen, Wei},
  title     = {Hierarchical Federated Learning with Integrated Sensing-Communication-Computation Over Space-Air-Ground Integrated Networks},
  booktitle = {Proc. IEEE Int. Conf. Commun. (ICC)},
  pages     = {1163--1168},
  year      = {2025},
  doi       = {10.1109/ICC52391.2025.11161256}
}

@article{wu2026ris_feel_iscc,
  author    = {Wu, S. and Liu, X. and Zheng, T. and Liu, Z. and Leung, V. C. M.},
  title     = {Reconfigurable Intelligent Surface-Assisted Federated Edge Learning With Integrated Sensing, Computation and Communication},
  journal   = {IEEE Trans. Mobile Comput.},
  year      = {2026},
  note = {early access},
  doi       = {10.1109/TMC.2026.3737851}
}

@article{wu2026dae_vae_iscc,
  author    = {Wu, Guofang and He, Yejun and Cao, Xiaowen and Guizani, Mohsen},
  title     = {{UAV}-Assisted Integrated Sensing, Communication and Computation Edge Inference: Task-Driven Joint Design With {DAE-VAE}},
  journal   = {IEEE Trans. Mobile Comput.},
  volume    = {25},
  number    = {12},
  pages     = {24546--24562},
  year      = {2026},
  doi       = {10.1109/TMC.2026.3736281}
}


%% file: references/future_cleaned.bib
@article{li2024mean,
  author = {Wang, Dezhi and Huang, Chongwen and He, Jiguang and Chen, Xiaoming and Wang, Wei and Zhang, Zhaoyang and Han, Zhu and Debbah, Mérouane},
  journal = {IEEE Trans. Wireless Commun.},
  number = {8},
  pages = {10430-10444},
  title = {Mean Field Game-Based Waveform Precoding Design for Mobile Crowd Integrated Sensing, Communication, and Computation Systems},
  volume = {23},
  year = {2024}
}

@article{lei2025secure_uav,
  author = {Lei, Hongjiang and Jiang, Congke and Park, Ki-Hong and Aboulhassan, Mohamed A. and Zhou, Sen and Pan, Gaofeng},
  journal = {IEEE Internet Things J.},
  number = {19},
  pages = {40851-40862},
  title = {On Secure {UAV}-Aided {ISCC} Systems},
  volume = {12},
  year = {2025}
}


%% file: references/gscholar-alert_cleaned.bib
@article{alert-ai-perf10994307,
  author  = {Chen, Weiwei and He, Yinghui and Yu, Guanding
             and Wang, Jianfeng and Luo, Haiyan},
  title   = {Sensing Framework Design and Performance
             Optimization With Action Detection for {ISCC}},
  journal = {IEEE Trans. Wireless Commun.},
  volume  = {24},
  number  = {10},
  pages   = {8361--8375},
  year    = {2025},
  doi     = {10.1109/TWC.2025.3566078}
}

@article{alert-ai-edge-uav-11095979,
  author = {Zhao, Hui and Luan, Mingan and Liyanage, Madhusanka and Chang, Zheng},
  journal = {IEEE Trans. Wireless Commun.},
  month = {},
  number = {},
  pages = {1272-1286},
  title = {Joint Optimization of Sensing, Communication, and Computing for Collaborative Multi-{UAV} Edge Computing System},
  volume = {25},
  year = {2026}
}

@article{alert-ai-uav-11023099,
  author  = {Yin, Baolin and Fang, Xuming and Wang, Xianbin},
  title   = {Robust Group Target Awareness Inference in
             Multi-{UAV}-Enabled {ISCC} Networks Based on
             Split Deep Reinforcement Learning},
  journal = {IEEE Trans. Wireless Commun.},
  volume  = {24},
  number  = {11},
  pages   = {9478--9492},
  year    = {2025},
  doi     = {10.1109/TWC.2025.3573654}
}

@article{alert-ai-Liu2025TowardIE,
  author = {Peng Liu and Zesong Fei and Xinyi Wang and Xiaoyang Li and Weijie Yuan and Yuanhao Li and Cheng Hu and Dusit Niyato},
  journal = {IEEE Trans. Wireless Commun.},
  pages = {6774--6789},
  title = {Toward Intelligent Edge Sensing for {ISCC} Network: Joint Multi-Tier {DNN} Partitioning and Beamforming Design},
  volume = {25},
  year = {2026},
  doi = {10.1109/TWC.2025.3626555}
}

@article{alert-ai-beam-xu2026edge,
  author = {Xu, Lingyun and Wang, Bowen and Li, Huiyong and Cheng, Ziyang},
  title = {Edge {AI} Inference in {ISCC} Networks: Sensing Accuracy Analysis and Precoding Design},
  journal = {IEEE Signal Process. Lett.},
  volume = {33},
  pages = {2026--2030},
  year = {2026},
  doi = {10.1109/LSP.2026.3683807}
}

@article{alert-ai-10959106,
  author = {Fu, Yang and Qin, Peng and Tang, Guoming and Zhao, Xiongwen},
  journal = {IEEE Trans. Veh. Technol.},
  month = {Sep.},
  number = {9},
  pages = {13909--13924},
  title = {Joint Design of Sensing, Communication, and Computation for Multi-{AAV}-Enabled Over-the-Air Federated Learning},
  volume = {74},
  year = {2025},
  doi = {10.1109/TVT.2025.3559147}
}

@article{alert-uav-11298206,
  author = {Chen, Yiyang and Liu, Wenchao and Zhang, Xuhui and Ren, Jinke and Xing, Huijun and Wang, Shuqiang and Shen, Yanyan and Tsang, Kim-Fung},
  title = {Full-Duplex Integrated Sensing, Communication, and Computation Over Low-Altitude Wireless Networks},
  journal = {IEEE Trans. Consum. Electron.},
  volume = {72},
  number = {1},
  pages = {925--937},
  year = {2026},
  doi = {10.1109/TCE.2025.3642992}
}

@article{yet-10628047,
  author = {Etiabi, Yaya and Njima, Wafa and Amhoud, El Mehdi},
  journal = {IEEE Internet Things J.},
  number = {22},
  pages = {36991-37007},
  title = {{FeMLoc}: Federated Meta-Learning for Adaptive Wireless Indoor Localization Tasks in {IoT} Networks},
  volume = {11},
  year = {2024}
}

@inproceedings{ahmed2024unified,
  author = {Ahmed, Abdullahi Isa and Etiabi, Yaya and Azim, Ali Waqar and Amhoud, El Mehdi},
  booktitle = {IEEE 35th Int. Symp. Personal, Indoor Mobile Radio Commun. (PIMRC)},
  organization = {IEEE},
  pages = {1--6},
  title = {A unified deep transfer learning model for accurate {IoT} localization in diverse environments},
  year = {2024}
}


%% file: references/introduction_cleaned.bib
@article{Wang2023vision,
  author = {Wang, Cheng-Xiang and You, Xiaohu and Gao, Xiqi and Zhu, Xiuming and Li, Zixin and Zhang, Chuan and Wang, Haiming and Huang, Yongming and Chen, Yunfei and Haas, Harald and Thompson, John S. and Larsson, Erik G. and Renzo, Marco Di and Tong, Wen and Zhu, Peiying and Shen, Xuemin and Poor, H. Vincent and Hanzo, Lajos},
  journal = {IEEE Commun. Surveys Tuts.},
  number = {2},
  pages = {905-974},
  title = {On the Road to {6G}: Visions, Requirements, Key Technologies, and Testbeds},
  volume = {25},
  year = {2023}
}

@article{dang2020six,
  author = {Shuping Dang and Osama Amin and Basem Shihada and Mohamed-Slim Alouini},
  journal = {Nature Electron.},
  pages = {20--29},
  title = {What should {6G} be?},
  volume = {3},
  year = {2020},
  doi = {10.1038/s41928-019-0355-6}
}

@article{saad2020vision,
  author = {Saad, Walid and Bennis, Mehdi and Chen, Mingzhe},
  journal = {IEEE Netw.},
  number = {3},
  pages = {134-142},
  title = {A Vision of {6G} Wireless Systems: Applications, Trends, Technologies, and Open Research Problems},
  volume = {34},
  year = {2020}
}

@misc{itu2030framework,
  author = {{ITU-R}},
  howpublished = {\url{https://www.itu.int/rec/R-REC-M.2160/en}},
  title = {{M.2160}: Framework and overall objectives of the future development of {IMT} for 2030 and beyond},
  year = {2023}
}

@misc{ISCC,
  author = {{FuTURE Forum}},
  howpublished = {\url{http://www.future-forum.org.cn/dl/201126/whitepaper/70C.pdf}},
  month = {Nov.},
  title = {Integration of sensing, communication and computing toward {6G}},
  year = {2020}
}

@article{Deng2024,
  author  = {Deng, Cailian and Fang, Xuming and Wang, Xianbin},
  title   = {Integrated Sensing, Communication, and Computation
             With Adaptive {DNN} Splitting in Multi-{UAV} Networks},
  journal = {IEEE Trans. Wireless Commun.},
  volume  = {23},
  number  = {11},
  pages   = {17429--17445},
  year    = {2024},
  doi     = {10.1109/TWC.2024.3453650}
}

@article{Sun2024,
  author = {Sun, Gangcan and Wu, Xue and Hao, Wanming and Zhu, Zhengyu and Li, Junjie and Chu, Zheng and Xiao, Pei},
  journal = {IEEE Trans. Veh. Technol.},
  number = {12},
  pages = {18719-18731},
  title = {Resource Management for Integrated Communications, Computing, and Sensing {(ICCS}) Networks},
  volume = {73},
  year = {2024}
}

@article{Letaief2022,
  author  = {Letaief, Khaled B. and Shi, Yuanming
             and Lu, Jianmin and Lu, Jianhua},
  title   = {Edge Artificial Intelligence for {6G}:
             Vision, Enabling Technologies, and Applications},
  journal = {IEEE J. Sel. Areas Commun.},
  volume  = {40},
  number  = {1},
  pages   = {5--36},
  year    = {2022},
  doi     = {10.1109/JSAC.2021.3126076}
}

@inproceedings{2024Dong,
  author = {Dong, Huanyu and Li, Peichun and Dai, Minghui and Wu, Yuan and Qian, Liping and Hei, Xiaojun},
  booktitle = {IEEE/CIC Int. Conf. Commun. China (ICCC)},
  pages = {54-59},
  title = {Coordinated Multi-Point Aided Integrated Sensing, Communication and Computation System: An Energy Efficient Design},
  year = {2024}
}

@article{Ren2025,
  author  = {Ren, Tianyu and Zhang, Miao and Wang, Fangyu
             and Zheng, Kaiyuan},
  title   = {Optimizing Trade-Offs in Reliability, Security,
             and Freshness for Integrated Communication,
             Sensing, and Over-the-Air Computing},
  journal = {IEEE Access},
  volume  = {13},
  pages   = {120807--120820},
  year    = {2025},
  doi     = {10.1109/ACCESS.2025.3588349}
}

@article{Jiang2025,
  author = {Jiang, Peng and Li, Ming and Liu, Rang and Wang, Wei and Liu, Qian},
  journal = {IEEE Trans. Commun.},
  number = {8},
  pages = {6469-6483},
  title = {Joint Waveform and Beamforming Design in {RIS}-{ISAC} Systems: A Model-Driven Learning Approach},
  volume = {73},
  year = {2025}
}

@article{Mason2025,
  author = {Mason, Federico and Pegoraro, Jacopo},
  journal = {IEEE Wireless Commun. Lett.},
  number = {3},
  pages = {821-825},
  title = {Using Deep Reinforcement Learning to Enhance Channel Sampling Patterns in Integrated Sensing and Communication},
  volume = {14},
  year = {2025}
}

@article{Wang2024,
  author = {Wang, Zihuan and Wong, Vincent W.S. and Schober, Robert},
  journal = {IEEE J. Sel. Topics Signal Process.},
  number = {5},
  pages = {933-949},
  title = {Integrated Sensing and Communications for End-to-End Predictive Beamforming Design in Vehicle-to-Infrastructure Networks},
  volume = {18},
  year = {2024}
}

@article{liu2022integrated,
  author = {Liu, Fan and Cui, Yuanhao and Masouros, Christos and Xu, Jie and Han, Tony Xiao and Eldar, Yonina C. and Buzzi, Stefano},
  journal = {IEEE J. Sel. Areas Commun.},
  number = {6},
  pages = {1728--1767},
  publisher = {IEEE},
  title = {Integrated sensing and communications: Towards dual-functional wireless networks for {6G} and beyond},
  volume = {40},
  year = {2022}
}

@article{yang2019federated,
  author  = {Yang, Kai and Jiang, Tao and Shi, Yuanming
             and Ding, Zhi},
  title   = {Federated Learning via Over-the-Air Computation},
  journal = {IEEE Trans. Wireless Commun.},
  volume  = {19},
  number  = {3},
  pages   = {2022--2035},
  year    = {2020},
  doi     = {10.1109/TWC.2019.2961673}
}

@article{chen2024edge,
  author = {Chen, Mingzhe and Yang, Zhaohui and Saad, Walid and Yin, Changchuan and Poor, H. Vincent and Cui, Shuguang},
  journal = {IEEE Trans. Wireless Commun.},
  number = {1},
  pages = {269-283},
  title = {A Joint Learning and Communications Framework for Federated Learning Over Wireless Networks},
  volume = {20},
  year = {2021}
}


%% file: references/iscc-for-ai_cleaned.bib
@inproceedings{Hu2025_PSS-Scopus-20250825-151225_0062,
  author    = {Hu, Qiaosheng and Zhang, Nannan and Wen, Dingzhu},
  title     = {Integrated Sensing-Communication-Computation Based
               Online Federated Learning with Limited Cache},
  booktitle = {Proc. IEEE Wireless Commun. Netw. Conf. (WCNC)},
  pages     = {1--6},
  year      = {2025},
  doi       = {10.1109/WCNC61545.2025.10978235}
}

@article{Jiao202519767_PSS-Scopus-20250825-151225_0064,
  author = {Jiao, Xiang and Zhu, Guangxu and Jiang, Wei and Chen, Li and Luo, Wu and Wen, Dingzhu},
  journal = {IEEE Internet Things J.},
  number = {12},
  pages = {19767--19781},
  title = {Sensing Communication Computation Integration for Federated Edge Learning With Controllable Model Dropout},
  volume = {12},
  year = {2025},
  doi = {10.1109/JIOT.2025.3541535}
}

@article{Zhuang20243205_PSS-Scopus-20250825-151225_0067,
  author  = {Zhuang, Zeming and Wen, Dingzhu and Shi, Yuanming
             and Zhu, Guangxu and Wu, Sheng and Niyato, Dusit},
  title   = {Integrated Sensing-Communication-Computation
             for Over-the-Air Edge {AI} Inference},
  journal = {IEEE Trans. Wireless Commun.},
  volume  = {23},
  number  = {4},
  pages   = {3205--3220},
  year    = {2024},
  doi     = {10.1109/TWC.2023.3306465}
}

@article{Yao2025_PSS-Scopus-20250825-151225_0072,
  author = {Yao, Jiacheng and Xu, Wei and Zhu, Guangxu and Huang, Kaibin and Cui, Shuguang},
  title = {Energy-Efficient Edge Inference in Integrated Sensing, Communication, and Computation Networks},
  journal = {IEEE J. Sel. Areas Commun.},
  volume = {43},
  number = {10},
  pages = {3580--3595},
  year = {2025},
  doi = {10.1109/JSAC.2025.3574612}
}

@inproceedings{Liang2023592_PSS-Scopus-20250825-151225_0173,
  author = {Liang, Yipeng and Chen, Qimei and Zhu, Guangxu and Jiang, Hao},
  booktitle = {Proc. IEEE Int. Conf. Commun. Workshops (ICC Workshops)},
  pages = {592 -- 598},
  title = {Theoretical Analysis and Performance Evaluation for Federated Edge Learning with Integrated Sensing, Communication and Computation},
  year = {2023}
}

@inproceedings{Zheng2024804_PSS-Scopus-20250825-151225_0197,
  author = {Zheng, Paul and Zhu, Yao and Hu, Yulin and Schmeink, Anke},
  booktitle = {Proc. IEEE Int. Conf. Commun. Workshops (ICC Workshops)},
  pages = {804 -- 809},
  title = {Over-the-Air Federated Learning Client Selection in Integrated Sensing, Computing and Communication},
  year = {2024}
}

@article{energy_iscc_edge,
  author = {Liu, Shu and Wen, Dingzhu and Li, Da and Chen, Qimei and Zhu, Guangxu and Shi, Yuanming},
  journal = {IEEE Trans. Mobile Comput.},
  month = {Dec},
  number = {12},
  pages = {14248-14262},
  title = {Energy-Efficient Optimal Mode Selection for Edge {AI} Inference via Integrated Sensing-Communication-Computation},
  volume = {23},
  year = {2024}
}

@article{wen2024task_oriented_edge_ai,
  author  = {Wen, Dingzhu and Liu, Peixi and Zhu, Guangxu
             and Shi, Yuanming and Xu, Jie and Eldar, Yonina C.
             and Cui, Shuguang},
  title   = {Task-Oriented Sensing, Computation, and Communication
             Integration for Multi-Device Edge {AI}},
  journal = {IEEE Trans. Wireless Commun.},
  volume  = {23},
  number  = {3},
  pages   = {2486--2502},
  year    = {2024},
  doi     = {10.1109/TWC.2023.3303232}
}


%% file: references/new.bib
@article{zhang2026ai_empowered_tscc,
  author  = {Zhang, Yuxin and Wang, Xingwei and Luo, Xuewen and Pei, Xinyue and Li, Fuliang and Han, Donghong and Li, Tianyu},
  title   = {A survey on {AI}-empowered task-oriented sensing, communication, and computation in {6G} networks},
  journal = {Comput. Sci. Rev.},
  volume  = {60},
  year    = {2026},
  note    = {Art. no. 100899},
  doi     = {10.1016/j.cosrev.2026.100899}
}

@article{liu2026integrated_sensing_edge_ai,
  author  = {Liu, Zhiyan and Chen, Xu and Wu, Hai and Wang, Zhanwei and Chen, Xianhao and Niyato, Dusit and Huang, Kaibin},
  title   = {Integrated Sensing and Edge {AI}: Realizing Intelligent Perception in {6G}},
  journal = {IEEE Commun. Surveys Tuts.},
  volume  = {28},
  pages   = {2725--2770},
  year    = {2026},
  doi     = {10.1109/COMST.2025.3592989}
}

@article{peng2026cellfree_iccs,
  author  = {Peng, Qihao and Luo, Qu and Chu, Zheng and Lin, Zihuai and Elkashlan, Maged and Xiao, Pei and Karagiannidis, George K. and Masouros, Christos},
  title   = {Latency-Aware Resource Allocation for Integrated Communications, Computation, and Sensing in Cell-Free {mMIMO} Systems},
  journal = {IEEE Trans. Wireless Commun.},
  volume  = {25},
  pages   = {11128--11142},
  year    = {2026},
  doi     = {10.1109/TWC.2026.3657950}
}

@article{n_prim_p018_2025_design_energy_efficient_data,
  author  = {Zhou, Ziqin and Li, Xiaoyang and Zhu, Guangxu and Zhou, Bingpeng and Xing, Hong and Huang, Kaibin},
  title   = {Integrated Sensing-Communication-Computation Design for Energy Efficient Data Processing},
  journal = {IEEE Trans. Netw. Sci. Eng.},
  volume  = {13},
  pages   = {4172--4186},
  year    = {2026},
  doi     = {10.1109/TNSE.2025.3605140}
}

@techreport{threedgpp2023tr38843_ai_ml_nr,
  author      = {{3GPP}},
  title       = {Study on Artificial Intelligence ({AI})/Machine Learning ({ML}) for {NR} Air Interface},
  institution = {{3GPP}},
  type        = {TR},
  number      = {38.843, V18.0.0},
  year        = {2023},
  month       = dec,
  note        = {Release 18},
  url         = {https://portal.3gpp.org/desktopmodules/Specifications/SpecificationDetails.aspx?specificationId=3983}
}

@inproceedings{wu2026early_exit_iscc,
  author    = {Guofang Wu and Yejun He and Xiaowen Cao},
  title     = {{UAV}-Assisted Integrated Sensing, Communication, and Computation Edge Inference Network With Early-Exit Mechanism},
  booktitle = {Proc. Int. Wireless Commun. Mobile Comput. Conf. (IWCMC)},
  pages     = {374--379},
  year      = {2026},
  doi       = {10.1109/IWCMC69287.2026.11579989}
}

@article{liu2026multitier_iscc,
  author  = {Peng Liu and Zesong Fei and Xinyi Wang and Qiao Qi and Zhaohui Yang and Meng Hua and Arumugam Nallanathan},
  title   = {Integrated Sensing, Communication, and Computing in Multi-Tier Systems: Joint Hybrid Beamforming Design and Computation Resource Allocation},
  journal = {arXiv preprint arXiv:2608.14086},
  year    = {2026}
}

@article{jiang2026isac_for_ai,
  author  = {Lai Jiang and Kaitao Meng and Murat Temiz and Christos Masouros},
  title   = {{ISAC} for {AI}: A Trade-off Framework Across Data Acquisition and Transfer in Federated Learning},
  journal = {arXiv preprint arXiv:2605.11915},
  year    = {2026}
}

@article{tian2026sense2com,
  author  = {Jin Tian and Yan Shi and Shanzhi Chen},
  title   = {{Sense2Com}: Coordinating Sensing, Communication, and Computation for {V2V} Cooperative Perception},
  journal = {Phys. Commun.},
  volume  = {78},
  year    = {2026},
  note    = {Art. no. 103214},
  doi     = {10.1016/j.phycom.2026.103214}
}

@article{zhang2025sagin_iscc,
  author  = {Yuexia Zhang and Yuanshuo Gang and Peng Wu and Guangteng Fan and Wenjun Xu and Bo Ai and Qingqing Wu},
  title   = {Integrated Sensing, Communication, and Computation in {SAGIN}: Joint Beamforming and Resource Allocation},
  journal = {IEEE Trans. Cogn. Commun. Netw.},
  volume  = {11},
  number  = {5},
  pages   = {3128--3143},
  year    = {2025},
  doi     = {10.1109/TCCN.2025.3577377}
}

@inproceedings{yang2023mmfi,
  author    = {Jianfei Yang and He Huang and Yunjiao Zhou and Xinyan Chen and Yuecong Xu and Shenghai Yuan and Han Zou and Chris Xiaoxuan Lu and Lihua Xie},
  title     = {{MM-Fi}: Multi-Modal Non-Intrusive {4D} Human Dataset for Versatile Wireless Sensing},
  booktitle = {Adv. Neural Inf. Process. Syst.},
  volume    = {36},
  pages     = {18756--18768},
  year      = {2023}
}

@inproceedings{lai2022fedscale,
  author    = {Fan Lai and Yinwei Dai and Sanjay Singapuram and Jiachen Liu and Xiangfeng Zhu and Harsha Madhyastha and Mosharaf Chowdhury},
  title     = {{FedScale}: Benchmarking Model and System Performance of Federated Learning at Scale},
  booktitle = {Proc. Int. Conf. Mach. Learn. (ICML)},
  pages     = {11814--11827},
  year      = {2022}
}

@inproceedings{banbury2021mlperf_tiny,
  author    = {Colby Banbury and Vijay Janapa Reddi and Peter Torelli and Nat Jeffries and Csaba Kiraly and Jeremy Holleman and Pietro Montino and David Kanter and Pete Warden and Danilo Pau and Urmish Thakker and Antonio Torrini and Jay Cordaro and Giuseppe Di Guglielmo and Javier Duarte and Honson Tran and Nhan Tran and Niu Wenxu and Xu Xuesong},
  title     = {{MLPerf Tiny Benchmark}},
  booktitle = {Proc. Neural Inf. Process. Syst. (NeurIPS) Datasets Benchmarks Track},
  volume    = {1},
  year      = {2021},
  url       = {https://datasets-benchmarks-proceedings.neurips.cc/paper_files/paper/2021/hash/da4fb5c6e93e74d3df8527599fa62642-Abstract-round1.html}
}

@article{bonati2023openran_gym,
  author  = {Leonardo Bonati and Michele Polese and Salvatore D'Oro and Stefano Basagni and Tommaso Melodia},
  title   = {{OpenRAN Gym}: {AI/ML} Development, Data Collection, and Testing for {O-RAN} on {PAWR} Platforms},
  journal = {Comput. Netw.},
  volume  = {220},
  year    = {2023},
  note    = {Art. no. 109502},
  doi     = {10.1016/j.comnet.2022.109502}
}

@article{lu2025cooperative_digital_twin,
  author  = {Binbin Lu and Xumin Huang and Yuan Wu and Li Ping Qian and Dusit Niyato and Cheng-Zhong Xu},
  title   = {Cooperative Perception Aided Digital Twin Model Update and Migration in Mixed Vehicular Networks},
  journal = {IEEE Trans. Intell. Transp. Syst.},
  volume  = {26},
  number  = {2},
  pages   = {2293--2308},
  year    = {2025},
  doi     = {10.1109/TITS.2024.3496121}
}

@article{liu2026semantic_sensor_selection,
  author  = {Zhiyan Liu and Kaibin Huang},
  title   = {Semantic-Relevance-Based Sensor Selection for Edge-{AI} Empowered Sensing Systems},
  journal = {IEEE Trans. Wireless Commun.},
  volume  = {25},
  pages   = {3721--3735},
  year    = {2026},
  doi     = {10.1109/TWC.2025.3605523}
}

@article{he2026sensem,
  author  = {Yinghui He and Xin Li and Jun Luo},
  title   = {Task-Oriented Integrated Sensing and Semantic Communications for Multi-Device Video Analytics},
  journal = {IEEE Trans. Mobile Comput.},
  volume  = {25},
  number  = {5},
  pages   = {7323--7337},
  year    = {2026},
  doi     = {10.1109/TMC.2025.3641899}
}


%% file: references/ref.bib
@article{wen2026airfeel,
  author  = {Dingzhu Wen and Sijing Xie and Xiaowen Cao and Yuanhao Cui and Jie Xu and Yuanming Shi and Shuguang Cui},
  title   = {Integrated Sensing, Communication, and Computation for Over-the-Air Federated Edge Learning},
  journal = {IEEE Trans. Wireless Commun.},
  volume  = {25},
  pages   = {2748--2762},
  year    = {2026},
  doi     = {10.1109/TWC.2025.3598997}
}

@article{zhao2024iov_iscc,
  author  = {Zhao, Junhui and Ren, Ruixing and Zou, Dan and Zhang, Qingmiao and Xu, Wei},
  title   = {{IoV}-Oriented Integrated Sensing, Computation, and Communication: System Design and Resource Allocation},
  journal = {IEEE Trans. Veh. Technol.},
  volume  = {73},
  number  = {11},
  pages   = {16283--16294},
  year    = {2024},
  doi     = {10.1109/TVT.2024.3422270}
}

@article{huang2025collaborative_sensing_offloading,
  author  = {Bangzhen Huang and Xuwei Fan and Shaolong Zheng and Ning Chen and Yifeng Zhao and Lianfen Huang and Zhibin Gao and Han-Chieh Chao},
  title   = {Collaborative Sensing-Aware Task Offloading and Resource Allocation for Integrated Sensing-Communication- and Computation-Enabled Internet of Vehicles ({IoV})},
  journal = {Sensors},
  volume  = {25},
  number  = {3},
  year    = {2025},
  note    = {Art. no. 723},
  doi     = {10.3390/s25030723}
}

@article{cheng2026d3pg_light_iscc,
  author  = {Qing Cheng and Wenwen Wu and Yebo Zhou},
  title   = {{D3PG-Light}: A Lightweight and Stable Resource Scheduling Framework for {UAV}-Integrated Sensing, Communication, and Computation Systems},
  journal = {Sensors},
  volume  = {26},
  number  = {6},
  year    = {2026},
  note    = {Art. no. 1829},
  doi     = {10.3390/s26061829}
}

@article{qi2026uav_iscc_efficiency,
  author  = {Honghao Qi and Muqing Wu},
  title   = {Computing Efficiency Optimization for {UAV}-Enabled Integrated Sensing, Computing, and Communication: A Memory-Based Deep Reinforcement Learning Approach},
  journal = {Drones},
  volume  = {10},
  number  = {3},
  year    = {2026},
  note    = {Art. no. 180},
  doi     = {10.3390/drones10030180}
}

@article{ji2026uav_emergency_iscc,
  author  = {Baofeng Ji and Zhan Huo and Boyuan Li and Yiwei Li and Huitao Fan and Zijian Li},
  title   = {Resource Allocation for {UAV}-Assisted Vehicular Networks with Integrated Sensing, Communication, and Computing in Emergency Scenarios},
  journal = {Phys. Commun.},
  volume  = {78},
  year    = {2026},
  note    = {Art. no. 103209},
  doi     = {10.1016/j.phycom.2026.103209}
}

@article{tang2025uav_feel,
  author  = {Yao Tang and Guangxu Zhu and Wei Xu and Man Hon Cheung and Tat-Ming Lok and Shuguang Cui},
  title   = {Integrated Sensing, Computation, and Communication for {UAV}-Assisted Federated Edge Learning},
  journal = {IEEE Trans. Wireless Commun.},
  volume  = {24},
  number  = {4},
  pages   = {2647--2662},
  year    = {2025},
  doi     = {10.1109/TWC.2024.3523381}
}

@article{liang2026fedavg_fedsgd_iscc,
  author  = {Liang, Yipeng and Chen, Qimei and Wen, Dingzhu and Zhu, Guangxu and Awan, Muhammad Kaleem and Zhao, Mingming and Jiang, Hao},
  title   = {Federated Learning with Integrated Sensing, Communication, and Computation: {FedSGD-ISCC} or {FedAVG-ISCC}},
  journal = {IEEE Trans. Veh. Technol.},
  year    = {2026},
  note    = {early access},
  doi     = {10.1109/TVT.2026.3701541}
}

@article{zhang2026online_fed_learning_limited_cache,
  author  = {Nannan Zhang and Qiaosheng Hu and Dingzhu Wen and Yuyi Mao and Tian-Ming Ma and Guanding Yu and Wei Wang},
  title   = {Integrated Sensing-Communication-Computation Based Online Federated Learning With Limited Cache in {OFDM} Systems},
  journal = {IEEE Trans. Cogn. Commun. Netw.},
  volume  = {12},
  pages   = {5578--5594},
  year    = {2026},
  doi     = {10.1109/TCCN.2026.3656294}
}

@article{wang2026decentralized_multitask_iscc,
  author  = {Chenye Wang and Zeming Zhuang and Dingzhu Wen and Yuanming Shi and Xin Wang},
  title   = {Decentralized Integration of Sensing-Communication-Computation for Multi-Task Edge {AI} Inference},
  journal = {IEEE Trans. Wireless Commun.},
  volume  = {25},
  pages   = {6553--6567},
  year    = {2026},
  doi     = {10.1109/TWC.2025.3625389}
}

@article{wen2026uav_assisted_edge_inference,
  author  = {Dingzhu Wen and Shuo Zhang and Guangxu Zhu and Yuan Liu and Yuanming Shi and Han Hu},
  title   = {{UAV}-Assisted Edge Inference With Integrated Sensing, Communication, and Computation},
  journal = {IEEE Trans. Wireless Commun.},
  volume  = {25},
  pages   = {13336--13350},
  year    = {2026},
  doi     = {10.1109/TWC.2026.3669999}
}

@article{cao2026vfeel,
  author  = {Xiaowen Cao and Dingzhu Wen and Suzhi Bi and Yuanhao Cui and Guangxu Zhu and Han Hu and Yonina C. Eldar},
  title   = {Joint Sensing, Communication, and Computation for Vertical Federated Edge Learning in Edge Perception Networks},
  journal = {IEEE Trans. Mobile Comput.},
  volume  = {25},
  number  = {8},
  pages   = {12956--12970},
  year    = {2026},
  doi     = {10.1109/TMC.2026.3674960}
}

@article{chen2026multimodal_edge_intelligence,
  author  = {Weiwei Chen and Yinghui He and Zhong Ye and Dingzhu Wen and Guanding Yu},
  title   = {Task-Oriented Multimodal Edge Intelligence via Integrated Sensing-Communication-Computation},
  journal = {IEEE Trans. Wireless Commun.},
  volume  = {25},
  pages   = {20071--20085},
  year    = {2026},
  doi     = {10.1109/TWC.2026.3710771}
}

@article{du2026flisc3,
  author  = {Du, An and Jia, Jie and Dustdar, Schahram and Morichetta, Andrea and Chen, Jian and Wang, Xingwei},
  title   = {{FLISC$^3$}: Federated Learning-Oriented Resource Optimization in {ISCC}-Enabled Edge Collaborative Networks},
  journal = {IEEE Trans. Serv. Comput.},
  volume  = {19},
  number  = {1},
  pages   = {364--379},
  year    = {2026},
  doi     = {10.1109/TSC.2025.3622026}
}

@inproceedings{yang2025movable_antennas_iscc,
  author    = {Yang, Jiayi and Guo, Shuai and Wang, Chengcheng and Wen, Dingzhu and Fu, Min and Zhou, Yong and Shi, Yuanming},
  title     = {Integrated Sensing-Communication-Computation for Movable Antennas Assisted Multi-Device Edge {AI} Inference},
  booktitle = {Proc. IEEE Global Commun. Conf. (GLOBECOM)},
  pages     = {2120--2125},
  year      = {2025},
  doi       = {10.1109/GLOBECOM59602.2025.11432560}
}

@article{hou2025air_iscc_uav,
  author  = {Hou, Peng and Zhu, Hongbin and Lu, Zhihui and Huang, Shih-Chia and Yang, Yang and Chai, Hongfeng},
  title   = {Learning-Based Over-the-Air Integrated Sensing, Communication and Computation in {UAV} Swarm-Enabled Intelligent Transportation Systems},
  journal = {IEEE Trans. Green Commun. Netw.},
  volume  = {9},
  number  = {3},
  pages   = {1414--1428},
  year    = {2025},
  doi     = {10.1109/TGCN.2024.3492028}
}

@inproceedings{xu2026task_priority_multimodal_iscc,
  author    = {Xu, Wen-Jian and Tian, Jie and Zhang, Yue and Li, Tiantian and Bian, Ji},
  title     = {Task-Priority-Driven {UAV}-Assisted Multimodal {ISCC} for Traffic Monitoring},
  booktitle = {Proc. 6th Int. Conf. Intell. Commun. Comput. (ICICC)},
  pages     = {176--180},
  year      = {2026},
  doi       = {10.1109/ICICC71012.2026.11637645}
}

@article{tang2025seco4,
  author  = {Tang, Qinqin and Yang, Yutian and Cui, Jiayi and Xie, Renchao and Huang, Tao and Chen, Tianjiao and Zhang, Ran and Xiong, Zehui},
  title   = {{SeCo4}: Co-Design of Sensing, Communication, and Computing for Intelligent Control in Industrial Cyber-Physical Systems},
  journal = {IEEE J. Sel. Areas Commun.},
  volume  = {43},
  number  = {10},
  pages   = {3362--3378},
  year    = {2025},
  doi     = {10.1109/JSAC.2025.3574623}
}

@inproceedings{yang2025intelligent_control_iiot,
  author    = {Yang, Yutian and Yin, Zihang and Tang, Qinqin and Liu, Yang and Cui, Jiayi and Xie, Renchao and Huang, Tao},
  title     = {Intelligent Control Integrating Sensing, Communication and Computing in Industrial Internet of Things},
  booktitle = {Proc. IEEE Int. Conf. Commun. (ICC)},
  pages     = {3424--3429},
  year      = {2025},
  doi       = {10.1109/ICC52391.2025.11161630}
}

@article{yu2025isac_fundamental_limits,
  author  = {Yu, Zhouyuan and Hu, Xiaoling and Liu, Chenxi and Peng, Mugen},
  title   = {Rethinking the Fundamental Performance Limits of Integrated Sensing and Communication Systems},
  journal = {IEEE Trans. Wireless Commun.},
  volume  = {24},
  number  = {11},
  pages   = {9311--9322},
  year    = {2025},
  doi     = {10.1109/TWC.2025.3572678}
}

@article{zhao2025task_semantic_ib,
  author  = {Zhao, Tantan and Li, Fan and Du, Hongyang and Sun, Li},
  title   = {Deep Reinforcement Learning- and Information Bottleneck-Enabled Task-Oriented Semantic Communication},
  journal = {IEEE J. Sel. Areas Commun.},
  volume  = {43},
  number  = {12},
  pages   = {4198--4214},
  year    = {2025},
  doi     = {10.1109/JSAC.2025.3623199}
}

@article{wang2025trustworthy_edge_intelligence,
  author  = {Wang, Xiaojie and Wang, Beibei and Wu, Yu and Ning, Zhaolong and Guo, Song and Yu, Fei Richard},
  title   = {A Survey on Trustworthy Edge Intelligence: From Security and Reliability to Transparency and Sustainability},
  journal = {IEEE Commun. Surveys Tuts.},
  volume  = {27},
  number  = {3},
  pages   = {1729--1757},
  year    = {2025},
  doi     = {10.1109/COMST.2024.3446585}
}

@article{huang2025trustworthy_iot,
  author  = {Huang, Boyang and Feng, Junhao and Jin, Xin},
  title   = {Sensing--Communication--Computing Integrated Resource Allocation for {AI}-Empowered Trustworthy {IoT}},
  journal = {IEEE Trans. Consum. Electron.},
  volume  = {71},
  number  = {1},
  pages   = {2003--2016},
  year    = {2025},
  doi     = {10.1109/TCE.2025.3546948}
}

@article{zhang2026covert_uav_iscc,
  author  = {Zhang, Qi and Gao, Wei and Liu, Chuan and Yao, Yu and Yan, Shihao and Shu, Feng and Jin, Shi},
  title   = {Covert Transmission for Active {RIS}-Aided Full-Duplex {UAV} Integrated Sensing, Communication, and Computation Systems},
  journal = {IEEE J. Sel. Areas Commun.},
  volume  = {44},
  pages   = {1992--2007},
  year    = {2026},
  doi     = {10.1109/JSAC.2025.3637785}
}

@techreport{threedgpp2026tr33777_isac_security,
  author      = {{3GPP}},
  title       = {Study on Security and Privacy Aspects of Integrated Sensing and Communication},
  institution = {{3GPP}},
  type        = {TR},
  number      = {33.777, V20.0.0},
  year        = {2026},
  month       = sep,
  note        = {Release 20},
  url         = {https://portal.3gpp.org/desktopmodules/Specifications/SpecificationDetails.aspx?specificationId=4456}
}

@techreport{threedgpp2024ts22137_isac,
  author      = {{3GPP}},
  title       = {Integrated Sensing and Communication},
  institution = {{3GPP}},
  type        = {TS},
  number      = {22.137, V19.1.0},
  year        = {2024},
  month       = apr,
  note        = {Release 19},
  url         = {https://portal.3gpp.org/desktopmodules/Specifications/SpecificationDetails.aspx?specificationId=4198}
}

@techreport{etsi2026gr_isc005,
  author      = {{ETSI}},
  title       = {Integrated Sensing and Communications ({ISAC}); Integration of Computing with {ISAC}},
  institution = {{ETSI}},
  type        = {GR},
  number      = {ISC 005, V1.1.1},
  year        = {2026},
  month       = oct,
  url         = {https://www.etsi.org/deliver/etsi_gr/ISC/001_099/005/01.01.01_60/gr_isc005v010101p.pdf}
}

@techreport{etsi2026gr_isc004,
  author      = {{ETSI}},
  title       = {Integrated Sensing and Communications ({ISAC}); Security, Privacy, Trustworthiness and Sustainability},
  institution = {{ETSI}},
  type        = {GR},
  number      = {ISC 004, V1.1.1},
  year        = {2026},
  month       = feb,
  url         = {https://www.etsi.org/deliver/etsi_gr/ISC/001_099/004/01.01.01_60/gr_ISC004v010101p.pdf}
}

@article{luo2026isac_testbeds,
  author  = {Luo, Xuewen and Lin, Qingfeng and Zhang, Ruoyu and Chen, Hsiao-Hwa and Wang, Xingwei and Huang, Min},
  title   = {{ISAC}---A Survey on Its Layered Architecture, Technologies, Standardizations, Prototypes, and Testbeds},
  journal = {IEEE Commun. Surveys Tuts.},
  volume  = {28},
  pages   = {485--526},
  year    = {2026},
  doi     = {10.1109/COMST.2025.3565534}
}

@inproceedings{tirchas2025ml_isac_experiment,
  author    = {Tirchas, Panagiotis and Karetsos, George T.},
  title     = {Machine Learning Enabled Integrated Sensing and Communication for Smart {6G} Environments},
  booktitle = {Proc. 28th Conf. Innov. Clouds, Internet Netw. (ICIN)},
  pages     = {161--165},
  year      = {2025},
  doi       = {10.1109/ICIN64016.2025.10942988}
}

@techreport{threedgpp2026tr2370014_isac,
  author      = {{3GPP}},
  title       = {Study on Integrated Sensing and Communication; Stage 2},
  institution = {{3GPP}},
  type        = {TR},
  number      = {23.700-14, V20.0.0},
  year        = {2026},
  month       = mar,
  note        = {Release 20},
  url         = {https://portal.3gpp.org/desktopmodules/Specifications/SpecificationDetails.aspx?specificationId=4399}
}

@techreport{threedgpp2026tr38765_isac,
  author      = {{3GPP}},
  title       = {Study on Integrated Sensing and Communication ({ISAC}) for {NR}},
  institution = {{3GPP}},
  type        = {TR},
  number      = {38.765, V20.0.0},
  year        = {2026},
  month       = jun,
  note        = {Release 20},
  url         = {https://portal.3gpp.org/desktopmodules/Specifications/SpecificationDetails.aspx?specificationId=4446}
}

@article{jiang2026large_ai_models,
  author  = {Jiang, Feibo and Pan, Cunhua and Dong, Li and Wang, Kezhi and Debbah, M{\'e}rouane and Niyato, Dusit and Han, Zhu},
  title   = {A Comprehensive Survey of Large {AI} Models for Future Communications: Foundations, Applications, and Challenges},
  journal = {IEEE Commun. Surveys Tuts.},
  volume  = {28},
  pages   = {4731--4764},
  year    = {2026},
  doi     = {10.1109/COMST.2026.3660844}
}

@article{qu2025edge_llm,
  author  = {Qu, Guanqiao and Chen, Qiyuan and Wei, Wei and Lin, Zheng and Chen, Xianhao and Huang, Kaibin},
  title   = {Mobile Edge Intelligence for Large Language Models: A Contemporary Survey},
  journal = {IEEE Commun. Surveys Tuts.},
  volume  = {27},
  number  = {6},
  pages   = {3820--3860},
  year    = {2025},
  doi     = {10.1109/COMST.2025.3527641}
}

@article{li2025agent_as_service,
  author  = {Li, Borui and Liu, Tianen and Wang, Weilong and Zhao, Chengqing and Wang, Shuai},
  title   = {Agent-as-a-Service: An {AI}-Native Edge Computing Framework for {6G} Networks},
  journal = {IEEE Netw.},
  volume  = {39},
  number  = {2},
  pages   = {44--51},
  year    = {2025},
  doi     = {10.1109/MNET.2024.3520987}
}

@article{zhou2026dt_mobile_ar,
  author  = {Zhou, Conghao and Gao, Jie and Hu, Shisheng and Cheng, Nan and Zhuang, Weihua and Shen, Xuemin},
  title   = {User-Centric Communication Service Provision for Edge-Assisted Mobile Augmented Reality},
  journal = {IEEE Trans. Mobile Comput.},
  volume  = {25},
  number  = {3},
  pages   = {3435--3449},
  year    = {2026},
  doi     = {10.1109/TMC.2025.3618147}
}

@inproceedings{tiwari2026dt_isac,
  author    = {Tiwari, Himanshu and Kar, Binayak and Tiwari, Priyanshu},
  title     = {Digital Twin-Assisted Belief-State Reinforcement Learning for Latency-Robust {ISAC} in {6G} Networks},
  booktitle = {Proc. IEEE Conf. Comput. Commun. (INFOCOM)},
  pages     = {1--6},
  year      = {2026},
  doi       = {10.1109/INFOCOM59046.2026.11571214}
}

@article{li2025embodied_nearfield,
  author  = {Li, Guoliang and Jin, Xibin and Wan, Yujie and Liu, Chenxuan and Zhang, Tong and Wang, Shuai and Xu, Chengzhong},
  title   = {Embodied Edge Intelligence Meets Near Field Communication: Concept, Design, and Verification},
 journal = {IEEE Netw.},
  volume  = {39},
  number  = {6},
  pages   = {78--86},
  year    = {2025},
  doi     = {10.1109/MNET.2025.3599899}
}

@article{chen2023neuromorphic_wireless_cognition,
  author  = {Chen, Jiechen and Skatchkovsky, Nicolas and Simeone, Osvaldo},
  title   = {Neuromorphic Wireless Cognition: Event-Driven Semantic Communications for Remote Inference},
  journal = {IEEE Trans. Cogn. Commun. Netw.},
  volume  = {9},
  number  = {2},
  pages   = {252--265},
  year    = {2023},
  doi     = {10.1109/TCCN.2023.3236940}
}

@inproceedings{chen2024neuromorphic_wakeup,
  author    = {Chen, Jiechen and Park, Sangwoo and Popovski, Petar and Poor, H. Vincent and Simeone, Osvaldo},
  title     = {Neuromorphic Semantic Communications with Wake-Up Radios},
  booktitle = {Proc. 25th IEEE Int. Workshop Signal Process. Adv. Wireless Commun. (SPAWC)},
  pages     = {91--95},
  year      = {2024},
  doi       = {10.1109/SPAWC60668.2024.10694141}
}

@article{wu2025neuromorphic_split,
  author  = {Wu, Dengyu and Chen, Jiechen and Rajendran, Bipin and Poor, H. Vincent and Simeone, Osvaldo},
  title   = {Neuromorphic Wireless Split Computing with Multi-Level Spikes},
  journal = {IEEE Trans. Mach. Learn. Commun. Netw.},
  volume  = {3},
  pages   = {502--516},
  year    = {2025},
  doi     = {10.1109/TMLCN.2025.3556634}
}

@inproceedings{li2025neuromorphic_multitask,
  author    = {Li, Xian and Chen, Zebin and Sun, Kaixuan and Bi, Suzhi},
  title     = {{NOMA} Digital Semantic Communication for Neuromorphic Multi-Task Edge Inference},
  booktitle = {Proc. IEEE/CIC Int. Conf. Commun. China Workshops (ICCC Workshops)},
  year      = {2025},
  doi       = {10.1109/ICCCWorkshops67136.2025.11148099}
}

@article{krikidis2026quantum_isac,
  author  = {Krikidis, Ioannis},
  title   = {Quantum Optical Integrated Sensing and Communication With Homodyne {BPSK} Detection},
  journal = {IEEE Wireless Commun. Lett.},
  volume  = {15},
  pages   = {1375--1379},
  year    = {2026},
  doi     = {10.1109/LWC.2026.3654015}
}

@article{astiti2026qdrl_wireless,
  author  = {Astiti, Ni Made Erma Pratiwi and Lee, Byung Moo},
  title   = {A Survey of Quantum Deep Reinforcement Learning for Resource Allocation in Future Wireless Networks},
  journal = {IEEE Commun. Surveys Tuts.},
  volume  = {28},
  pages   = {109--148},
  year    = {2026},
  doi     = {10.1109/COMST.2025.3638435}
}

@article{wang2024xrf55,
  author  = {Wang, Fei and Lv, Yizhe and Zhu, Mengdie and Ding, Han and Han, Jinsong},
  title   = {{XRF55}: A Radio Frequency Dataset for Human Indoor Action Analysis},
  journal = {Proc. ACM Interact. Mob. Wearable Ubiquitous Technol.},
  volume  = {8},
  number  = {1},
  pages   = {21:1--21:34},
  year    = {2024},
  doi     = {10.1145/3643543}
}

@article{alkhateeb2023deepsense6g,
  author  = {Alkhateeb, Ahmed and Charan, Gouranga and Osman, Tawfik and Hredzak, Andrew and Morais, Joao and Demirhan, Umut and Srinivas, Nikhil},
  title   = {{DeepSense 6G}: A Large-Scale Real-World Multi-Modal Sensing and Communication Dataset},
  journal = {IEEE Commun. Mag.},
  volume  = {61},
  number  = {9},
  pages   = {122--128},
  year    = {2023},
  doi     = {10.1109/MCOM.006.2200730}
}

@inproceedings{alkhateeb2019deepmimo,
  author    = {Alkhateeb, Ahmed},
  title     = {{DeepMIMO}: A Generic Deep Learning Dataset for Millimeter Wave and Massive {MIMO} Applications},
  booktitle = {Proc. Inf. Theory Appl. Workshop (ITA)},
  year      = {2019},
  month     = feb,
  address   = {San Diego, CA, USA}
}

@article{pegoraro2025disc,
  author  = {Pegoraro, Jacopo and Saucedo, Pablo and Lacruz, Jesus Omar and Rossi, Michele and Widmer, Joerg},
  title   = {{DISC}: A Dataset for Integrated Sensing and Communications in {mmWave} Systems},
  journal = {IEEE Commun. Mag.},
  volume  = {63},
  number  = {10},
  pages   = {94--100},
  year    = {2025},
  doi     = {10.1109/MCOM.002.2400687}
}

@article{cheng2023m3sc,
  author  = {Cheng, Xiang and Huang, Ziwei and Bai, Lu and Zhang, Haotian and Sun, Mingran and Liu, Boxun and Li, Sijiang and Zhang, Jianan and Lee, Minson},
  title   = {{$M^3$SC}: A Generic Dataset for Mixed Multi-Modal ({MMM}) Sensing and Communication Integration},
  journal = {China Commun.},
  volume  = {20},
  number  = {11},
  pages   = {13--29},
  year    = {2023},
  doi     = {10.23919/JCC.fa.2023-0268.202311}
}

@article{cheng2025synthsom,
  author  = {Cheng, Xiang and Huang, Ziwei and Yu, Yong and Bai, Lu and Sun, Mingran and Han, Zengrui and Zhang, Ruide and Li, Sijiang},
  title   = {{SynthSoM}: A Synthetic Intelligent Multi-Modal Sensing-Communication Dataset for Synesthesia of Machines ({SoM})},
  journal = {Sci. Data},
  volume  = {12},
  year    = {2025},
  note    = {Art. no. 819},
  doi     = {10.1038/s41597-025-05065-x}
}

@inproceedings{beuster2023uav_testbed,
  author    = {Beuster, Julia and Andrich, Carsten and D{\"o}bereiner, Michael and Schieler, Steffen and Engelhardt, Maximilian and Schneider, Christian and Thom{\"a}, Reiner},
  title     = {Measurement Testbed for Radar and Emitter Localization of {UAV} at 3.75 {GHz}},
  booktitle = {Proc. 17th Eur. Conf. Antennas Propag. (EuCAP)},
  pages     = {1--5},
  year      = {2023},
  doi       = {10.23919/EuCAP57121.2023.10133118}
}

@article{bhat2026mmhsense,
  author  = {Bhat, Nabeel Nisar and Karnaukh, Maksim and Vandenbroeke, Stein and Lemoine, Wouter and Struye, Jakob and Kumar, Siddhartha and Moghaddam, Mohammad Hossein and Lacruz, Jesus Omar and Widmer, Joerg and Berkvens, Rafael and Famaey, Jeroen},
  title   = {{mmHSense}: Multi-Modal and Distributed {mmWave} {ISAC} Datasets for Human Sensing},
  journal = {IEEE Access},
  volume  = {14},
  pages   = {69961--69971},
  year    = {2026},
  doi     = {10.1109/ACCESS.2026.3691174}
}

@inproceedings{huang2026sensing_dataset_protocol,
  author    = {Huang, Jiawei and Zhang, Di and Cui, Yuanhao and Cao, Xiaowen and Han, Tony Xiao and Jing, Xiaojun and Masouros, Christos},
  title     = {A Sensing Dataset Protocol for Benchmarking and Multi-Task Wireless Sensing},
  booktitle = {Proc. IEEE Int. Conf. Commun. (ICC)},
  pages     = {1--7},
  year      = {2026},
  doi       = {10.1109/ICC59461.2026.11588048}
}

@article{zheng2026signal_level_iscc,
  author  = {Zheng, Paul and Zhu, Yao and Yuan, Xiaopeng and Hu, Yulin and Schmeink, Anke},
  title   = {Analytically Characterized Optimal Power Control for Signal-Level-Integrated Sensing, Computing, and Communication in Federated Learning},
  journal = {IEEE Trans. Wireless Commun.},
  volume  = {25},
  pages   = {25941--25958},
  year    = {2026},
  doi     = {10.1109/TWC.2026.3747602}
}

@article{li2023iscco_mimo,
  author  = {Li, Xiaoyang and Liu, Fan and Zhou, Ziqin and Zhu, Guangxu and Wang, Shuai and Huang, Kaibin and Gong, Yi},
  title   = {Integrated Sensing, Communication, and Computation Over-the-Air: {MIMO} Beamforming Design},
  journal = {IEEE Trans. Wireless Commun.},
  volume  = {22},
  number  = {8},
  pages   = {5383--5398},
  year    = {2023},
  doi     = {10.1109/TWC.2022.3233795}
}


%% file: references/references_cleaned.bib
@article{aircomp_00_2007,
  author = {Nazer, Bobak and Gastpar, Michael},
  journal = {IEEE Trans. Inf. Theory},
  number = {10},
  pages = {3498--3516},
  title = {Computation Over Multiple-Access Channels},
  volume = {53},
  year = {2007},
  doi = {10.1109/TIT.2007.904785}
}

@techreport{mec_00_2014,
  author = {Patel, Milan and Hu, Yunchao and H{\'e}d{\'e}, Patrice and Joubert, Jerome and Thornton, Chris and Naughton, Brian and others},
  institution = {{Mobile-Edge Computing initiative}},
  month = sep,
  number = {Issue 1},
  title = {Mobile-Edge Computing: Introductory Technical White Paper},
  year = {2014}
}

@techreport{mcs_01_2006,
  author = {Burke, Jeffrey and Estrin, Deborah and Hansen, Mark and Parker, Andrew and Ramanathan, Nithya and Reddy, Sasank and Srivastava, Mani},
  institution = {Center for Embedded Networked Sensing (CENS), University of California, Los Angeles},
  number = {CENS Technical Report},
  title = {Participatory Sensing},
  year = {2006}
}

@article{mcs_02_2011,
  author = {Ganti, Raghu K. and Ye, Fan and Lei, Hui},
  journal = {IEEE Commun. Mag.},
  number = {11},
  pages = {32-39},
  title = {Mobile crowdsensing: current state and future challenges},
  volume = {49},
  year = {2011}
}

@article{isac_00_2011,
  author = {Sturm, Christian and Wiesbeck, Werner},
  journal = {Proc. IEEE},
  number = {7},
  pages = {1236-1259},
  title = {Waveform Design and Signal Processing Aspects for Fusion of Wireless Communications and Radar Sensing},
  volume = {99},
  year = {2011}
}


%% file: references/related_surveys_cleaned.bib
@article{S05, author = {Wang, Xiaojie and Guo, Qi and Ning, Zhaolong and Guo, Lei and Wang, Guoyin and Gao, Xinbo and Zhang, Yan}, title = {Integration of Sensing, Communication, and Computing for Metaverse: A Survey}, journal = {ACM Comput. Surv.}, volume = {56}, number = {10}, note = {Art. no. 245}, year = {2024}, doi = {10.1145/3659946}}

@article{S01,
  author  = {Wen, Dingzhu and Zhou, Yong and Li, Xiaoyang
             and Shi, Yuanming and Huang, Kaibin
             and Letaief, Khaled B.},
  title   = {A Survey on Integrated Sensing,
             Communication, and Computation},
  journal = {IEEE Commun. Surveys Tuts.},
  volume  = {27},
  number  = {5},
  pages   = {3058--3098},
  year    = {2025},
  doi     = {10.1109/COMST.2024.3521498}
}

@article{S02, author = {Zhu, Guangxu and Lyu, Zhonghao and Jiao, Xiang and Liu, Peixi and Chen, Mingzhe and Xu, Jie and Cui, Shuguang and Zhang, Ping}, title = {Pushing {AI} to wireless network edge: an overview on integrated sensing, communication, and computation towards {6G}}, journal = {Sci. China Inf. Sci.}, volume = {66}, number = {3}, pages = {130301}, year = {2023}, doi = {10.1007/s11432-022-3652-2}}

@article{S06, author = {Zhao, M. K. and Huang, Y. S. and Li, Xuan}, title = {Federated Learning for {6G}: A Survey From Perspective of Integrated Sensing, Communication and Computation}, journal = {ZTE Commun.}, volume = {21}, number = {2}, pages = {25--33}, year = {2023}, doi = {10.12142/ZTECOM.202302005}}

@article{S11, author = {Zhang, Likang and Du, Qinghe and Lu, Lei and Zhang, Shijiao}, title = {Overview of the Integration of Communications, Sensing, Computing, and Storage as Enabling Technologies for the Metaverse over {6G} Networks}, journal = {Electronics}, volume = {12}, number = {17}, pages = {3651}, year = {2023}, doi = {10.3390/electronics12173651}}

@article{S04,
  author = {Mahyar Nemati and Yun Hee Kim and Jinho Choi},
  journal = {IEEE Access},
  pages = {11772--11788},
  publisher = {Institute of Electrical and Electronics Engineers Inc.},
  title = {Toward Joint Radar, Communication, Computation, Localization, and Sensing in {IoT}},
  volume = {10},
  year = {2022},
  doi = {10.1109/ACCESS.2022.3146830}
}

@article{S03, author = {Dai, Minghui and Li, Yang and Li, Peichun and Wu, Yuan and Qian, Liping and Lin, Bin and Su, Zhou}, title = {A Survey on Integrated Sensing, Communication, and Computing Networks for Smart Oceans}, journal = {J. Sens. Actuator Netw.}, volume = {11}, number = {4}, pages = {70}, year = {2022}, doi = {10.3390/jsan11040070}}

@article{ai_iscc,
  author  = {Wu, Guofang and He, Yejun and Cao, Xiaowen
             and Yuen, Chau},
  title   = {{AI}-Enabled Integrated Sensing, Communication,
             and Computation Survey: Techniques, Status,
             and Perspectives},
  journal = {IEEE Internet Things J.},
  volume  = {12},
  number  = {16},
  pages   = {32676--32700},
  year    = {2025},
  doi     = {10.1109/JIOT.2025.3579940}
}

@article{survey_ioV_iscc, author = {Li, Changle and Dong, Mengyuan and Fu, Yuchuan and Yu, F. Richard and Cheng, Nan}, title = {Integrated Sensing, Communication, and Computation for {IoV}: Challenges and Opportunities}, journal = {IEEE Commun. Surveys Tuts.}, volume = {28}, pages = {1136--1168}, year = {2026}, doi = {10.1109/COMST.2025.3612388}}


%% file: references/taxonomy_cleaned.bib
@article{he2023integrated,
  author  = {He, Yinghui and Yu, Guanding
             and Cai, Yunlong and Luo, Haiyan},
  title   = {Integrated Sensing, Computation, and Communication:
             System Framework and Performance Optimization},
  journal = {IEEE Trans. Wireless Commun.},
  volume  = {23},
  number  = {2},
  pages   = {1114--1128},
  year    = {2024},
  doi     = {10.1109/TWC.2023.3285869}
}

@article{mao2017survey,
  author = {Mao, Yuyi and You, Changsheng and Zhang, Jun and Huang, Kaibin and Letaief, Khaled B},
  journal = {IEEE Commun. Surveys Tuts.},
  number = {4},
  pages = {2322--2358},
  publisher = {IEEE},
  title = {A survey on mobile edge computing: The communication perspective},
  volume = {19},
  year = {2017},
  doi = {10.1109/COMST.2017.2745201}
}

@article{lyu2018multiuser,
  author = {Lyu, Xiao and Tian, Hui and Sengul, Cigdem and Zhang, Ping},
  journal = {IEEE Trans. Veh. Technol.},
  number = {4},
  pages = {3435--3447},
  publisher = {IEEE},
  title = {Multiuser joint task offloading and resource optimization in proximate clouds},
  volume = {66},
  year = {2017}
}
